\documentclass[preprint,superscriptaddress,nofootinbib]{revtex4}
\usepackage{doi}
\usepackage{hyperref}

\usepackage{graphicx, hyperref, amsmath, amssymb, slashed, color, bbm}
\usepackage{subfigure}
\usepackage{multirow}
\usepackage[normalem]{ulem}
\usepackage{verbatim}
\newenvironment{code}{\linespread{0.8}\footnotesize\verbatim}{\endverbatim\normalsize}

\newcommand{\nc}{\newcommand}

\nc{\beq}{\begin{equation}}
\nc{\eeq}{\end{equation}}
\nc{\barray}{\begin{eqnarray}}
\nc{\earray}{\end{eqnarray}}
\nc{\barrayn}{\begin{eqnarray*}}
\nc{\earrayn}{\end{eqnarray*}}
\nc{\bcenter}{\begin{center}}
\nc{\ecenter}{\end{center}}
\nc{\ket}[1]{| #1 \rangle}
\nc{\bra}[1]{\langle #1 |}
\nc{\0}{\ket{0}}
\nc{\mc}{\mathcal}
\nc{\er}[1]{(\ref{eq:#1})}
\nc{\onehalf}{\frac{1}{2}}
\nc{\partialbar}{\bar{\partial}}
\nc{\psit}{\widetilde{\psi}}
\nc{\Tr}{\mbox{Tr}}
\nc{\hc}{\mbox{H.c.}}
\nc{\ev}{\;\mathrm{eV}}
\nc{\mev}{\;\mathrm{MeV}}
\nc{\gev}{\;\mathrm{GeV}}
\nc{\tev}{\;\mathrm{TeV}}

\def\chii0{\chi_i^0}
\def\chij0{\chi_j^0}

\newcommand{\gsim}{\lower.7ex\hbox{$\;\stackrel{\textstyle>}{\sim}\;$}}
\newcommand{\lsim}{\lower.7ex\hbox{$\;\stackrel{\textstyle<}{\sim}\;$}}
\nc{\ttbar}{t\bar t}
\nc{\TAFB}{A_{FB}^t}
\nc{\lepAFB}{A_{FB}^\ell}
\nc{\TAC}{A_{C}^t}
\nc{\Lag}{\mathcal{L}}
\nc{\Proj}{\mathcal{P}}

\begin{document}

\title{Flavor and Collider Signatures of Asymmetric Dark Matter}

\author{Ian-Woo Kim}
\affiliation{Michigan Center for Theoretical Physics, University of Michigan, Ann Arbor, MI 48109, USA}
\affiliation{CERN, Theory Division, 1211 Geneva 23, Switzerland}
\author{Kathryn M. Zurek}
\affiliation{Michigan Center for Theoretical Physics, University of Michigan, Ann Arbor, MI 48109, USA}

\begin{abstract}

We consider flavor constraints on, and collider signatures of, Asymmetric Dark Matter (ADM) via higher dimension operators.   In the supersymmetric models we consider, R-parity violating (RPV) operators carrying $B-L$ interact with $n$ dark matter (DM) particles $X$ through an interaction of the form $W = X^n {\cal O}_{B-L}$, where ${\cal O}_{B-L} = q \ell d^c, u^c d^c d^c,~ \ell \ell e^c$.  This interaction ensures that the lightest ordinary supersymmetric particle (LOSP) is unstable to decay into the $X$ sector, leading to a higher multiplicity of final state particles and reduced missing energy at a collider.  Flavor-violating processes place constraints on the scale of the higher dimension operator, impacting whether the LOSP decays promptly.  While the strongest limitations on RPV from $n-\bar n$ oscillations and proton decay do not apply to ADM, we analyze the constraints from meson mixing, $\mu-e$ conversion, $\mu \rightarrow 3 e$ and $b \rightarrow s \ell^+ \ell^-$.  We show that these flavor constraints, even in the absence of flavor symmetries, allow parameter space for prompt decay to the $X$ sector, with additional jets and leptons in exotic flavor combinations. We study the constraints from existing 8 TeV LHC SUSY searches with {\it (i)} 2-6 jets plus missing energy, and {\it (ii)} 1-2 leptons, 3-6 jets plus missing energy, comparing the constraints on ADM-extended supersymmetry with the usual supersymmetric simplified models.  

\end{abstract}
\preprint{MCTP-13-29}
\preprint{CERN-PH-TH/2013-230}

\maketitle

\tableofcontents

\section {Introduction}

The notion that Dark Matter (DM) may be related to the baryon asymmetry originates from a time almost as early as the weakly interacting massive particle (WIMP) paradigm itself \cite{Nussinov:1985xr,Gelmini:1986zz}.  In these models, a mechanism sets the DM and baryon asymmetries such that $n_X - n_{\bar X} \sim n_b - n_{\bar b}$, where $n_X,~n_{\bar X}$ are the DM and anti-DM number densities, and $n_b,~n_{\bar b}$ are the baryon and anti-baryon asymmetries.  Since the ratio of DM to baryon densities is observed to be $\rho_{DM}/\rho_B \sim 5$, this suggests $m_X \sim 5 m_p \simeq 5$ GeV, where $m_X$ is the DM mass and $m_p$ is the proton mass.  Thus in these models, the natural mass scale for the DM is around 1-10 times the proton mass, significantly below the weak scale.  

The idea that the DM and baryon densities have a common mechanism setting their densities is a simple and compelling framework.  The challenge for a model of DM that relates the DM and baryon asymmetries is, however, that it must satisfy the many requirements from our observations of the weak scale and below.  Many of the earliest models, especially those making use of electroweak sphalerons \cite{Barr:1990ca,Kaplan:1991gb,Gudnason:2006yj}, had become highly constrained by these observations, particularly those from LEP, making models of DM relating the DM and baryon asymmetries observationally less than compelling.  

Employing ideas from hidden sector model building \cite{Strassler:2006im}, the Asymmetric Dark Matter (ADM) paradigm \cite{Kaplan:2009ag} showed how to evade these constraints by making use of higher dimension operators ${\cal O}_{B-L}$ which carry no Standard Model (SM) gauge charge but carry $B-L$.  These operators are connected to the DM sector via higher dimension operators
\beq
{\cal O}_{ADM} = \frac{{\cal O}_{B-L} {\cal O}_X}{M^{n+m-4}},
\label{ADMops}
\eeq 
where ${\cal O}_{B-L}$ has dimension $m$ and ${\cal O}_X$ has dimension $n$.  The operators in Eq.~\ref{ADMops} share a primordial matter-anti-matter asymmetry between the visible and DM sectors, realizing the relationship $n_X - n_{\bar X} \sim n_b - n_{\bar b}$.  For a review and list of references of DM models employing the higher dimension operators, see \cite{Zurek:2013wia}.

ADM can be embedded within supersymmetry (SUSY), which stabilizes the ADM particle via $R$-parity, and limits the types of operators in the superpotential.  The simplest (lowest dimension) superpotential operators for ${\cal O}_{B-L}$ are the R-parity violating (RPV) operators 
\beq
W_{B-L} = \ell H,~~u^c d^c d^c,~~q \ell d^c,~~~\ell \ell e^c,
\label{RPV} 
\eeq
where $\ell$ is a SM lepton doublet, $H$ the Higgs doublet, $u^c,~d^c$ right-handed anti-quarks, $e^c$ a right-handed charged anti-lepton, and  $q$ is a quark doublet.  The simplest form of superpotential operators for ${\cal O}_X$ is $X$, so that the simplest ADM interactions take the form
\beq
W_{\rm ADM} = X \ell H,~~\frac{X u_i^c d_j^c d_k^c}{M_{ijk}},~~\frac{X q_i \ell_j d_k^c}{M_{ijk}},~~~\frac{X \ell_i \ell_j e_k^c}{M_{ijk}}, 
\label{basicADM}
\eeq
where now we have explicitly included a flavor index $i,~j,~k$ on the generic scale of the operator $M$.

\begin{figure}[t]
\centering
\subfigure[]{\includegraphics[width=5cm]
{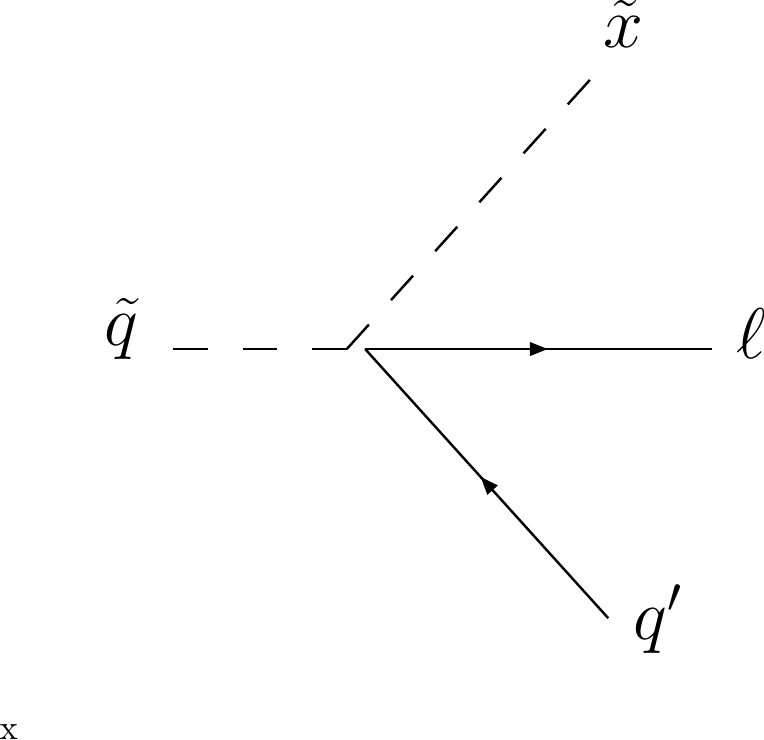}}
\qquad\qquad
\subfigure[]{\includegraphics[width=5cm]{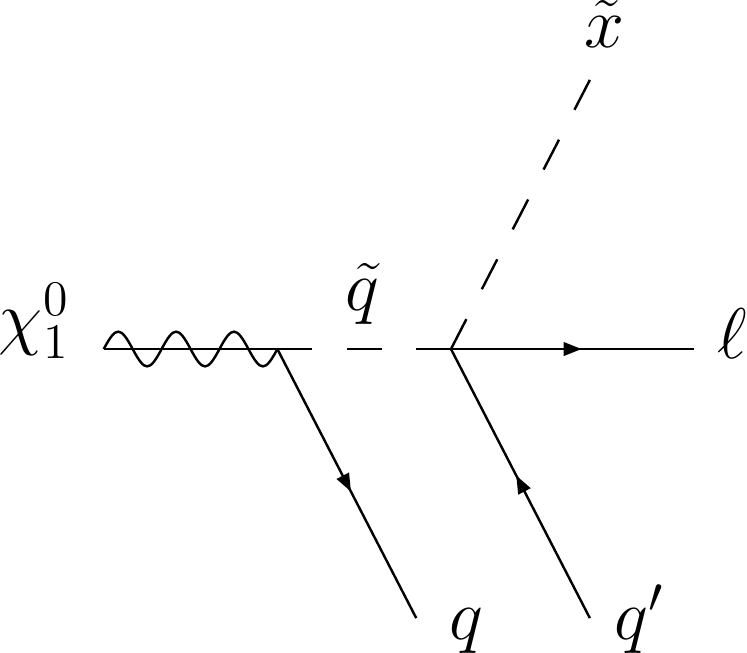}}
\caption{\label{fig:feynman_xqld_decay}  Decay of a squark LOSP directly through the interaction of Eq.~(\ref{basicADM}), and decay of a neutralino LOSP through an off-shell squark for $q \ell d^c$ models. Here, the quark flavors $q$ and $q'$ are generically different. $\tilde{x}$ denotes the scalar component of the ADM supermultiplet $X$. Decay of a slepton LOSP and a neutralino LOSP through an off-shell squark is also given by the same diagrams trading a squark and a lepton with a slepton and a quark, respectively. } 
\end{figure}

These interactions are centrally important for the collider phenomenology of ADM-extended SUSY models.  The interactions in Eq.~(\ref{basicADM}) induce decay of the lightest ordinary supersymmetric particle (LOSP) to the DM particle, through the processes shown in Fig.~\ref{fig:feynman_xqld_decay}.  This implies that, in comparison to the Minimal Supersymmetric Standard Model (MSSM), the missing energy is reduced while the multiplicity of final state particles increases, so that experimental sensitivity to ADM models can be very different in SUSY searches at the LHC. 
A number of theories, such as Hidden Valleys \cite{Strassler:2006im,Strassler:2006qa}, MeV DM \cite{Hooper:2008im}, RPV \cite{Graham:2012th} and Stealth SUSY \cite{Fan:2011yu,Fan:2012jf}, have already aimed to evade SUSY constraints by reducing the missing energy and increasing the number of final state particles.  While ADM models have similar structure in their collider signatures, they also have a potentially wider range of flavor signatures. 

Whether such signatures are realized at a collider depends on whether the LOSP is unstable to decay to the $X$ sector before the LOSP exits the detector.  The lifetime of the LOSP is set by its nature ({\it e.g.} squark, neutralino or slepton), by the supersymmetric spectrum, and, most importantly, by the scale $M$ of the operator.  The scale $M$ can be strongly constrained by flavor physics, in a way similar to RPV.
Taken alone, without additional flavor structure, the RPV operators in Eq.~(\ref{RPV}) are known to have disastrous effects in, {\it e.g.}, proton decay and neutron--anti-neutron ($n-\bar{n}$) oscillations \cite{Zwirner:1984is}.  

There are, however, several important differences between ADM operators and RPV operators.  First of all, with the presence of ${\cal O}_X$, $R$-parity is no longer violated, if the operator ${\cal O}_X$ itself carries $R$-parity of -1.  This new R-parity stabilizes the lightest R-parity odd scalar, $\tilde{x}$, of supermultiplet $X$.  Second, DM now effectively carries baryon or lepton number, so that globally $B$ and $L$ are not violated.  That forbids $n-\bar{n}$ oscillations as well as proton decay (when the $X$ fermion is heavier than the proton).  For certain types of $X$ sectors, the DM can induce proton decay, but it must be catalyzed by the DM, and for this to happen frequently enough to be observable, the scale $M$ must be quite low, around a TeV \cite{Davoudiasl:2010am,Davoudiasl:2011fj}.  Thus the worst of the usual constraints on RPV is lifted for ADM.  

Depending on the flavor structure of the model and the UV completion, however, the scales $M_{ijk}$ in Eq.~(\ref{basicADM}) are still constrained by meson oscillations, by flavor changing processes such as $b \rightarrow s \ell^+ \ell^-$, and by various types of lepton flavor violation such as $\mu \rightarrow e$ conversion and $\mu \rightarrow 3 e$.   

The flavor structure, and the corresponding constraints on the scale of the operator, thus has important implications for the collider signatures of ADM.   As we will see, for example, the lifetime of a pure Bino neutralino at the LHC through the operator $X q \ell d^c$ and an intermediate right-handed $d$-squark is roughly 
\beq
c\tau \sim (200 \mbox{~mm}) \times \left(\frac{M}{100 \mbox{ TeV}}\right)^2 \left(\frac{m_{\tilde{q}}}{500 \mbox{ GeV}}\right)^4 \left(\frac{100 \mbox{ GeV}}{m_{\chi^0}}\right)^{7},
\label{lifetime}
\eeq 
where $m_{\tilde{q}}$ is the mass of the intermediate squark and $M$ is the scale of the $X q \ell d^c$ operator.
Depending on the constraints on $M$ and $m_{\tilde{q}}$, $\chi^0$ may be collider stable though cosmologically unstable.  Therefore, it is important to consider constraints from a displaced secondary vertex search for generic ADM models. Previously, some lifetime estimates have been made using naive dimensional analysis~\cite{Kaplan:2009ag,Chang:2009sv}, but it is desirable to refine the displaced vertex analysis.

The goal of this paper is to study the flavor structure and constraints on ADM and its implications for collider searches for SUSY.  We compute the flavor constraints on the scale $M$ of the operator, relate these constraints to the lifetime of the LOSP, and derive constraints on the ADM-extended MSSM from standard SUSY searches.  Unlike many recent efforts to lift constraints on RPV  operator  coefficients through flavor structures \cite{Csaki:2011ge,Ruderman:2012jd}, we will assume no flavor symmetry, but rather examine the range of possible signatures that could arise in a general flavor structure. Note that the flavor constraints we place on DM in ADM models will have applications to many models with flavorful DM \cite{Kile:2011mn,Kamenik:2011nb,Agrawal:2011ze,Kumar:2013hfa,Kile:2013ola}, because the UV completion of the ADM models we consider contain some of the same interactions. 

The outline of our paper is as follows.  In Sec.~\ref{sec:flavor}, we carry out a thorough analysis of the flavor structure of all three ADM models ($q \ell d^c,~u^c d^c d^c,~\ell \ell e^c$) for the simplest UV completions (except for the $\ell H$ model which is essentially a model with a right-handed neutrino).   We extract constraints on the general scale $M$ of the ADM operator from various flavor processes. 
We highlight the results in Sec.~\ref{sec:flavor}, and provide details of our flavor analysis in Appendix A.   Next,  in Sec.~\ref{sec:displaced}, we examine the implications of the flavor constraints on $M$ for the LOSP lifetime at the LHC.  We give details in Appendix B of exact expressions for the lifetime of the LOSP through three and four body decays (for which Eq.~(\ref{lifetime}) is only an approximate proxy).   We show that prompt, displaced, and collider stable signatures are all possible consistent with flavor constraints, even in the absence of a flavor symmetry.  Then, in Sec.~\ref{sec:LHC}, we carry out a detailed analysis of the constraints on this model from existing searches assuming prompt LOSP decays.  We compare the constraints in the standard SUSY searches against those for ADM for 8 TeV LHC analyses utilizing {\it (i)} 0 lepton plus 2-6 jets plus missing energy, and {\it (ii)} 1-2 leptons plus 3-6 jets plus missing energy.  We thus lay firm groundwork for a more exhaustive analysis of SUSY ADM signatures at the LHC in the future, before concluding in Sec.~\ref{sec:conclusion}.

\section{Operators and Their Flavor Constraints}
\label{sec:flavor}

We begin our study by discussing the UV completions for each higher dimension operator in Eq.~(\ref{basicADM}), assessing the impact from flavor constraints  on the scale $M$ of the operators in $W_{\rm ADM}$.
As we will see in Section \ref{sec:displaced}, a careful computation of the lifetime of the LOSP shows that only when the scale of the operator in Eq.~(\ref{basicADM}) is $M \gtrsim 100 \mbox{ TeV}$ will the decay of a neutralino LOSP be collider stable, or displaced, at the LHC (though the details depend on the supersymmetric spectrum of the model). 
 Thus for phenomenological study of prompt decays at the LHC, we are mostly interested in flavor constraints that require $M \gtrsim m_M/\lambda^2 \gtrsim 10-100$ TeV, where $M$ is determined from a UV completion by $m_M$, the mass of the mediator being integrated out to generate the operator, and $\lambda^2$, a product of couplings of that mediator to SM states and the DM.  
We summarize the results for constraints derived from meson oscillations, $\mu-e$ conversion, $\mu \rightarrow 3 e$ and $b \rightarrow s \ell^+ \ell^-$ including $B_s \rightarrow \ell^+ \ell^-$ in this section, and refer the reader to Appendix A for the details of our computations.  
$K-\bar{K}$ mixing provides, in many cases (except for the $X \ell \ell e^c$ operator), the strongest constraint.
 
We emphasize that we take a conservative approach without assuming a flavor structure, since there are many ways to relax flavor constraints by imposing a flavor structure on the model.  
For example, since both meson oscillations and lepton flavor constrain products of couplings to different generations, if the couplings to one of the generations is much larger than to the other generations the constraints will be considerably relaxed.  In the case of meson oscillations the usefulness of this change is somewhat limited, however, since rotating from a flavor basis to the mass basis will induce couplings to the other generations which are generically not small unless the flavor and mass bases are closely aligned (which would constitute a tuning in the absence of a flavor symmetry). In such cases, a flavor symmetry can alleviate these constraints. Therefore, our results on flavor constraints and the corresponding discussion of displaced vertices from LOSP decay must be taken as conservative.  Even without the assumption of a flavor symmetry, we will find that prompt flavor violating decays of the LOSP are still possible at the LHC. 
In addition, deriving constraints in the absence of a flavor symmetry leaves open the interesting possibility for exotic flavor signatures at the LHC.

\subsection{$X q \ell d^c$}

We begin by analyzing the $X q \ell d^c$ operator, assuming only one flavor of DM:
\beq
W_{ADM} = \frac{X q_i  \ell_j  d_k^c }{M_{ijk}}.
\label{qld}
\eeq
There are three UV completions at the renormalizable level: 
\barray
W^{(D)} &=& \lambda^i_{XD} X d_i^c D + \lambda^{i j}_{D} D^c q_i \ell_j + m_D D D^c, 
\label{eq:modelDinQLD} \\
W^{(L)} &=& \lambda^i_{X L} X \ell_i L^c + \lambda^{i j}_L L q_i d_j^c  + m_L L L^c 
\label{eq:modelLinQLD} \\
W^{(Q)} &=& \lambda^i_{XQ} X q_i Q^c + \lambda^{i j}_Q Q \ell_i d_j^c + m_Q Q Q^c, 
\label{eq:modelQinQLD}
\earray
where $i,~j,~k$ are generation indices.  Note that the effective scale of $W$ is determined by 
\begin{eqnarray}
M_{ijk} = m_M/(\lambda_{X M}^i \lambda_M^{j k}) \label{eq:effscale}
\end{eqnarray}
 for a mediator $M = D,~L,~Q$ with mass $m_M$. This relation also holds for the UV completion, given an appropriate mediator $M$, of the $X u^c d^c d^c$ and $X \ell \ell e^c$ operators, as will be shown in the next subsections.  In these expressions, as throughout this paper, a lowercase letter indicates a SM field and an uppercase letter represents exotic heavy states, which are integrated out to generate the higher dimension operator. Note that we define fields in the mass eigenstate basis here.  
For simplicity, we consider only one flavor of DM, 
 as well as a single pair of heavy mediator fields $(D, D^c)$, $(L, L^c)$ or $(Q, Q^c)$.  If we extend this simple model with multiple DM flavors or multiple mediator states, we have more freedom in assigning a flavor structure that could lift some of the flavor constraints that we study, but we do not pursue this direction.  We also assume only one of the UV completions is dominant, and we will label the UV completion by the state which is being integrated out.  Our results do not qualitatively change if we consider mixed UV completions.

We consider the constraints on $X q \ell d^c$ derived from $K-\bar K,~D-\bar D,~B-\bar B$ mixing, $\mu-e$ conversion, $\mu \rightarrow 3 e$ and $b \rightarrow s \ell^+ \ell^-$ in turn.   

\begin{figure}
\centering
\subfigure[]{\includegraphics[width=2.5cm]{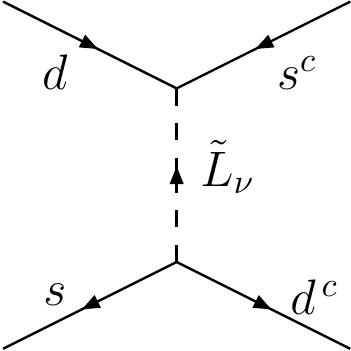}}
\quad
\subfigure[]{\includegraphics[width=2.5cm]{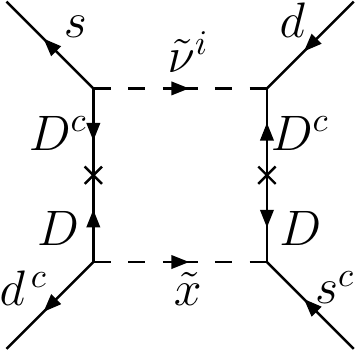}}
\quad
\subfigure[]{\includegraphics[width=2.5cm]{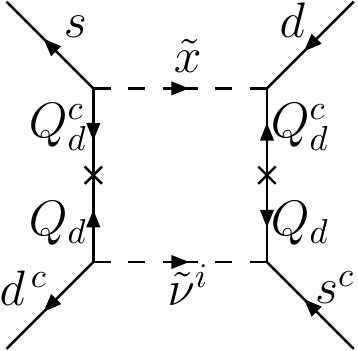}}
\quad
\subfigure[]{\includegraphics[width=2.5cm]{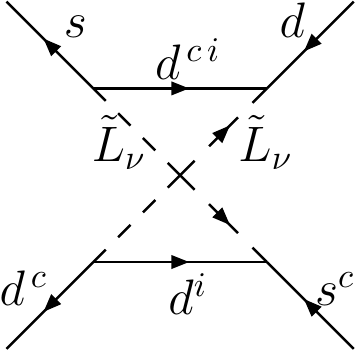}}
\quad
\subfigure[]{\includegraphics[width=2.5cm]{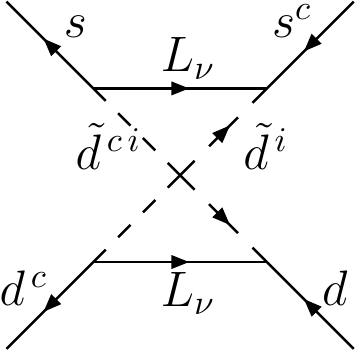}}

\subfigure[]{\includegraphics[width=2.5cm]{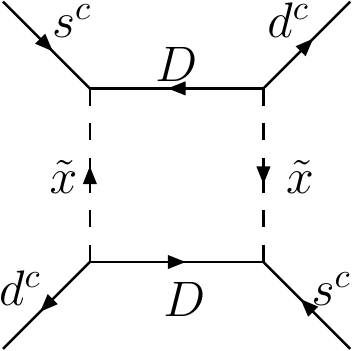}}
\quad
\subfigure[]{\includegraphics[width=2.5cm]{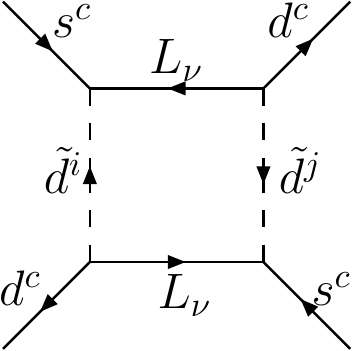}}
\quad
\subfigure[]{\includegraphics[width=2.5cm]{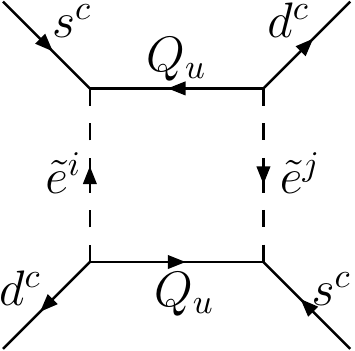}}

\subfigure[]{\includegraphics[width=2.5cm]{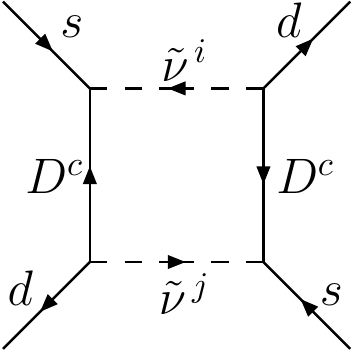}}
\quad
\subfigure[]{\includegraphics[width=2.5cm]{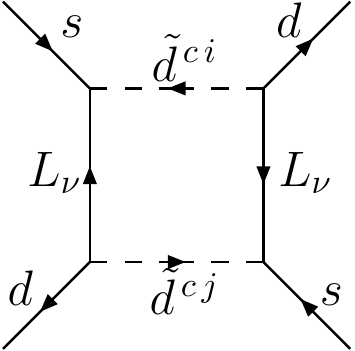}}
\quad
\subfigure[]{\includegraphics[width=2.5cm]{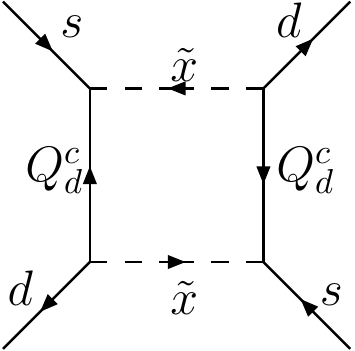}}

\caption{ \label{fig:meson mixing} Diagrams contributing to $K - \bar{K}$ mixing in $X q \ell d^c$ models. Diagrams (a) - (e) contribute to $(\bar{s}_R d_L)(\bar{s}_L d_R)$.  Diagrams (f) - (h) contribute to 
$(\bar{s}_R \gamma^\mu d_R)^2$. Diagrams (i) - (k) contribute to $(\bar{s}_L \gamma^\mu d_L)^2$. For $(\bar{s}_{L,R} \gamma^\mu d_{L,R})^2$, we only show a representative diagram for each UV completion.   
Here, we use 2-component spinor notation to reduce confusion. }
\end{figure}

\subsubsection{Meson Mixing}

Both tree and loop level processes give rise to meson mixing through the UV completions for the operator in Eq.~(\ref{qld}).  Sample processes are shown in Fig.~\ref{fig:meson mixing}.  While the tree level processes in principle give rise to a stronger constraint on the mediator mass, they do not constrain the DM coupling $\lambda_{XM}$ to the UV particle $M$, nor do they constrain all UV completions (the $D$ and $Q$ UV completions are untouched by the tree level constraint).  An exhaustive compilation of the couplings constrained by $K-\bar K$, $D-\bar D$ and $B-\bar B$ mixing is given in Table~\ref{table: flavor constraints} in Appendix A; we highlight the conclusions here. 

Meson mixing is most strongly constraining for the operator $(\bar{s}_R d_L)(\bar{s}_L d_R)/\Lambda^2$, where $K-\bar K$ mixing gives $\Lambda \gtrsim 2 \times 10^4$ TeV \cite{Isidori:2010kg}.  For the $L$ UV completion, Fig.~\ref{fig:meson mixing}a will generate $K-\bar{K}$ mixing at tree level,
$
\frac{\lambda^{12}_L \lambda^{12}_L}{m_{\tilde{L}}^2}(\bar{s}_R d_L)(\bar{s}_L d_R). 
$
For $B_d$ and $B_s$ meson mixing, we also have similar tree-level diagrams generating left-right operators, which have stringent constraints as summarized in Appendix A, Table~\ref{table: flavor constraints}.  Nevertheless, we note that only the $L$ UV completion is constrained for a very limited combination of couplings by meson mixing at tree level. 

Loop diagrams, on the other hand, probe a wider array of flavor-changing couplings, since any of the superpartner flavors may appear in the loop.  In some cases, they also probe precisely the combination of couplings that enters into $M_{ijk}$ which ultimately determines whether decays are prompt or displaced at the LHC.
Contributions to ${\cal O}_A = (\bar{s}_R \gamma^\mu d_R)^2/\Lambda_A^2$, $(\bar{s}_L \gamma^\mu d_L)^2/\Lambda_A^2$ and ${\cal O}_B (\bar{s}_L d_R) (\bar{s}_R d_L)/\Lambda_B^2$ occur, and the loop functions which characterize the constraints are detailed in Appendix A.  In the limit that the fermions and the scalars in the loop have a common mass $m_F$ and $m_\phi$ respectively, the amplitude simplifies considerably:
\begin{eqnarray}
\frac{1}{\Lambda_A^2} \sim \frac{\lambda^4}{64 \pi^2}\left(\frac{m_F^2+m_\phi^2}{(m_F^2 - m_\phi^2)^2} - \frac{2 m_F^2 m_\phi^2}{(m_F^2 - m_\phi^2)^3}\log\left(\frac{m_F^2}{m_\phi^2}\right)\right) \\ \nonumber
\frac{1}{\Lambda_B^2} \sim\frac{\lambda^4 m_F^2}{16 \pi^2}\left(-\frac{2}{(m_F^2 - m_\phi^2)^2} + \frac{m_F^2 + m_\phi^2}{(m_F^2 - m_\phi^2)^3}\log\left(\frac{m_F^2}{m_\phi^2}\right)\right),
\end{eqnarray}  
where $\lambda^2$ represents the appropriate combination of couplings shown in Table~\ref{table: flavor constraints} for amplitudes having structure corresponding to operators $A$ and $B$.   The constraint on $\Lambda_B \gtrsim 2 \times 10^4 \mbox{ TeV}$ is strongest and corresponds to a limit on the parameters of the UV completion $m_M/\lambda^2 \gtrsim 1000$ TeV.  While it is not a universal constraint on all the couplings to all generations, as can be seen in Table~\ref{table: flavor constraints}, it is the most severe constraint on $M_{ijk}$.

\subsubsection{$\mu-e$ conversion, $\mu \rightarrow 3 e$, $B_s \rightarrow \ell^+ \ell^-$ and $b \rightarrow s \ell^+ \ell^-$}

\begin{figure}
\centering
\subfigure[]{\includegraphics[width=2.5cm]{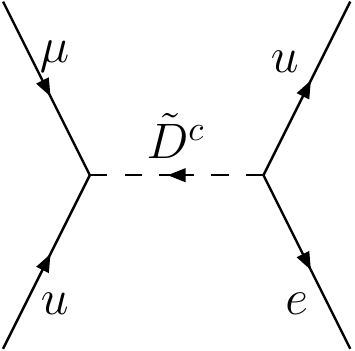}}
\quad
\subfigure[]{\includegraphics[width=2.5cm]{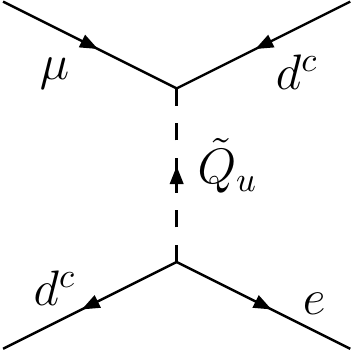}}
\quad
\subfigure[]{\includegraphics[width=2.5cm]{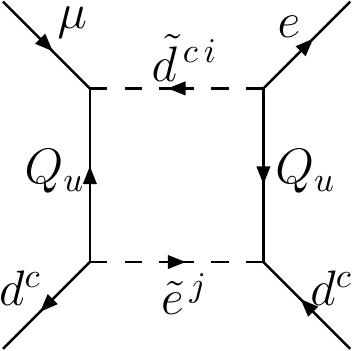}}
\quad
\subfigure[]{\includegraphics[width=2.5cm]{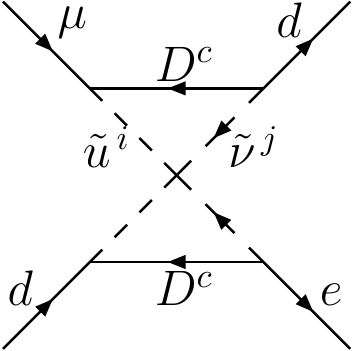}}
\quad
\subfigure[]{\includegraphics[width=2.5cm]{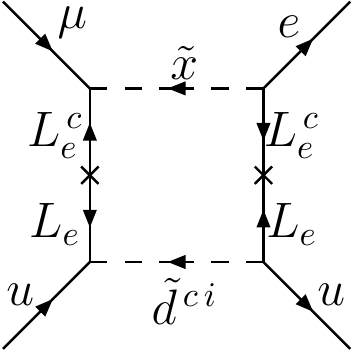}}
\quad

\caption{ \label{fig:mu e conversion} Diagrams contributing to $\mu-e$ conversion in $X q \ell d^c$ models. We show only a typical one-loop diagram for each UV completion. For a more complete set of diagrams, see Appendix A.}
\end{figure}

Lepton flavor violation may also constrain the UV completions of the ADM operators with heavy states $Q,~L,~D$.
 The strongest constraints are derived from $\mu-e$ conversion, and are summarized in Appendix~\ref{mu-e conversion} in Table~\ref{table: mu e conversion}.  At tree level, the $D$ UV completion and the $Q$ UV completion have contributions to $\mu-e$ conversion through diagrams shown in Fig. \ref{fig:mu e conversion}, resulting in the operators 
\barray
&&-\frac{1}{2} \frac{\lambda^{11}_D \lambda^{12}_D}{m^2_{\tilde{D}}} 
( \bar{e}_L \gamma^\rho \mu_L) (\bar{u}_L \gamma_\rho u_L )\,, \qquad \mbox {for $(D)$,} 
\label{eq:mu-e_conversion_at_tree}\\
&&-\frac{1}{2} \frac{\lambda^{11}_Q \lambda^{21}_Q}{m^2_{\tilde{Q}}}
( \bar{e}_L \gamma^\rho \mu_L) (\bar{d}_R \gamma_\rho d_R )\,, \qquad \mbox {for $(Q)$,} \nonumber
\earray
where we rearrange spinors using the Fierz identities, $(\bar{e}_L u^c_R)(\bar{u}^c_R \mu_L) = -\frac{1}{2} (\bar{e}_L \gamma^\rho \mu_L)(\bar{u}_L \gamma_\rho u_L)$ and $(\bar{e}_L d_R)(\bar{d}_R \mu_L) = -\frac{1}{2} (\bar{e}_L \gamma^\rho \mu_L) (\bar{d}_R \gamma_\rho d_R)$.  
The branching ratio of $\mu-e$ conversion is obtained for the various nuclei, and can be 
translated into the value for Al, ${\rm Br}_{\mu N \rightarrow e N} (Z=13) \leq 10^{-12}$ (see Appendix~\ref{mu-e conversion}).  We then derive the constraint 
\barray
\frac{m_{\tilde{D}}}{\sqrt{\lambda_D^{11} \lambda_D^{12}}} \geq 290~{\rm TeV},\qquad
\frac{m_{\tilde{Q}}}{\sqrt{\lambda_Q^{11} \lambda_Q^{21}}} \geq 210~{\rm TeV}.
\earray 
The number of coefficients constrained by the tree level process is, however, limited.
On the other hand, loop level contributions, as also shown in Fig.~\ref{fig:mu e conversion}, constrain all three UV completions for various combinations of couplings. These are detailed exhaustively in Table~\ref{table: mu e conversion} in Appendix~\ref{mu-e conversion}.  At loop level, the constraints on $M$ from $\mu-e$ conversion are at the level of 10-100 TeV, and therefore not important from the point of view of displaced vertices at the LHC.  $\mu \rightarrow e \gamma$ and $\mu \rightarrow 3 e$ appear only at loop level and also are not strong constraints.  We detail the constraints from $\mu \rightarrow 3 e$ in Appendix A, Table~\ref{table:mu_to_eee}.  While not significant for the $q \ell d^c$ model, $\mu \rightarrow 3 e$ will become important for the $\ell \ell e^c$ model.

\begin{figure}
\centering
\subfigure[]{\includegraphics[width=2.5cm]{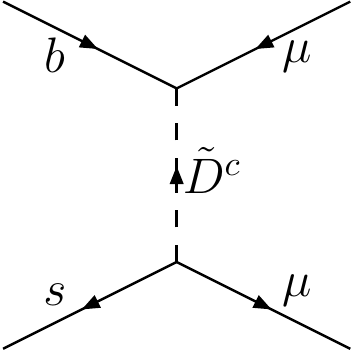}}
\quad
\subfigure[]{\includegraphics[width=2.5cm]{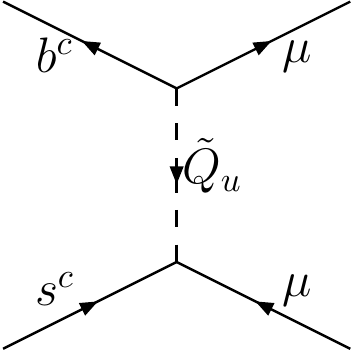}}
\quad
\subfigure[]{\includegraphics[width=2.5cm]{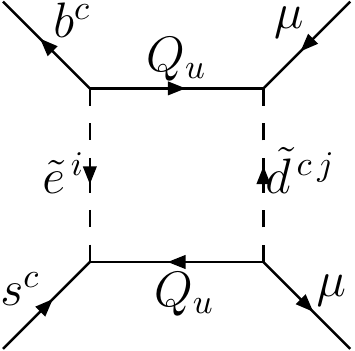}}
\quad
\subfigure[]{\includegraphics[width=2.5cm]{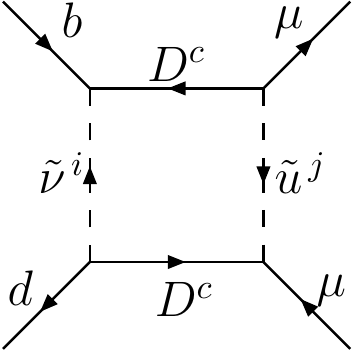}}
\quad
\subfigure[]{\includegraphics[width=2.5cm]{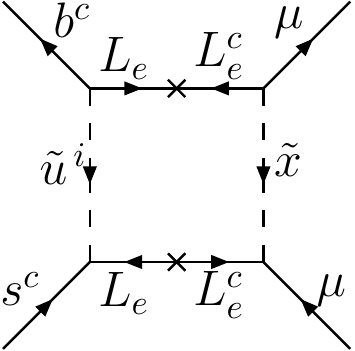}}

\caption{\label{fig:Bstomumu} Diagrams contributing to $B_s \rightarrow \mu^+\mu^-$ in the $q \ell d^c$ model. We show only a representative one-loop diagram for each UV completion.  The additional box diagrams are shown in Fig. \ref{fig:additional_box_diagram_for_b_to_s_transition}. For other $b$-$s$ transitions such as $b \rightarrow s \ell^+ \ell^-$, one can easily obtain contributing Feynman diagrams by properly changing the external states in the above diagrams. } 
\end{figure}

At tree level, we also have contributions to $b-s$ conversion with a pair of leptons, as for example in $B_s \rightarrow \mu^+ \mu^-$.  Processes are shown in Fig.~\ref{fig:Bstomumu} (a) and (b),
\barray
&& -\frac{1}{2} \frac{\lambda^{32}_D \lambda^{22}_D}{m^2_{\tilde{D}}} 
( \bar{\mu}_L \gamma^\rho \mu_L )(\bar{s}_L \gamma_\rho b_L)\,, \qquad \mbox {for $(D)$,} \\
&& -\frac{1}{2} \frac{\lambda^{23}_Q \lambda^{22}_Q}{m^2_{\tilde{Q}}}
( \bar{\mu}_L \gamma^\rho \mu_L ) ( \bar{s}_R \gamma_\rho b_R)\,, \qquad \mbox {for $(Q)$,} \nonumber 
\earray
where we again rearrange spinors using the Fierz identities, similarly as in Eq.~(\ref{eq:mu-e_conversion_at_tree}).  Currently, the experimental bound for $B_s \rightarrow \mu^+ \mu^-$ is ${\rm Br}(B_s \rightarrow \mu^+ \mu^-) < 4.2\times 10^{-9}$ \cite{ATLAS:2012ola,Aad:2012pn,Chatrchyan:2012rga,Aaij:2012ac}\footnote{Recently, the first evidence of $B_s \to \mu^+ \mu^-$ is found at LHCb and the result is consistent with SM\cite{Aaij:2012nna}, but in our analysis, we use the combined constraints only from the LHC data at 7 TeV\cite{ATLAS:2012ola}.}.  

We can also constrain the scale of four-fermion effective operators through the process $b \rightarrow s \ell^+ \ell^-$ \cite{Altmannshofer:2012az}. The tree level constraints lead to (see Appendix A, Table~\ref{table:b_s_transition_constraint} for details): 
\barray
& & m_D / \sqrt{\lambda^{3\ell}_D \lambda^{2\ell}_D} > 32~{\rm TeV} ~~\mbox{for strongest,} \quad  > 11~{\rm TeV}~~\mbox{for weakest,} \nonumber \\
& & m_Q / \sqrt{\lambda^{2\ell}_Q \lambda^{2\ell}_Q } > 45~{\rm TeV} ~~\mbox{for strongest,} \quad > 11~{\rm TeV}~~\mbox{for weakest,}
\earray
where $\ell=1,~2$ denotes electron and muon, respectively, for the lepton final states, and we show both the strongest constraint and the weakest constraint since the constraint varies depending on the sign of the coupling, and whether it is real or imaginary. 
While only the $D$ and $Q$ UV completions contribute at tree level, all UV completions contribute at one loop, as shown in Fig. \ref{fig:Bstomumu} (c), (d) and (e), though the loop suppression implies that this constraint will be weak. The details can be found in Appendix \ref{sec:appendix_b_s_transition}.

\subsubsection{Summary of Constraints for $X q \ell d^c$}
 
 There are many combinations of couplings constrained in Tables~\ref{table: flavor constraints}-\ref{table:mu_to_eee}, but it is important to see the over-arching patterns.
 \begin{itemize}
 \item The strongest constraints are on the operator $(\bar s_L d_R)(\bar d_L s_R)/\Lambda^2$, which give rise roughly to a constraint  $M \gtrsim 1000 \mbox{ TeV}$ for the UV completions via $D$ and $Q$.  Since $M$ is the quantity which enters into the lifetimes in Fig.~\ref{fig:feynman_xqld_decay}, it directly enters into the discussion of displaced vertices in the next section. These constraints can be eased and $M$ lowered if one or both of the quarks in the decay of $\tilde{\chi}_0 \rightarrow q q \ell$ is third generation.  Note that the constraints are equally strong on all lepton flavors.
 \item The UV completions via $L$ are less constrained.  The strongest constraint on $M$ is derived from the geometric mean of the $K-\bar K$ mixing and $\mu \rightarrow e \gamma$, which results in $M \gtrsim 10 \mbox{ TeV}$.  The constraints can be relaxed somewhat if the lepton in  $\tilde{\chi}_0 \rightarrow q q \ell$ is $\tau$ or one of the quarks is third generation. 
 \end{itemize}

In Sec.~\ref{sec:displaced} we give precise formulae for the LOSP lifetime as a function of $M$, thus mapping the flavor constraints onto displaced vertex signatures for ADM.  Before examining the collider signatures, however, we complete our discussion of the flavor constraints with an examination of the other ADM operators.

\subsection{$X u^c d^c d^c$}

Considering next the $X u^c d^c d^c$ operator, the UV completions for this operator are
\barray
W^{(U)} &=& \lambda^i_{XU} X u_i^c U + \frac{1}{2} \lambda^{i j}_U U^c d_i^c d_j^c + m_U U U^c, \\
W^{(D)} &=& \lambda^i_{XD} X d_i^c D + \lambda^{ij}_D D^c u_i^c d_j^c + m_D D D^c,\nonumber 
\earray
where $i,~j$ are flavor indices.

Similar to the case of $X q \ell d^c$, the combinations of the couplings which are constrained are shown in Table~\ref{table: flavor constraints}.  
Because all fields involved are right-handed, the strongest constraint from $(\bar s_L d_R)(\bar d_L s_R)$ is eliminated, and more modest constraints on $m_M/\lambda^2$ between 10 and 100 TeV result.   In addition, when the operator is completed via $D$, $M_{i j k} = m_D / \lambda_{XD}^i \lambda_D^{j k}$, which enters into the LOSP lifetime, is directly constrained, though only in particular generational combinations.  Note in addition that $\lambda_{XU}^3$ is the only coupling which remains unconstrained.  A variety of other processes from $q_i \rightarrow q_j q \bar q$ meson decays will constrain $\lambda^2/m_M^2$, similar to $\mu-e$ conversion or $B_s \rightarrow \mu^+ \mu^-$ constraints on the $q \ell d^c$ model.  These constraints are, however, rather weak.  Since no constraints on $M_{ijk}$ from exceed 100 TeV, prompt decays of the LOSP are unconstrained by flavor.

\subsection{$X \ell \ell e^c$}

Lastly, we consider the UV completions for $X \ell \ell e^c$,
\barray
W^{(L)} &=& \lambda^i_{XL} X \ell_i L^c + \lambda^{i j}_L L \ell_i e_j^c + m_L L L^c, 
\label{eq:modelLinXLLE}\\
W^{(E)} &=& \lambda^i_{XE} X e^c_i E + \frac{1}{2} \lambda^{i j}_E E^c \ell_i \ell_j + m_E E E^c, 
\label{eq:modelEinXLLE} 
\earray
where $i$ and $j$ are again flavor indices.

\subsubsection{$\mu \rightarrow 3 e$}

\begin{figure}
\centering
\subfigure[]{\includegraphics[width=3cm]{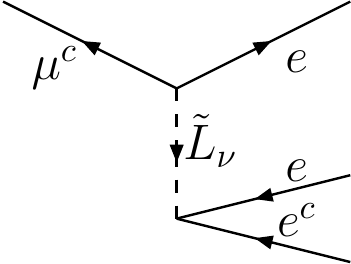}}
\quad
\subfigure[]{\includegraphics[width=3cm]{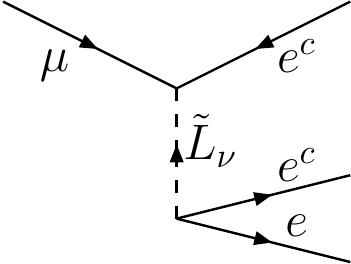}}
\quad
\subfigure[]{\includegraphics[width=2.5cm]{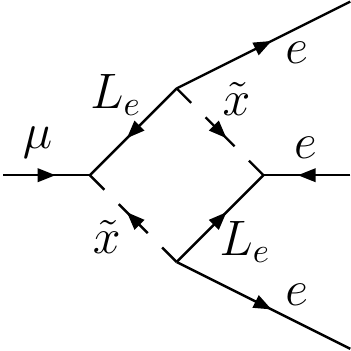}}
\quad
\subfigure[]{\includegraphics[width=2.5cm]{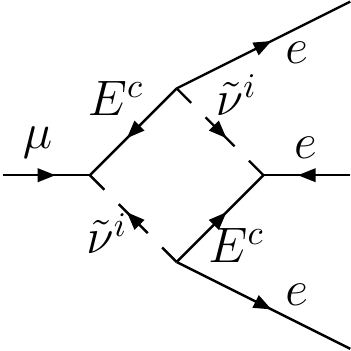}}

\caption{\label{fig:tree_level_mu_to_eee} Diagrams contributing to $\mu \rightarrow e^+ e^- e^-$ in $\ell \ell e^c$ models. We show only a typical diagram for loop contributions for each UV completion. }
\end{figure}

The $L$ UV completion of the $\ell \ell e^c$ model has tree level contribution to the $\mu \rightarrow e e e$ process as shown in Fig. \ref{fig:tree_level_mu_to_eee}, which leads to effective operators:
\barray
- \frac{1}{2} \frac{\lambda^{11}_L \lambda^{12}_L}{m^2_{\tilde{L}}} 
(\bar{e}_L \gamma^\rho e_L )(\bar{e}_R \gamma_\rho \mu_R) 
- \frac{1}{2} \frac{\lambda^{21}_L \lambda^{11}_L}{m^2_{\tilde{L}}}
(\bar{e}_R \gamma^\rho e_R )(\bar{e}_L \gamma_\rho \mu_L)\,.
\earray
The branching ratio ${\rm Br} (\mu \rightarrow 3e)$ is smaller than $10^{-12}$, and thus the mass 
 $m_{\tilde{L}}$ and $\lambda$'s involved are constrained to be
\barray
\frac{m_{\tilde{L}}}{\sqrt{\tilde\lambda^2}} \geq 87~{\rm TeV}\,,
\qquad\mbox{where $\tilde{\lambda}^2 = \lambda_L^{11} \sqrt{(\lambda^{12}_L)^2 + (\lambda^{21}_L)^2}$}.
\earray 
Loop processes are also constrained and we detail their contributions in Table~\ref{table:mu_to_eee}.  The conclusion of the detailed results in the Appendix A is that no process constrains $M > 100$ TeV, and therefore, the LOSP decays at the LHC will be prompt.

\section{Prompt Versus Displaced Vertex LOSP Decays at Colliders}
\label{sec:displaced}

In this section, we connect the  flavor constraints on the scale $M_{ijk}$ summarized in the previous section to the lifetime of the LOSP decaying through the ADM operator Eq.~(\ref{basicADM}) in the processes of  Fig.~\ref{fig:feynman_xqld_decay}\footnote{This figure shows explicitly the decay for the $X q \ell d^c$ operator, though for the $u^c d^c d^c$ and $\ell \ell e^c$ models, the decay processes are similar.}.  LOSPs that participate in $W_{\rm ADM}$, such as squarks or sleptons, can decay directly into two SM particles and the ADM through the $W_{\rm ADM}$ operator, as in the left panel of Fig.~\ref{fig:feynman_xqld_decay}.   If the LOSP does not appear in $W_{\rm ADM}$ directly ({\em e.g.} neutralinos or charginos), they will decay through an off-shell squark or slepton, as in the right panel of Fig.~\ref{fig:feynman_xqld_decay}.  Three-body decay and four-body decay of the LOSP lead to completely different lifetime scales and thus result in very different constraints from displaced vertex measurement at the LHC. In Appendix \ref{sec:decay_formula}, we derive the LOSP decay width for general three- and four-body decay as shown in Fig. \ref{fig:feynman_general_decay} with various group representations for participating particles.  We summarize the results of Appendix \ref{sec:decay_formula} here.

For three-body LOSP decay of a squark or a slepton, the secondary vertex displacement $c \tau$ is of the form 
\begin{eqnarray}
(c \tau)^{-1} = F^{\rm (3-body)} \times \left(\frac{m_{\rm LOSP}}{100~{\rm GeV}} \right)^3 \times 
\left(\frac{100~{\rm TeV}}{M_{ijk}} \right)^2 ~(\rm mm^{-1}), \label{eq:displaced_vertex_losp_3_body_generic}
\end{eqnarray}
where $m_{\rm LOSP}$ is the LOSP mass and $F^{\rm (3-body)}$ is the coefficient that can be calculated from Eq.~(\ref{eq:Gamma_3_body}). Here, we ignore the SM particle masses, which, in particular, excludes top quark final state cases.  Note that we also ignore the ADM mass in Eq.~(\ref{eq:displaced_vertex_losp_3_body_generic}) since the squark mass and the slepton mass must be much larger than a typical ADM mass around 10 GeV due to other direct collider constraints. We use the millimeter unit for the displacement since the detectors at the LHC can roughly resolve the displaced vertex up to a millimeter. 

Assuming that the LOSP decays through only one dominant coupling $1/M_{ijk}$ that does not involve the third generation\footnote{The third generation complicates the general discussion because the top quark mass cannot be ignored and the third generation squarks generally have a large mixing. We leave the third generation specific scenarios for the future work.}, we list the 3-body LOSP decays for each superpotential operator and obtain $F^{\rm (3-body)}$ for each case in Table~\ref{table:LOSP3BodyDecay}. One can easily see that the displacement is generically prompt for ${\mathcal O}(100~{\rm GeV})$ LOSP mass for the $M_{ijk}$ scale around the flavor constraints in the previous section. To show it clearly, we list the scale of $M_{ijk}$ that gives a displaced vertex at 1 mm with a 1 TeV LOSP in Table~\ref{table:LOSP3BodyDecay}.   

\begin{table}
\begin{itemize}
\item For $X q\ell d^c$: 

\begin{tabular}{cccc} 
LOSP & ~~~~~~~~~~Decay Mode~~~~~~~~~~~ & ~~~~$(F^{\rm (3-body)})^{-1}$ (mm)~~~~ & $\Lambda_*$ (TeV) \\
\hline  
Left-handed $u$-squark & $\tilde{u}_i \to d_j e^+_k \tilde{x}^* $ & $4.71\times 10^{-5}$  & { $4.61 \times 10^5$ } \\ 
Left-handed $d$-squark & $\tilde{d}_i \to d_j \bar{\nu}_k \tilde{x}^* $ & $4.71\times 10^{-5}$ & { $4.61\times 10^5$} \\ 
Right-handed $d$-squark & $\tilde{d}^c_i \to \bar{u}_j e^+_k \tilde{x}^*, \bar{d}_j \bar{\nu}_k \tilde{x}^*  $ & $2.36\times 10^{-5}$ & { $6.51\times 10^5$} \\ 
Left-handed slepton & $\tilde{e}^-_i \to \bar{u}_j d_k \tilde{x}^* $ & $1.57\times 10^{-5}$ & { $7.96 \times 10^5$} \\ 
Sneutrino v&  $\tilde{\nu}_i \to \bar{d}_j d_k \tilde{x}^*$ & $1.57\times 10^{-5}$  & { $7.96\times 10^5$} \\
\hline \hline
\end{tabular}

\item For $X u^c d^c d^c$: 

\begin{tabular}{cccc}
LOSP & ~~~~~~~~~~Decay Mode~~~~~~~~~~~ & ~~~~$(F^{\rm (3-body)})^{-1}$ (mm)~~~~  & $\Lambda_*$ (TeV) \\
\hline 
Right-handed $u$-squark & $\tilde{u}^c_i \to d_j d_k \tilde{x}^* $ ($j \neq k$) & $2.36\times 10^{-5}$ & {$6.51\times 10^5$} \\ 
Right-handed $d$-squark & $\tilde{d}^c_i \to u_j d_k \tilde{x}^* $ ($i \neq k$) & $2.36\times 10^{-5}$ & {$6.51 \times 10^4$} \\ 
\hline \hline 
\end{tabular} 

\item For $\ell \ell e^c$: 

\begin{tabular}{cccc}
LOSP & ~~~~~~~~~~Decay Mode~~~~~~~~~~~ & ~~~~$(F^{\rm (3-body)})^{-1}$ (mm)~~~~ & $\Lambda_*$ (TeV) \\
\hline 
Left-handed slepton & $\tilde{e}^-_i \to \bar{\nu}_j e^-_k \tilde{x}^* $ ($i \neq j$)  &  $4.71\times 10^{-5}$ & { $4.61\times 10^5$} \\
Right-handed slepton & $\tilde{e}^{c+}_i \to \bar{\nu}_j e^+_k \tilde{x}^*,  e^+_j  \bar{\nu}_k \tilde{x}^*  $ ($j \neq k$) & $2.36\times 10^{-5}$ & { $6.51\times 10^5$} \\ 
Sneutrino & $\tilde{\nu}_i \to e^+_j e^-_k \tilde{x}^*  $ ($i \neq j$) & $4.71\times 10^{-5}$ & {$4.61\times 10^5$} \\ 
\hline \hline 
\end{tabular}

\end{itemize}
\caption{\label{table:LOSP3BodyDecay} 3-body decay modes, for various LOSP choices, and their lifetime factor $F^{\rm (3-body)}$ (from Eq.~(\ref{eq:displaced_vertex_losp_3_body_generic})) in $q \ell d^c$, $u^c d^c d^c$ and $\ell \ell e^c$ models.  $\Lambda_*$ is the scale of $M_{ijk}$ that gives rise to a displaced vertex at 1 mm with $m_{\rm LOSP} = 1$ TeV.   }

\end{table}

For four-body LOSP decay, the displacement is given by the following expression if we assume that a contribution from a single intermediate particle $\phi$ dominates: 
\begin{eqnarray}
 (c \tau)^{-1}  &=&  F^{\rm (4-body)} \times \left( \frac{100~{\rm TeV}}{M_{ijk}} \right)^2 
\times \left( \frac{500~{\rm GeV}}{m_\phi} \right)^4 
\times \left( \frac{m_{\rm LOSP}}{100~{\rm GeV}} \right)^7 \times 
\label{eq:displaced_vertex_losp_4_body_generic}  \\  
&& \qquad \times \frac{1}{x^5} \left[ 
 (10 x^3 - 120 x^2 - 120 x) + 60 (1-x) (2-x) \log (1-x) 
\right]   ({\rm mm}^{-1}), \nonumber 
\end{eqnarray}
where $\phi$ is the intermediate squark or slepton with the mass $m_\phi$ and $x = (m_{\rm LOSP} / m_\phi )^2$. The coefficient $F^{\rm (4-body)}$ can be determined by Eq.~(\ref{eq:Gamma_4_body}).  Note that the expression in the second line in Eq.~(\ref{eq:displaced_vertex_losp_4_body_generic}) is reduced to $\sim (1 + x) $ in the limit of $x \ll 1$.  

\begin{figure}
\centering 
\subfigure[~$M = 1000$ TeV]{\includegraphics[width=7.5cm]{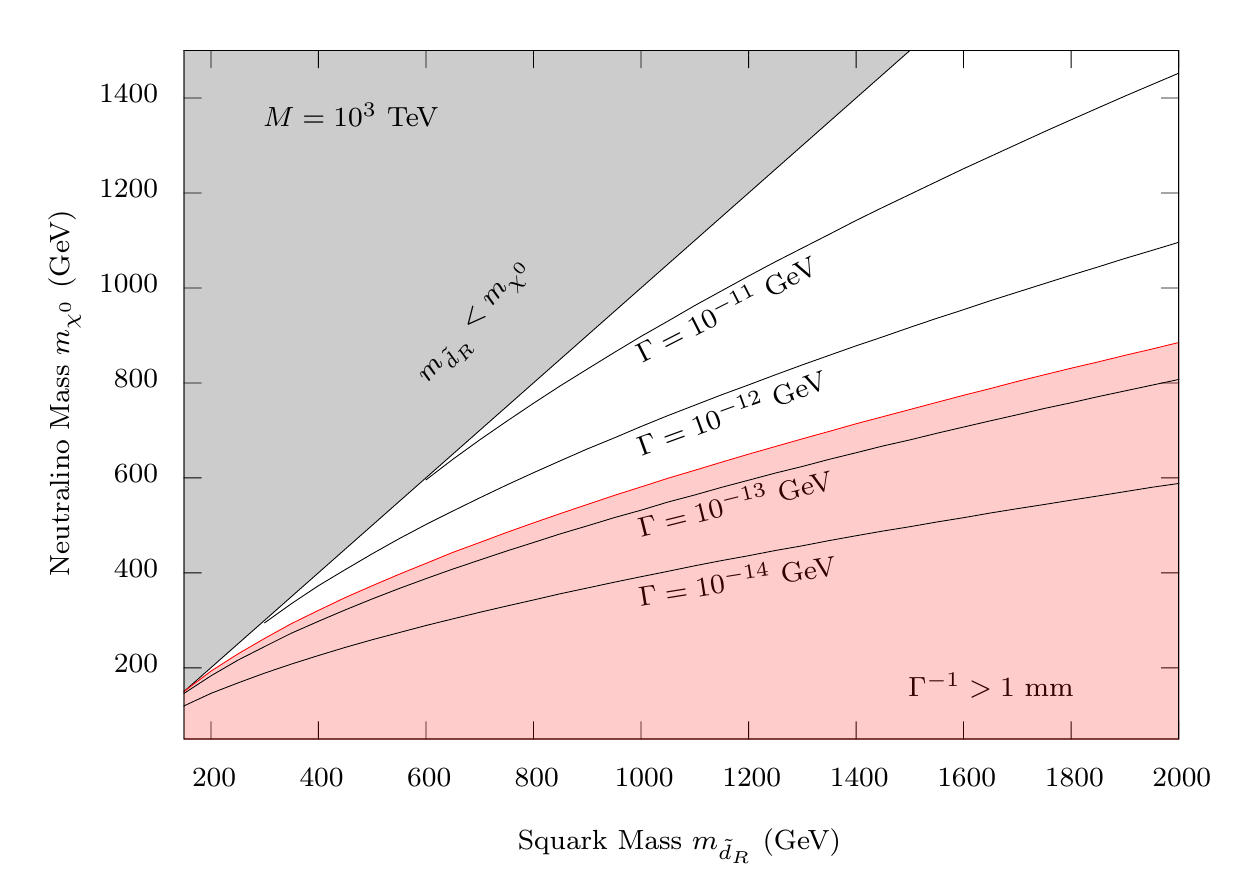}}
\quad 
\subfigure[~$M = 10^4$ TeV]{\includegraphics[width=7.5cm]{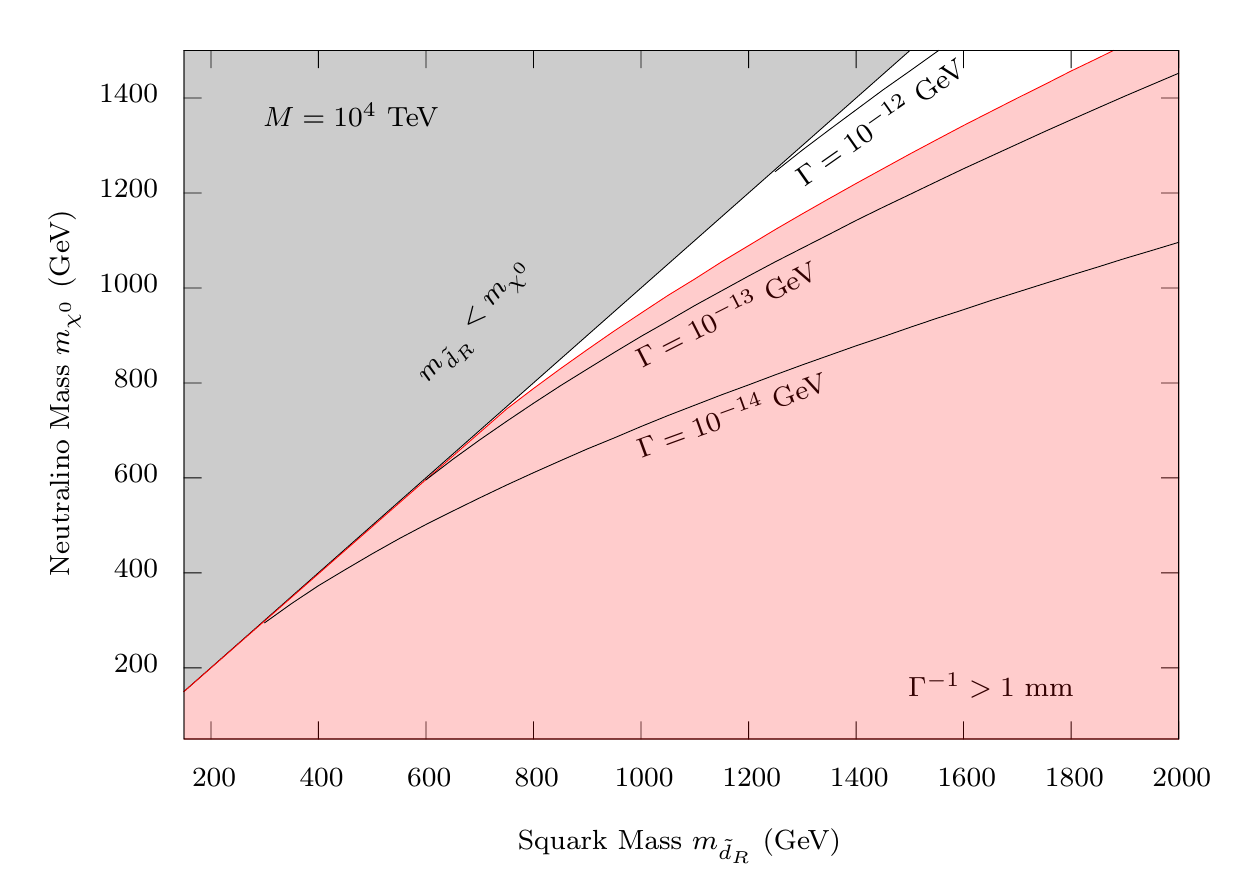}}
\caption{Neutralino decay width in $(m_{\tilde{d}_R}, m_{\chi^0})$ plane in the $q \ell d^c$ model for (a) $M = 1000$ TeV and (b) $M = 10^4$ TeV, where $M$ is the effective mass scale of the dominant $X q \ell d^c$ operator.  In the lower-right shaded region (red), the neutralino will leave a displaced vertex at the LHC, defined by where the lifetime is longer than a millimeter.   } 
\label{fig:neutralino decay}
\end{figure}

We have many possibilities for such 4-body LOSP decay in the ADM models. Since gauginos and Higgsino do not participate in the operators $X q \ell d^c$, $X u^c d^c d^c$ and $X \ell \ell e^c$, the neutralino, chargino and gluino LOSP will decay through intermediate squarks or sleptons/sneutrinos. While the gluino LOSP decay is simply determined from QCD interactions through intermediate squarks, the neutralino LOSP and chargino LOSP depend on the details of the mixing. In general, several off-shell intermediate particle exchanges can contribute with similar size. An exhaustive study for this is beyond the scope of this paper. Instead, we only consider special cases to show typical constraints.  

In Fig.~\ref{fig:neutralino decay}, we consider the case with pure Bino (neutralino) LOSP with one light right-handed $d$-squark $\tilde{d}^c$. We assume that only the first generation coupling $1/M_{111}$ for the $X q_i \ell_j d^c_k$ operator is dominant. In this scenario, we obtain $(F^{\rm (4-body)})^{-1} = 2.04 \times 10^2$ mm.  Fig.~\ref{fig:neutralino decay}a shows the neutralino decay width contour in $(m_{\tilde{d}^c},m_{\chi^0})$ plane with $M \equiv M_{111} = 1000$ TeV and Fig.~\ref{fig:neutralino decay}b shows one for $M = 10^4$ TeV. While displaced vertices result over a significant fraction of the parameter space, the decays are prompt over much of the parameter space even for high choices of $M$, naively consistent with the flavor constraints even in the absence of flavor symmetries.  

In the case of displaced decays, by searching for the displaced vertex, we can clearly identify DM creation inside the detector and probe the nature of the DM directly at the LHC. Thus, displaced vertex searches are very important for ADM searches at the LHC.   In the case of prompt decays, however, one basic question is how ADM models fare when subjected to the usual supersymmetric searches.  In the next section we compare the constraints from two standard searches for SUSY against those obtained in ADM when the LOSP is unstable to decay.

\section{LHC Constraints}
\label{sec:LHC}

\begin{figure}
\centering
\subfigure[~gluino-gluino]{\includegraphics[height=3.5cm]{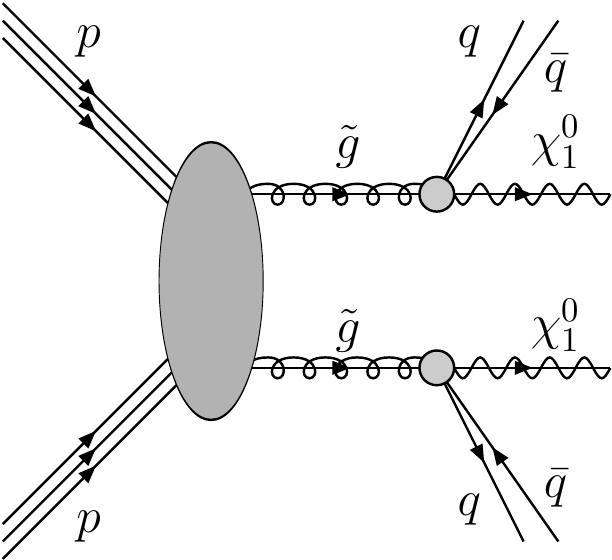}}
\quad
\subfigure[~squark-gluino]{\includegraphics[height=3.5cm]{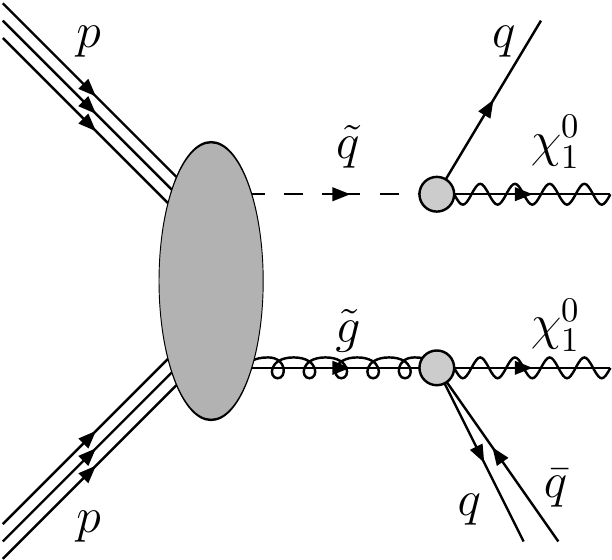}}
\quad
\subfigure[~squark-squark]{\includegraphics[height=3.5cm]{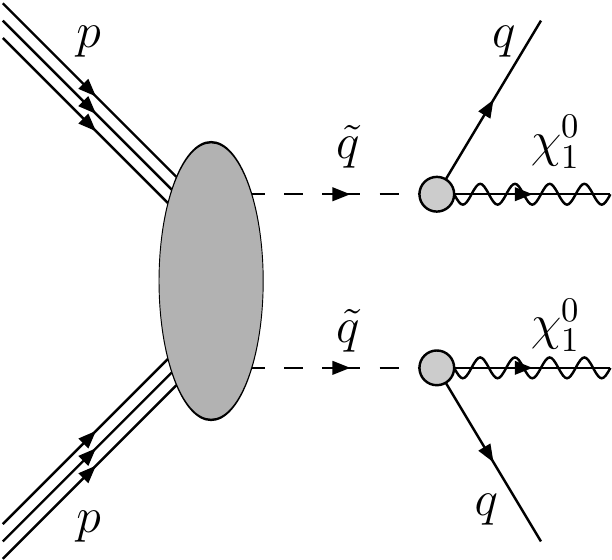}}
\caption{ Relevant processes for ATLAS 0-lepton+2-6 jet+MET analysis for Simplified Model {\bf Sim0}. }
\label{fig:collider_for_0lep_simplified_SUSY}
\end{figure}

\begin{figure}
\centering
\subfigure[~gluino-gluino {({\bf Sim1g})}]{\includegraphics[height=3.5cm]{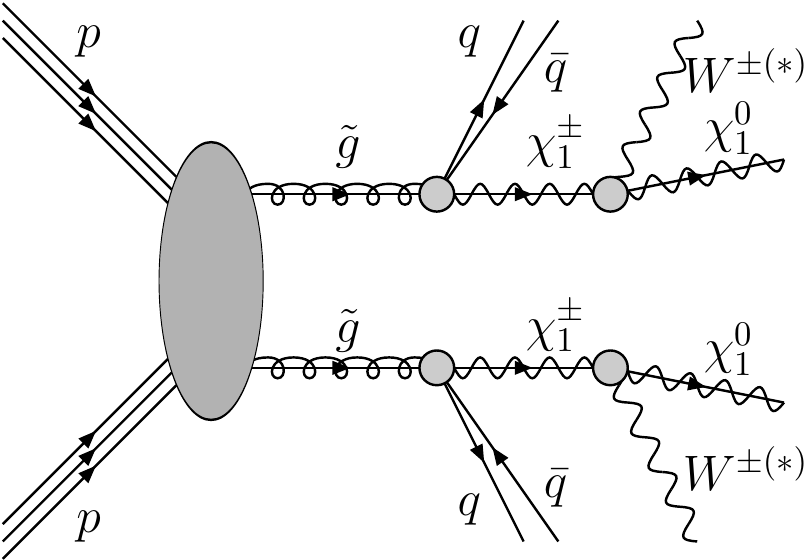}}
\quad
\subfigure[~squark-squark {({\bf Sim1q})}]{\includegraphics[height=3.5cm]{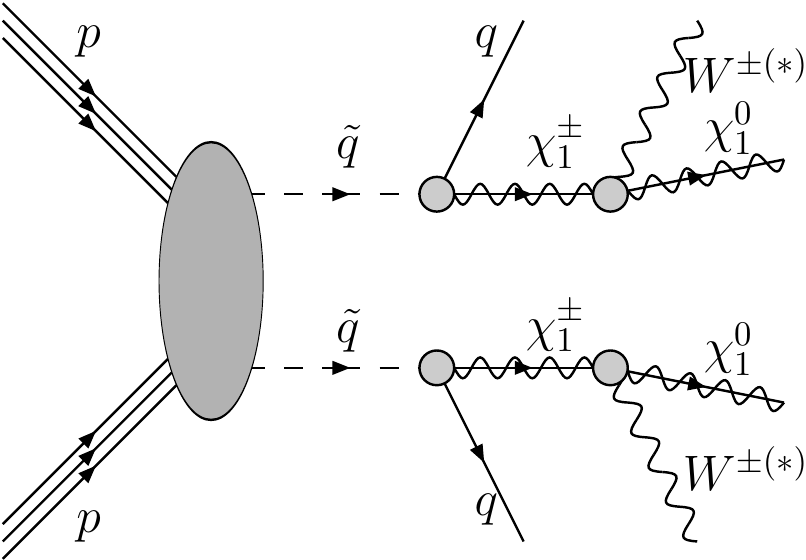}}
\caption{Relevant processes for ATLAS 1-2 lepton + 3-6 jet + MET analysis for Simplified Models {\bf Sim1g} and {\bf Sim1q}. }
\label{fig:collider_for_1_2lep_simplified_SUSY}
\end{figure}

In order to compare the standard searches for SUSY against those obtained in ADM, we consider two ATLAS analyses with 20.3 fb$^{-1}$ of data at 8 TeV.  We have chosen the ATLAS, instead of CMS, analyses in this study since the collaboration quotes the 95\% confidence limit, $S^{95}_{\rm exp}$, on the number of events from new physics, once the cuts of the analysis have been applied.  This allows us to simulate the SM plus new physics and easily extract the constraint by simply taking the difference with a simulation having the SM only.  
We utilize
\begin{enumerate}
\item an analysis with a lepton veto, 2-6 hard jets and high missing transverse energy (MET) $E_T^{\rm miss}$\cite{ATLAS-CONF-2013-047}.  We will refer to this analysis as``0 lepton+2-6 jet+MET analysis'' (or ``0 lepton analysis'' for short); 
\item an analysis with 1 or 2 leptons, 3-6 hard jets and high $E_T^{\rm miss}$\cite{ATLAS-CONF-2013-062}.   We will refer to this analysis as the ``1-2 lepton+3-6 jet+MET analysis''  (or ``1-2 lepton analysis'' for short).
\end{enumerate}

Both of these analyses are the most standard SUSY searches for typical gluino or 1st/2nd generation squark pair production modes in R-parity conserving SUSY scenarios.  We aim to compare the ADM models with the ordinary SUSY models, represented by Simplified Models \cite{Alwall:2008ag,Alves:2011sq}, with the relevant processes shown in Figs.~\ref{fig:collider_for_0lep_simplified_SUSY},~\ref{fig:collider_for_1_2lep_simplified_SUSY}. The Simplified Models are designed for ease of model-independent comparison among different $R$-parity conserving SUSY scenarios.

\begin{figure}
\centering
\subfigure[~gluino-gluino]{\includegraphics[height=3.5cm]{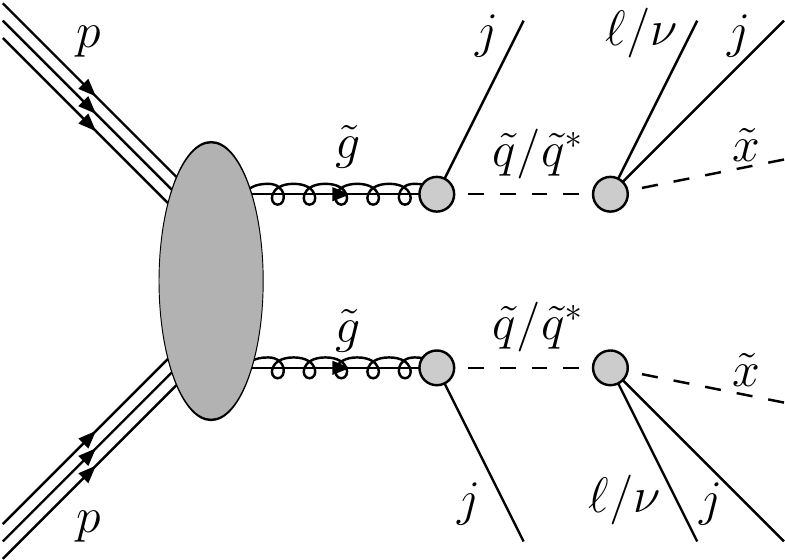}}
\quad
\subfigure[~squark-gluino]{\includegraphics[height=3.5cm]{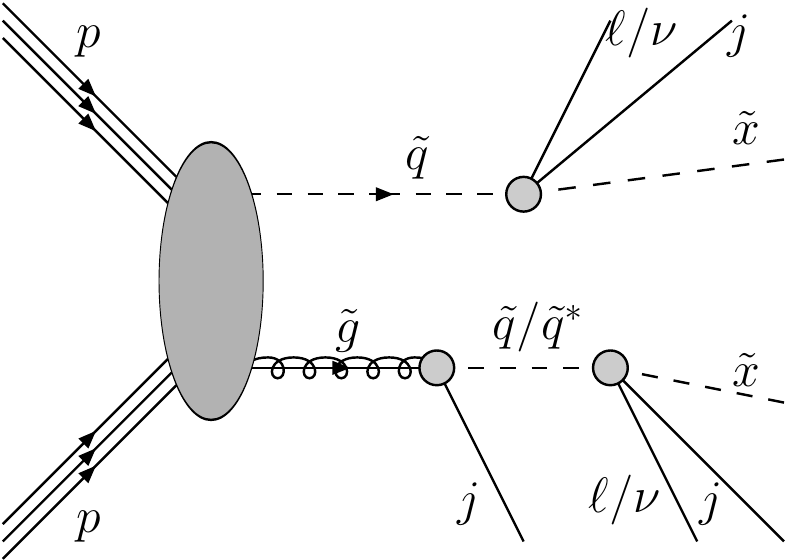}}
\quad
\subfigure[~squark-squark]{\includegraphics[height=3.5cm]{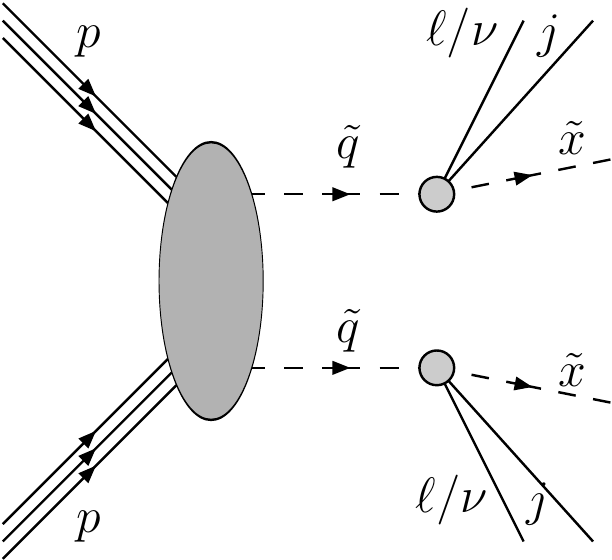}}
\caption{  Relevant processes for the squark LOSP case in $q \ell d^c$ model. Here $\ell/\nu$ implies lepton or neutrino which is almost equally produced in squark decay.
The $u^c d^c d^c$ model has the same diagrams with a lepton/neutrino replaced by a jet in the final decays of squarks. }
\label{fig:collider_for_0lep_xudd_squarklsp}
\end{figure}

\begin{figure}
\centering
\subfigure[~gluino-gluino]{\includegraphics[height=3.5cm]{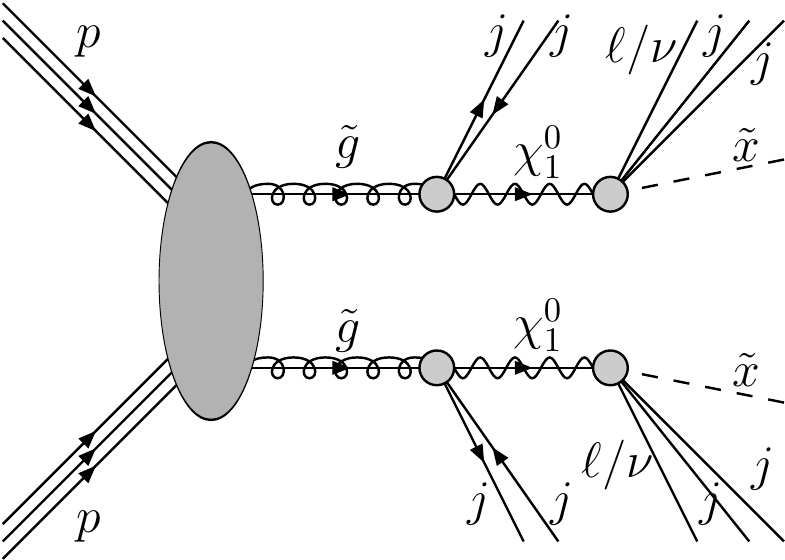}}
\quad
\subfigure[~squark-gluino]{\includegraphics[height=3.5cm]{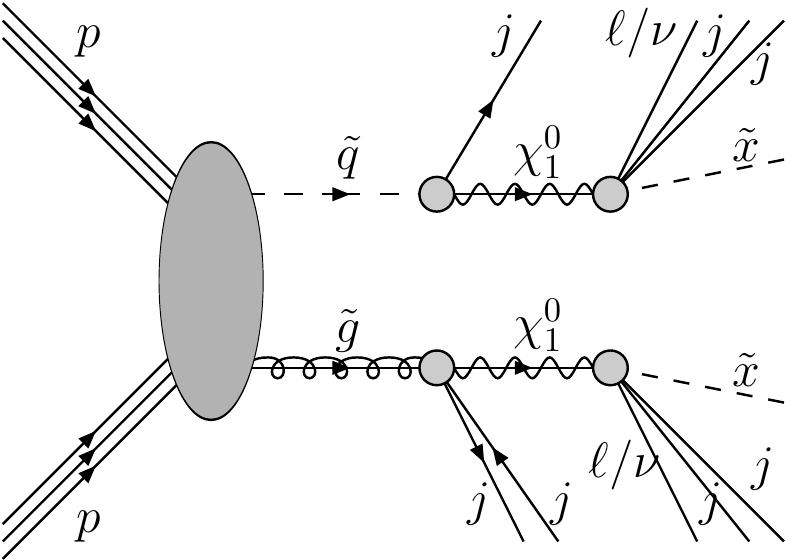}}
\quad
\subfigure[~squark-squark]{\includegraphics[height=3.5cm]{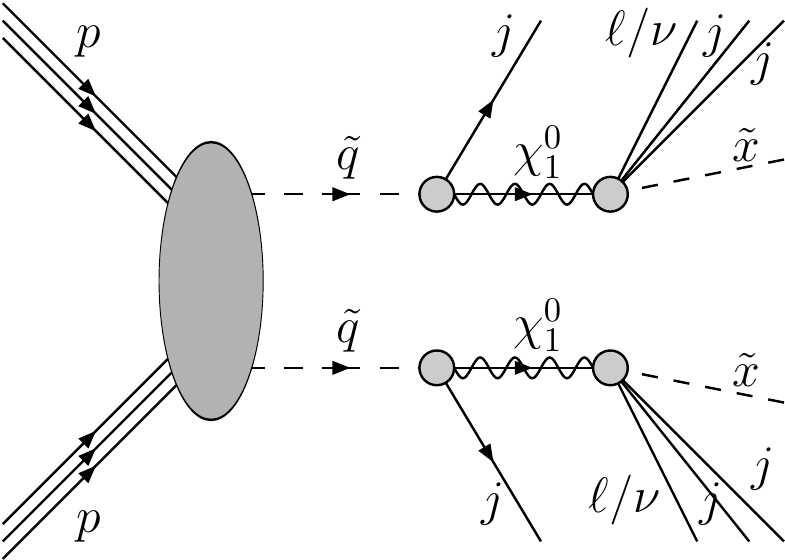}}
\caption{  Relevant processes for the neutralino LOSP case in $q \ell d^c$ model. $\ell/\nu$ implies lepton or neutrino. The 4-body decay of neutralino decay is through off-shell squark as shown in Fig. \ref{fig:feynman_xqld_decay} (b).  The $u^c d^c d^c$ model has the same diagrams with a lepton/neutrino replaced by a jet in the neutralino decay. }
\label{fig:collider_for_0lep_xudd_neutralinolsp}
\end{figure}

In the case of ADM, both the 0-lepton and 1-2 lepton analyses are well-targeted to the $q \ell d^c$ model, as shown in the processes of Figs.~\ref{fig:collider_for_0lep_xudd_squarklsp},~\ref{fig:collider_for_0lep_xudd_neutralinolsp}.  For $u^c d^c d^c$, the 0-lepton+2-6 jet+MET search is effective through the processes again shown in Figs.~\ref{fig:collider_for_0lep_xudd_squarklsp},~\ref{fig:collider_for_0lep_xudd_neutralinolsp}, where additional jets from the LOSP decay are traded for a reduced missing energy cut.  
Other ATLAS and CMS analyses may also be relevant for constraining certain ADM models (such as the ATLAS and CMS high jet multiplicity analyses \cite{Aad:2013wta,CMS-PAS-SUS-13-012} for the $u^c d^c d^c$ model).  We have not explored these constraints here, instead choosing a representative sample which utilizes the most standard types of SUSY analyses.  
In addition, we do not consider gluino and slepton/sneutrino LOSPs, or the constraints on $\ell \ell e^c$ operators.  A more exhaustive analysis including these other cases is very interesting for future work.

\subsection{Analyses}

We briefly review the 8 TeV ATLAS 0 lepton+2-6 jet+MET analysis and 1-2 lepton+3-6 jet+MET analyses  and  how these analyses may constrain ADM $q \ell d^c$ and $u^c d^c d^c$ models, in comparison to the Simplified Models that are utilized in the original ATLAS analysis. We also summarize the definition of the observables and the notation used in the analyses in Appendix \ref{sec:analysis_observable}.   

\subsubsection{0 lepton+2-6 jet+MET analysis}

The ATLAS 0 lepton+2-6 jet+MET analysis with 20.3 fb$^{-1}$ at $\sqrt{s} = 8$ TeV is summarized in Table \ref{table:ATLAS_0lep_2_6jet_MET}. The analysis is designed to maximize the discovery potential for gluino and squark pair production with decays to neutralinos and jets. Events with signal leptons are vetoed.  Events are classified into 10 non-exclusive channels: AL, AM, BM, BT, CM, CT, D, EL, EM and ET,  where A, B, C, D and E imply the number of jets $N = 2,3,4,5$ and 6, respectively and L, M and T imply loose, medium and tight cut on the effective mass scale, respectively.

For comparision, we consider the Simplified Model process shown in Fig. \ref{fig:collider_for_0lep_simplified_SUSY}.   The Simplified Model has the gluino $\tilde{g}$, the lightest neutralino $\chi^0_1$ and all the left-handed squarks $\tilde{q}_{i\,L}$ and right-handed squarks $\tilde{q}_{i\,R}$ of the first and second generation with degenerate mass. In this model, the only SUSY particle production channel is gluino/squark pair production through the SUSY QCD processes. The gluino decays through $\tilde{g} \rightarrow q \tilde{q}^{(*)} \rightarrow q q \chi^0_1$ with 100\% branching ratio (BR), where the intermediate squark $\tilde{q}^{(*)}$ can be either on-shell or off-shell depending on mass parameters, and a squark directly decays into the neutralino and a quark. To distinguish this from other simplified models we consider for the 1-2 lepton analysis, we denote this model ``{\bf Sim0}.'' 

The most important features that will be relevant for distinguishing the constraints on the ADM model versus the Simplified Model are: {\it (i)} $E_T^{\rm miss} > 160 \mbox{ GeV}$, which we will see rather dramatically reduces the acceptance of the ADM models; {\it (ii)} $N_{\rm jet}$ with $p_T  > 60 \mbox{ GeV}$, which improves the acceptance for the ADM models with a large number of jets; {\it (iii)} $m_{\rm eff}$ and $E_T^{\rm miss}/m_{\rm eff}$, both of which improve the acceptance of the ADM model over the Simplified Model.  Overall, we will find that the $E_T^{\rm miss}$ cut is severe enough that in most cases the constraint on the ADM models will be much weaker than for the Simplified Model.  Our discussion will also show, however, that better searches could easily be implemented replacing the hard missing energy cut with a higher multiplicity of hard jets or leptons.  Thus, it is desirable to compare the ADM model with the conventional SUSY models by performing similar LHC analyses with higher multiplicity (such as \cite{Aad:2013wta,CMS-PAS-SUS-13-012}). We postpone this study for the future work. 

\begin{table}
\centering
\begin{tabular}{|c||c|c||c|c||c|c||c||c|c|c|}
\hline
\multirow{3}{*}{Requirement} & \multicolumn{10}{|c|}{Channel} \\
\cline{2-11}
 & \multicolumn{2}{|c||}{A (2 jets)} & \multicolumn{2}{|c||}{B (3 jets)} & \multicolumn{2}{|c||}{C (4 jets)}
 & D (5 jets) & \multicolumn{3}{|c|}{E (6 jets)} \\
\cline{2-11}
& L & M & M & T & M & T & - & L & M & T \\
\hline\hline 
Common & \multicolumn{10}{|c|}{$E_T^{\rm miss} > 160$ GeV,\quad  $p_T (j_1) > 130$ GeV,\quad
$p_T (j_2) > 60$ GeV } \\ 
\hline 
$p_T (j_i) (i\geq3)$ & \multicolumn{10}{|c|}{ $> 60$ GeV for $i=3..N$ for $N$-jet channel} \\
\hline
$\Delta \phi({\rm jet}, E^{\rm miss}_T) >$ & \multicolumn{4}{|c||}{ 0.4 ($i = [1,2,(3)]$)} & \multicolumn{6}{|c|}{0.4 ($i=[1,2,3]$),\quad 0.2 for $p_T(j_i) > 40$ GeV} \\ 
\hline 
$E_T^{\rm miss} / m_{\rm eff} (Nj)>$ &0.2 & -$^{(a)}$ & 0.3 & 0.4 & 0.25 & 0.25 & 0.2 & 0.15 & 0.2 & 0.25  \\
\hline
$m_{\rm eff}$(\rm incl.) [GeV] $>$ & 1000 & 1600 & 1800 & 2200 & 1200 & 2200 & 1600 & 1000 & 1200 & 1500 \\ 
\hline\hline 
$S^{95}_{\rm exp}$ & ~1135.0~ & ~42.7~ & ~17.0~ & ~5.8~ & ~72.9~ & ~3.3~ & ~13.6~ & ~57.3~ & ~21.4~ & ~6.5~ \\ 
Error & $^{+332.7}_{-291.5}$ & $^{+15.5}_{-11.4}$ & $^{+6.6}_{-4.6}$ & $^{+2.9}_{-1.8}$ & $^{+23.6}_{-18.0}$ & $^{+2.1}_{-1.2}$ & $^{+5.1}_{-3.5}$ & $^{+20.0}_{-14.4}$ & $^{+7.6}_{-5.8}$ & $^{+3.0}_{-1.9}$ \\
\hline
\end{tabular}
\caption{\label{table:ATLAS_0lep_2_6jet_MET} A Summary of ATLAS 0 lepton+2-6 jet+MET analysis at 8 TeV, 20.3 fb$^{-1}$. This table is an excerpt from Table 1 and Table 4 in \cite{ATLAS-CONF-2013-047}. }
\end{table}

\subsubsection{1-2 lepton+3-6 jet+MET analysis}

The ATLAS 1-2 lepton+3-6 jet+MET analysis with 20.3 fb$^{-1}$ at $\sqrt{s} = 8$ TeV is summarized in Table \ref{table:ATLAS_1_2lep_3_6jet_MET_part1} and in Table \ref{table:ATLAS_1_2lep_3_6jet_MET_part2}. This analysis effectively selects gluino and squark pair production events with a lepton or two from decays of charginos or sleptons. The analysis is divided into soft and hard lepton channels. Signal leptons with $p_T < 25$ GeV are regarded as soft and in turn have 7 classes: soft single lepton 1 b-jet Low-mass/High-mass, soft single lepton 2 b-jets Low-mass/High-mass, soft single lepton 3-jet/5-jet and soft dimuon channel. Hard lepton channels have 3 classes: 3-jet, 5-jet and 6-jet, with each class having inclusive/binned channels, and electron/muon subchannels according to the lepton identity.  Thus there are 12 channels in total for the hard lepton case. We summarize the requirements and the observed 95\% C.L. limit of this analysis from the ATLAS experiment in Tables~\ref{table:ATLAS_1_2lep_3_6jet_MET_part1},~\ref{table:ATLAS_1_2lep_3_6jet_MET_part2}. 
 
For the 1-2 lepton analysis, we compare the $q\ell d^c$ model with the Simplified Models by varying the relative ratio between colored SUSY particle masses and the LOSP mass. To this end, we use two Simplified Models as shown in Fig. \ref{fig:collider_for_1_2lep_simplified_SUSY}, which are referred to as ``one-step'' Simplified Models in the ATLAS analysis \cite{Aad:2012ms}. The first model, shown in Fig. \ref{fig:collider_for_1_2lep_simplified_SUSY}a, which we call ``{\bf Sim1g},'' has the gluino $\tilde{g}$, the lightest chargino $\chi^\pm_1$ and the lightest neutralino $\chi^0_1$. Production is gluino pairs, with the gluino decaying via $\tilde{g} \rightarrow q \bar{q}' \chi^{\pm}_1 \rightarrow q \bar{q}' W^{(*)} \chi^0_1$ with 100\% branching, where $q$ and $q'$ are quarks with different isospin and $W^{(*)}$ is the on-shell (off-shell) $W$ boson, depending on the mass gap between $\chi^\pm_1$ and $\chi^0_1$. The second model shown in Fig. \ref{fig:collider_for_1_2lep_simplified_SUSY}b, which we call ``{\bf Sim1q},'' has the left-handed squark $\tilde{q}_L$, the lightest chargino $\chi^\pm_1$ and the lightest neutralino $\chi^0_1$. Note that only left-handed squarks are involved since $\chi^\pm_1$ and $\chi^0_1$ are assumed to be mostly Wino-like. Now the production is only through squark pairs with the squark decaying through $\tilde{q}_L \rightarrow q' \chi^{\pm}_1 \rightarrow q' W^{(*)} \chi^0_1$.  For simplicity\footnote{Admittedly, this choice is far from general. We simply follow the ATLAS analysis here for comparison with $q\ell d^c$ model in $m_{\tilde{g}}(m_{\tilde{q}}) - m_{\chi^0_1}$ scan. }, we fix the ratio among the colored superparticle ($\tilde{g}$/$\tilde{q}_L$), $\chi^\pm_1$ and $\chi^0_1$ 
\barray
m_{\chi^\pm_1} = \frac{m_{\tilde{g}(\tilde{q})} - m_{\chi^0_1}}{2}.
\earray

Similarly to the 0 lepton analysis, we will find $E_T^{\rm miss}$ to be a key variable in distinguishing the ADM model from the Simplified Models, though both the $p_T$ cut on the hardest lepton and jet will play an important role.  Note, however, that the $E_T^{\rm miss}$ cut here is stronger than in the 0 lepton analysis in order to filter the SM $W$ and top-quark events. For some soft channels, $b$-tagging is employed, and thus the $b$-tagging efficiency affects the event acceptance. In the ATLAS analysis, different $b$-tagging efficiency has been applied by adjusting a $b$-tagging parameter for different channels. However, in our analysis, we simply rely on the detector simulator we use; since the efficiency difference is at the $\sim 10 \%$ level and cross section differences between two adjacent scan points are much higher, our results will not be significantly changed because of the $b$-tagging method.

\begin{table}
\centering
\begin{tabular}{|c||c|c||c|c||c|c||c|}
\hline 
 Class & \multicolumn{2}{|c||}{Soft 1-$\ell$ 1-b} & \multicolumn{2}{|c||}{Soft 1-$\ell$ 2-b} & \multicolumn{2}{|c||}{Soft 1-$\ell$}  & Soft 2-muon \\
\hline
 Subclass & L & H & L & H & 3-j & 5-j & 2-j \\ 
\hline \hline
\multirow{4}{*}{Lepton}& \multicolumn{6}{|c||}{$N_{\ell} = 1$,  $10 (6) \leq p_T^{\ell} \leq 25$ } &  2 muons, $6 \leq p_T^{\mu} \leq 25$ \\ 
                       & \multicolumn{6}{|c||}{}  &  $m_{\mu\mu} > 15$ \\ 
 & \multicolumn{6}{|c||}{-} & $|m_{\mu\mu}-m_Z| >10$ \\
\cline{2-8}
 &  \multicolumn{7}{|c|}{$p_T^{{\rm add.}\,\ell} < 7 (6)$} \\
\hline \hline
$N_{\rm jet} $ & \multicolumn{2}{|c||}{$\geq 3$} & \multicolumn{2}{|c||}{$\geq 2$} & [3,4] & $\geq 5$ & $\geq 2$ \\
\hline 
$p_T^j$ & $>180,40,40$ & $>180,25,25$ & \multicolumn{2}{|c||}{$>60,60$, $< 50$} & \multicolumn{2}{|c||}{$>180,25,25,\dots$} & $>70,25,25\dots$ \\
\hline
$N_{\rm b-tag}$& \multicolumn{2}{|c||}{$\geq 1$, but not leading} & \multicolumn{2}{|c||}{2} & \multicolumn{2}{|c||}{-} & 0 \\
\hline \hline
$E_T^{\rm miss} >$ & 250 & 300 & 200 & 300 & 400 & 300 & 170 \\
\hline
$m_T >$  & \multicolumn{2}{|c||}{$100$} & \multicolumn{2}{|c||}{-}  & \multicolumn{2}{|c||}{100} & 80\\
\hline
$E_T^{\rm miss}/m_{\rm eff}^{\rm incl}$ & \multicolumn{2}{|c||}{$>0.35$} & \multicolumn{2}{|c||}{-} & \multicolumn{2}{|c||}{$>0.3$} & - \\
\hline
$\Delta R_{\rm min}({\rm jet}, \ell)$ & \multicolumn{2}{|c||}{$>1.0$} & \multicolumn{2}{|c||}{-} & $>1.0$ & - & $>1.0$ \\
\hline \hline
$\Delta \phi_{\rm min}$ & \multicolumn{2}{|c||}{-} & \multicolumn{2}{|c||}{$>0.4$} & \multicolumn{3}{|c|}{-} \\ 
\hline
$m_{CT} >$ & \multicolumn{2}{|c||}{-} & 150 & 200 & \multicolumn{3}{|c|}{-} \\
\hline
$H_{T2}$ & \multicolumn{2}{|c||}{-} & $<50$ & - & \multicolumn{3}{|c|}{-} \\
\hline \hline
$S^{95}_{\rm exp}$ & 6.9$^{+3.0}_{-2.0}$ & 6.3$^{+1.9}_{-1.1}$ & 13.2$^{+5.9}_{-4.1}$ & 5.3$^{+2.4}_{-1.4}$ & 6.3$^{+2.7}_{-1.8}$ & 10.0$^{+3.6}_{-3.0}$ & 5.9$^{+2.1}_{-1.0}$ \\ 
\hline
\end{tabular}
\caption{\label{table:ATLAS_1_2lep_3_6jet_MET_part1} A summary of ATLAS 1-2 lepton+3-6 jet+MET analysis at 8 TeV, 20.3 fb$^{-1}$, Part 1: Soft lepton events. Here, the leading lepton $p_T$ is confined to be less than 25 GeV. Note that the dimensionful numbers in the table are in GeV units. L and H denote low-mass and high-mass channels, respectively. }
\end{table}

\begin{table}
\begin{tabular}{|c||c|c|c|c||c|c|c|c||c|c|c|c|}
\hline
 & \multicolumn{12}{|c|}{Hard 1-$\ell$ } \\
\hline
Class & \multicolumn{4}{|c||}{3 jet} & \multicolumn{4}{|c||}{5 jet} & \multicolumn{4}{|c|}{6 jet} \\  
\hline
Subclass & \multicolumn{2}{|c|}{Inclusive} & \multicolumn{2}{|c||}{Binned} & \multicolumn{2}{|c|}{Inclusive} & \multicolumn{2}{|c||}{Binned} & \multicolumn{2}{|c|}{Inclusive} & \multicolumn{2}{|c|}{Binned} \\ 
\hline
$\ell$ type & $e$ & $\mu$ & $e$ & $\mu$ & $e$ & $\mu$ & $e$ & $\mu$ & $e$ & $\mu$ & $e$ & $\mu$ \\ 
\hline\hline
Lepton & \multicolumn{12}{|c|}{$N_{\ell} = 1$, $p_T^{\ell} > 25$, $p_T^{{\rm add.} \ell} < 10$} \\
\hline \hline
$N_{\rm jet}$ & \multicolumn{4}{|c||}{$ \geq 3$} & \multicolumn{4}{|c||}{$\geq 5$} & \multicolumn{4}{|c|}{$\geq 6$} \\ 
\hline 
$p_T^{\rm jet}$ & \multicolumn{4}{|c||}{$> 80,80,30$} & \multicolumn{4}{|c||}{$>80,50,40,40,40$} & \multicolumn{4}{|c|}{$>80,50,40,40,40,40$} \\
\hline 
$p_T^{\rm add. jets}$ & \multicolumn{2}{|c|}{-} & \multicolumn{2}{|c||}{$<40$} & \multicolumn{2}{|c|}{-} & \multicolumn{2}{|c||}{$<40$} & \multicolumn{2}{|c|}{-} & \multicolumn{2}{|c|}{$<40$} \\ 
\hline\hline
$E_T^{\rm miss} >$ & \multicolumn{2}{|c|}{500} & \multicolumn{2}{|c||}{300} & \multicolumn{4}{|c||}{300} & \multicolumn{2}{|c|}{350} & \multicolumn{2}{|c|}{250} \\ 
\hline
$m_T >$ & \multicolumn{4}{|c||}{150} & \multicolumn{2}{|c|}{200} & \multicolumn{2}{|c||}{150} & \multicolumn{4}{|c|}{150} \\ 
\hline 
$E_T^{\rm miss}/m_{\rm eff}^{\rm excl.}$ & \multicolumn{4}{|c||}{$>0.3$} & \multicolumn{8}{|c|}{-} \\ 
\hline 
$m^{\rm incl}_{\rm eff}$ & \multicolumn{2}{|c|}{1400} & \multicolumn{2}{|c||}{800} & \multicolumn{2}{|c|}{1400} & \multicolumn{2}{|c||}{800} & \multicolumn{4}{|c|}{600} \\
\hline  \hline
$S^{95}_{\rm exp}$ & ~5.7~ & ~5.1~ & ~20.2~ & ~15.6~ & ~5.4~ & ~4.7~ & ~12.6~ & ~7.6~ & ~4.4~ & ~4.1~ & ~7.8~ & ~7.1~ \\
Error & $^{+2.2}_{-1.5}$ & $^{+2.0}_{-1.5}$ & $^{+8.3}_{-4.8}$ & $^{+5.8}_{-3.8}$ & $^{+2.3}_{-1.5}$ & $^{+1.9}_{-1.2}$ & $^{+3.2}_{-2.7}$ & $^{+2.8}_{-2.4}$ & $^{+1.9}_{-0.8}$ & $^{+1.3}_{-1.1}$ & $^{+3.1}_{-2.4}$ & $^{+3.4}_{-1.4}$ \\
\hline
\end{tabular}
\caption{\label{table:ATLAS_1_2lep_3_6jet_MET_part2} A summary of the ATLAS 1-2 lepton+3-6 jet+MET analysis at 8 TeV, 20.3 fb$^{-1}$, Part 2: Hard lepton events. The leading lepton $p_T$ must be higher than 25 GeV. The dimensionful numbers in the table are in GeV units. }
\end{table}

\subsection{Event Generation}

We use {\texttt{MadGraph5 v1.5.8}} for the Matrix-Element (ME) event generation \cite{Alwall:2011uj}. The generated events are reweighted to match the Next-to-Leading-Order (NLO) cross section. We employ \texttt{Prospino 2.1} to obtain the cross section of gluino and squark pair production at NLO \cite{Beenakker:1996ed,Prospino21}. 

Since the processes under consideration consist of cascades of multiple decay chains through on-shell states with very narrow decay width, it is desirable to divide a single process into one 2-to-2 process and multiple decay subprocesses for each on-shell particle in the process, to generate events for them separately, and to merge all of the sub-parts into a single process by doing the appropriate Lorentz transformation and color flow matching\footnote{In this paper, we do not consider spin correlation.}.  We created a utility called {\texttt{evchain}} for doing the job automatically \cite{evchain}. A detailed description of {\texttt{evchain}} is presented in Appendix~\ref{sec:analysispipeline}.
We use {\texttt{PYTHIA6}} for the $q\ell d^c$ model and {\texttt{PYTHIA8}} for the $u^c d^c d^c$ model for parton shower (PS) and hadronization\footnote{{\texttt{PYTHIA}} 6 does not support the color-triplet vertex ($\epsilon_{ijk}$) as an acceptable color flow structure.}\cite{Sjostrand:2006za,Sjostrand:2007gs}. We generate only leading order SUSY events without employing Matrix Element/Parton Shower matching, assuming the LO parton-showered distribution scaled with NLO $K$-factor approximates the true distribution well for large $\sqrt{s}$ for typical gluino/squark production.  Because we do not use a matched sample for the signal events, our result should be interpreted with care in the compressed mass spectrum where the $p_T$ of additional QCD jets can be comparable with the $p_T$ of jets from superparticle decay, significantly changing the $p_T$ distribution of the leading jets. Such points need further focused study with appropriate matching. 
For detector simulation, we modify {\texttt{PGS4}}  to enable anti-$k_T$ jet reconstruction, and we rely on the $b$-tagging efficiency implemented in {\texttt{PGS4}}\cite{PGS4}.  

For the 0 lepton analysis, we scan mass parameters in the gluino-common squark mass plane by fixing the neutralino mass $m_{\chi^0_1}$. For the ADM model, we fix the mass of the ADM to be 10 GeV (a well-motivated value), and we consider four different cases: squark LOSP (with the neutralino decoupled), and neutralino LOSP with $m_{\chi^0_1}=$100, 300 and 500 GeV. The Simplified Model {\bf Sim0} is scanned in the same ($m_{\tilde{g}},m_{\tilde{q}}$) mass plane with the neutralino mass $m_{\chi^0_1}=$ 10, 100, 300 and 500 GeV, where 10 GeV is chosen for comparison with the ADM model with squark LOSP.  The gluino and squark mass parameters are scanned by generating 10,000 events for each parameter, from 100 GeV to 3000 GeV with 100 GeV spacing. For a squark LOSP, we additionally impose the condition $m_{\tilde{g}} > m_{\tilde{q}}$.  For high cross section regions where $m_{\tilde{g}}$ or $m_{\tilde{q}}$ is below 1000 GeV, we scale the number of events as needed to reduce statistical errors.  

For the 1-2 lepton analysis, in which only the $q \ell d^c$ model is relevant, we additionally scan the mass parameters in the plane of $(m_{\tilde{g}},m_{\chi^0_1})$ and $(m_{\tilde{q}},m_{\chi^0_1})$ with decoupled squarks and gluino, respectively. The $(m_{\tilde{g}},m_{\chi^0_1})$ scan is compared with the Simplified Model {\bf Sim1g} and the $(m_{\tilde{q}},m_{\chi^0_1})$ scan is compared with the Simplified Model {\bf Sim1q}, so that we generate events for those Simplified Models in the same scanning.  Due to reduced experimental sensitivity, the scan region is confined to 1500 GeV for $m_{\tilde{g}}$, to 1300 GeV for $m_{\tilde{q}}$, and to 1000 GeV for $m_{\chi^0_1}$.  We reduce the grid spacing to 50 GeV for this scan.  We also show the 1-2 lepton constraint for the $q \ell d^c$ model in the $(m_{\tilde{g}},m_{\tilde{q}})$ plane, but we compare the result only with the 0 lepton analysis constraint for the Simplified Model {\bf Sim0}. Again, the ADM mass is fixed to be 10 GeV.

\subsection{Results}

We discuss our results for the $q \ell d^c$ model, followed by the $u^c d^c d^c$.  For the former model, we apply both the 0 lepton and 1-2 lepton analyses, while for the latter we apply the 0 lepton analysis only.  In each case, we consider a squark LOSP decay into the ADM sector first (which is topologically most similar to the Simplified Model for comparison), before constraining a neutralino LOSP decay into the ADM sector.

\subsubsection{$W = X q \ell d^c$}

\vspace{0.1in}

\underline{\textit{Squark LOSP}}

\vspace{0.1in}

\begin{figure}
\centering
\subfigure[~0 lepton+ 2-6 jets + MET]{\includegraphics[width=7.5cm]{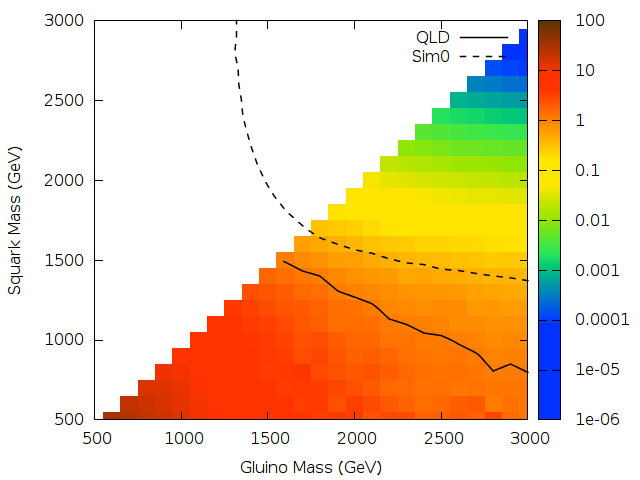} }
\quad
\subfigure[~1-2 lepton + 3-6 jets + MET]{\includegraphics[width=7.5cm]{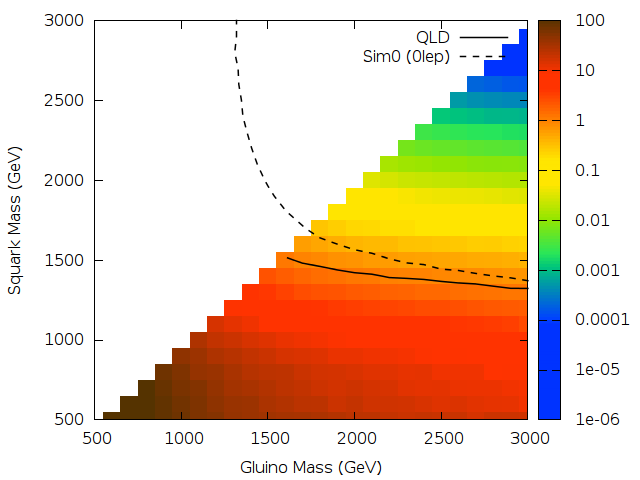} }
 \caption{ATLAS 0 lepton+2-6 jet+MET / 1-2 lepton + 3-6 jet + MET analyses  for the $q \ell d^c$ model with squark LOSP (solid line) in ($m_{\tilde g},m_{\tilde{q}}$) plane compared with the Simplified Model {\bf Sim0} of Fig.~\ref{fig:collider_for_0lep_simplified_SUSY} (dashed line).  Here the constraint on the Simplified Model {\bf Sim0} is taken from the result of the 0 lepton analysis for both figures. The neutralino mass $m_{\chi^0_1}$ for {\bf Sim0} is 10 GeV for comparison with the ADM mass in the $q \ell d^c$ model.}
\label{fig:QLD_squarklosp_scan}
\end{figure}

We first present the squark LOSP case of the $q \ell d^c$ model via the diagrams of Fig.~\ref{fig:collider_for_0lep_xudd_squarklsp}.   We assume the first two generation squarks are nearly degenerate in mass, but have a large enough mass splitting that the heavier squarks decay promptly to very soft (undetectable) jets and leptons and a lighter squark until the lightest squark is reached at the bottom of the cascade. We implement this by putting a 5 GeV mass splitting between the lightest squark and the others. The LOSP squark finally decays to the ADM with a quark and a lepton/neutrino. Hence, additional jets and leptons appear in the event, but the missing energy is reduced. 

The result of the ATLAS 0 lepton + 2-6 jet + MET and the ATLAS 1-2 lepton + 3-6 jet + MET analyses at $\sqrt{s}$ = 8 TeV with luminosity of 20.3 fb$^{-1}$ for the squark LOSP $q \ell d^c$ case are shown in Fig.~\ref{fig:QLD_squarklosp_scan}. The color level shows the maximum of $S_i / S^{95}_{{\rm exp},i}$ for all channels $i$, where $S_i$ is the number of events for the channel from our event generation at a given point, and $S^{95}_{{\rm exp},i}$ from the analysis given in Table \ref{table:ATLAS_0lep_2_6jet_MET} (for Fig. ~\ref{fig:QLD_squarklosp_scan}a) and Table \ref{table:ATLAS_1_2lep_3_6jet_MET_part1} and \ref{table:ATLAS_1_2lep_3_6jet_MET_part2} (for Fig. ~\ref{fig:QLD_squarklosp_scan}b).  Thus the contour at 1 (shown as the dashed or solid lines) can be interpreted roughly as the 95\% C.L. exclusion\footnote{A correct interpretation of the confidence level by combining such multiple \textit{non-exclusive} channels must be taken with care, and it is beyond the scope of this paper.  }.  In the plots, we show the Simplified Model {\bf Sim0} exclusion contour by performing the same 0 lepton analysis. The neutralino mass for the {\bf Sim0} model is 10 GeV.

\begin{figure}
\centering
\subfigure[~$E_T^{\rm miss}$]{\includegraphics[width=7.5cm,height=5.5cm]{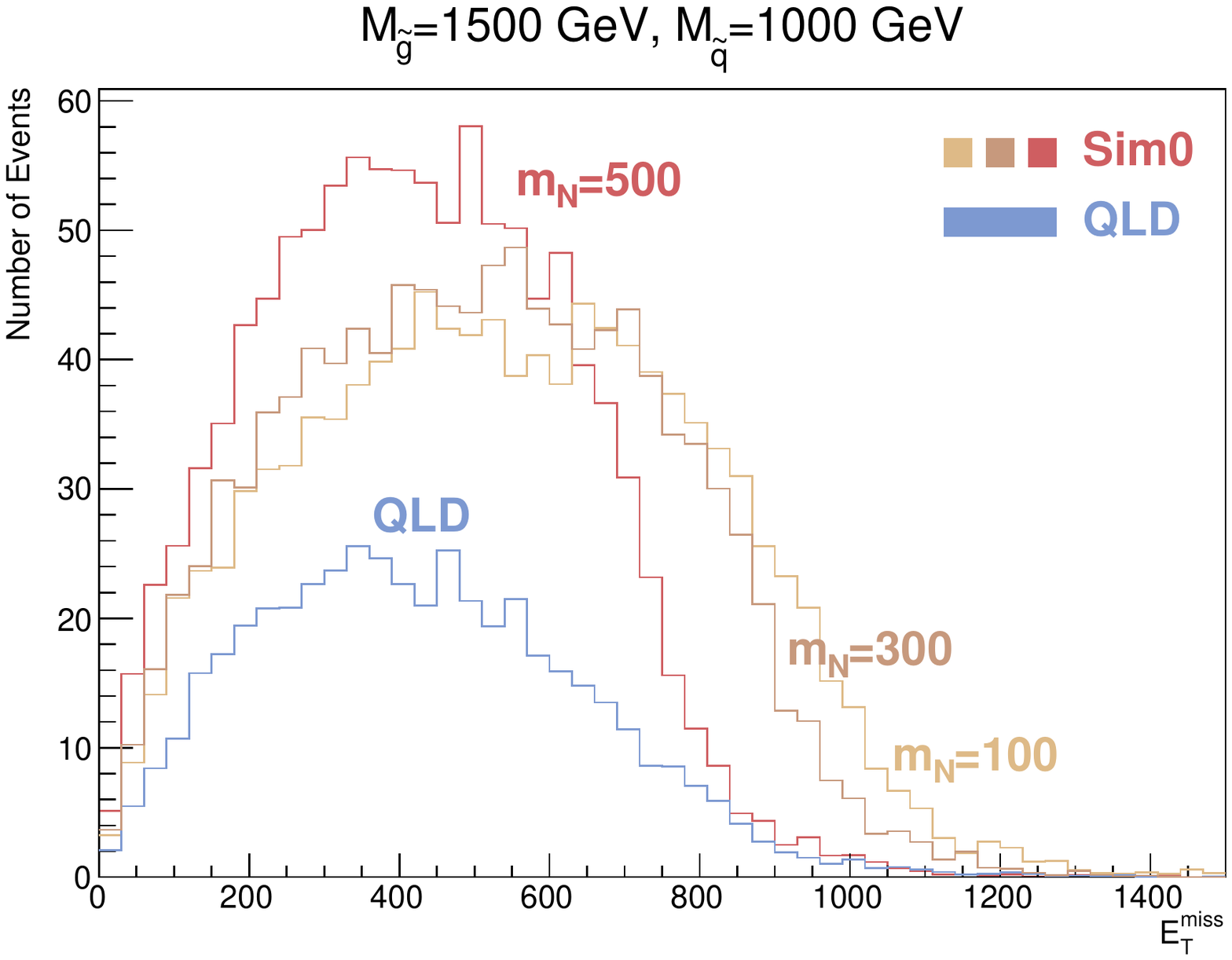}}
\quad
\subfigure[~Hardest lepton $p_T$]{\includegraphics[width=7.5cm,height=5.5cm]{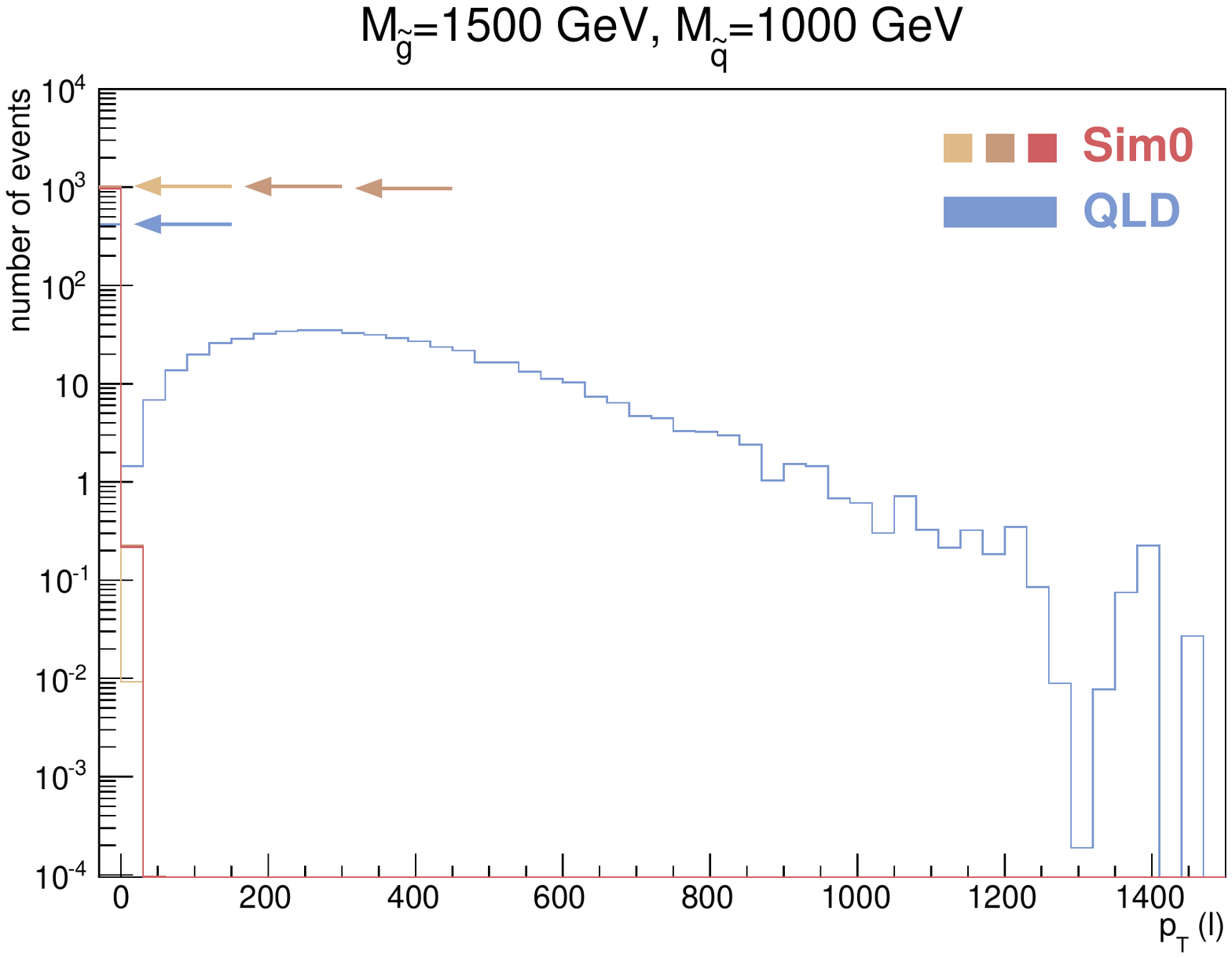}}
< \caption{The missing transverse energy (MET) distribution and the $p_T$ distribution of the hardest lepton of $X q \ell d^c$ model with squark LOSP (blue histogram) and Simplified Models {\bf Sim0} (red histograms). For the Simplified Models, we show three different neutralino masses $m_{\chi^0_1} =$ 100, 300 and 500 GeV. 
< For the lepton $p_T$, the first bin shows the number of events that passes the lepton veto cut of 0 lepton analysis.  We indicate the first bin using arrows in the right panel. The color scheme for the neutralino mass is the same for both graphs. }
\label{fig:QLD_SquarkLOSP_distribution}
\end{figure}

The 0 lepton analysis result shows that the constraint is weaker for the $q \ell d^c$ model, while the 1-2 lepton analysis constraint for the $q \ell d^c$ model is similarly matched with the Simplified Model {\bf Sim0} 0 lepton analysis constraint.  
The reason why the constraint from the 0 lepton analysis on the $q \ell d^c$ model is weaker is simply because half of the LOSP squarks decay into a charged lepton, which is vetoed in the analysis.

To see this, we show the MET distribution and the $p_T$ distribution of the hardest lepton in Fig.~\ref{fig:QLD_SquarkLOSP_distribution} at a mass parameter point $(m_{\tilde{g}},m_{\tilde{q}})=$ (1500 GeV, 1000 GeV). The MET distribution in Fig.~\ref{fig:QLD_SquarkLOSP_distribution}a is obtained after applying signal object identification/isolation, the lepton veto, and the two hardest jet $p_T$ cuts: $p_T(j_1) > 130$ GeV, $p_T(j_2) > 60$ GeV, from the 0 lepton analysis. The $p_T$ distribution of the hardest lepton in Fig.~\ref{fig:QLD_SquarkLOSP_distribution}b is obtained after applying the same cuts except the lepton veto cut, instead applying the MET cut: $E_T^{\rm miss} > 160$ GeV. One can easily see the MET distribution in Fig. \ref{fig:QLD_SquarkLOSP_distribution}a is not very different for the Simplified Model {\bf Sim0} than for the $q \ell d^c$ model, though the rate is different due to the lepton veto as one can see in the lepton $p_T$ distribution in Fig. \ref{fig:QLD_SquarkLOSP_distribution}b: the Simplified Model {\bf Sim0} has 100\% no-lepton events, while the $qld^c$ model has 45\% no-lepton events.

\begin{figure}
\centering
\subfigure[~0 lepton analysis for  $m_{\chi^0_1}=100$ GeV]{\includegraphics[width=7.5cm]{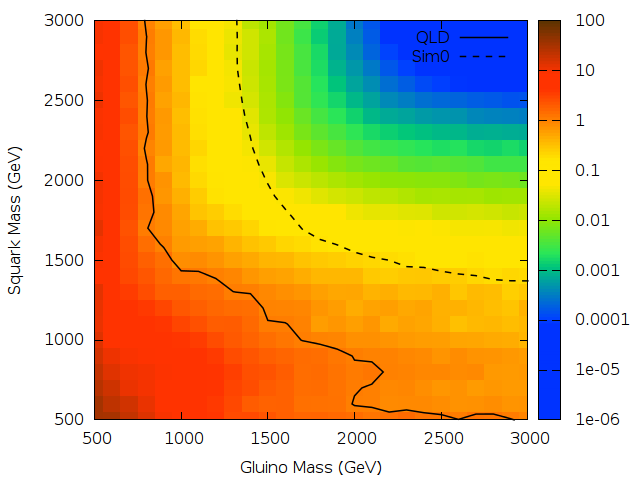}}
\quad
\subfigure[~1-2 lepton analysis for $m_{\chi^0_1} = 100$ GeV]{\includegraphics[width=7.5cm]{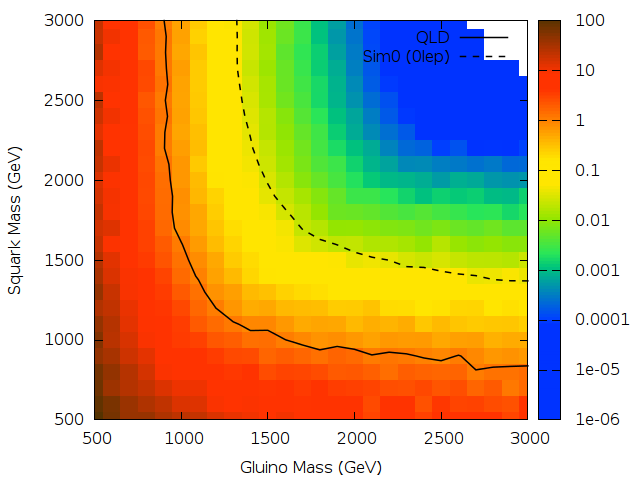}}

\subfigure[~0 lepton analysis for  $m_{\chi^0_1}=300$ GeV]{\includegraphics[width=7.5cm]{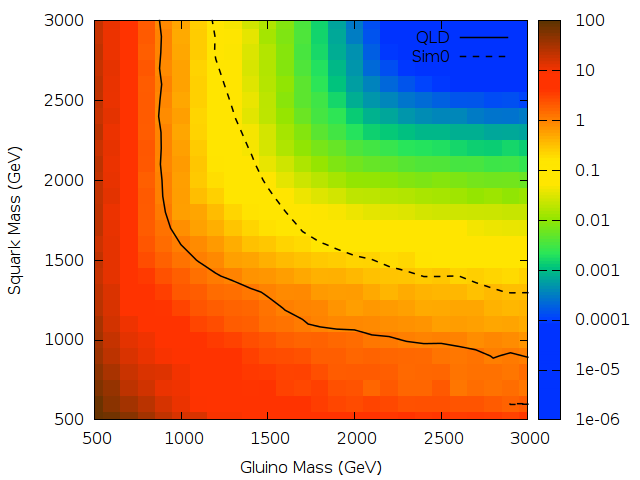} }
\quad
\subfigure[~1-2 lepton analysis for $m_{\chi^0_1} = 300$ GeV]{\includegraphics[width=7.5cm]{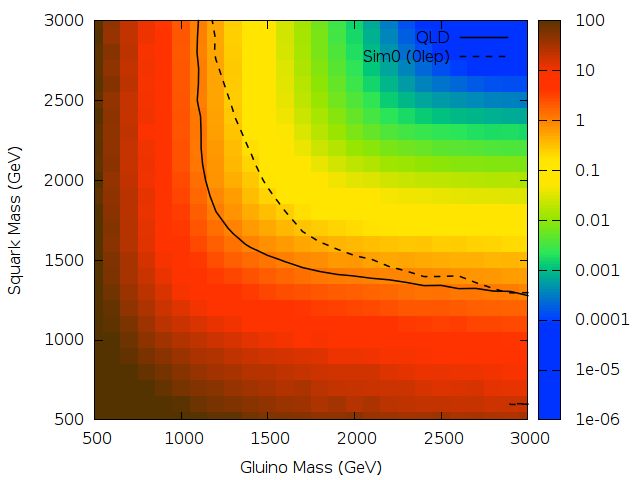} }

\subfigure[~0 lepton analysis for $m_{\chi^0_1} = 500$ GeV]{\includegraphics[width=7.5cm]{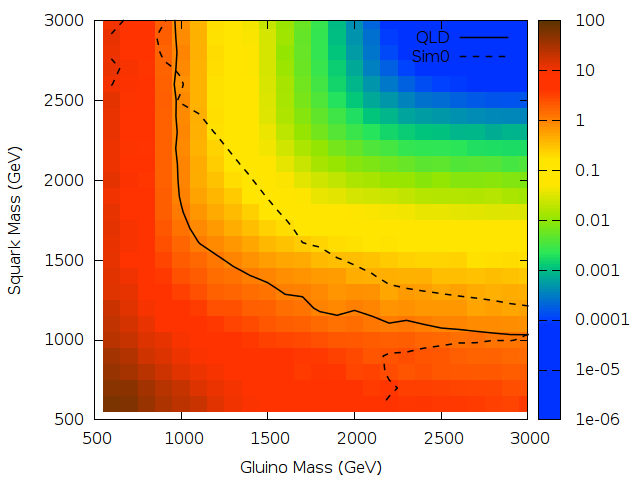}}
\quad
\subfigure[~1-2 lepton analysis for $m_{\chi^0_1} = 500$ GeV]{\includegraphics[width=7.5cm]{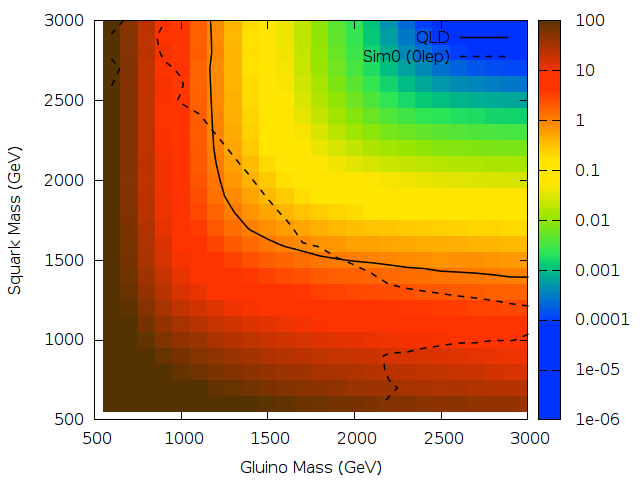}}
\caption{ Constraint from ATLAS 0 lepton ({\it left}) and 1-2 lepton ({\it right}) analyses  on the $q \ell d^c$ model with neutralino LOSP (solid line) in the ($m_{\tilde g},m_{\tilde{q}}$) plane, compared with the Simplified Model {\bf Sim0} (dashed line).  Here the constraint on the {\bf Sim0} model of Fig.~\ref{fig:collider_for_0lep_simplified_SUSY} is taken from the 0 lepton analysis result for both left- and right-hand plots.  The $X q \ell d^c$ model with a neutralino LOSP decays through Fig.~\ref{fig:collider_for_0lep_xudd_neutralinolsp}, with $m_{\chi^0_1} =$ 100, 300 and 500 GeV.  }
\label{fig:QLD_neutlosp_scan}
\end{figure}

\begin{figure}
\centering
\subfigure[$E_T^{\rm miss}$]{\includegraphics[width=7.5cm,height=5.5cm]{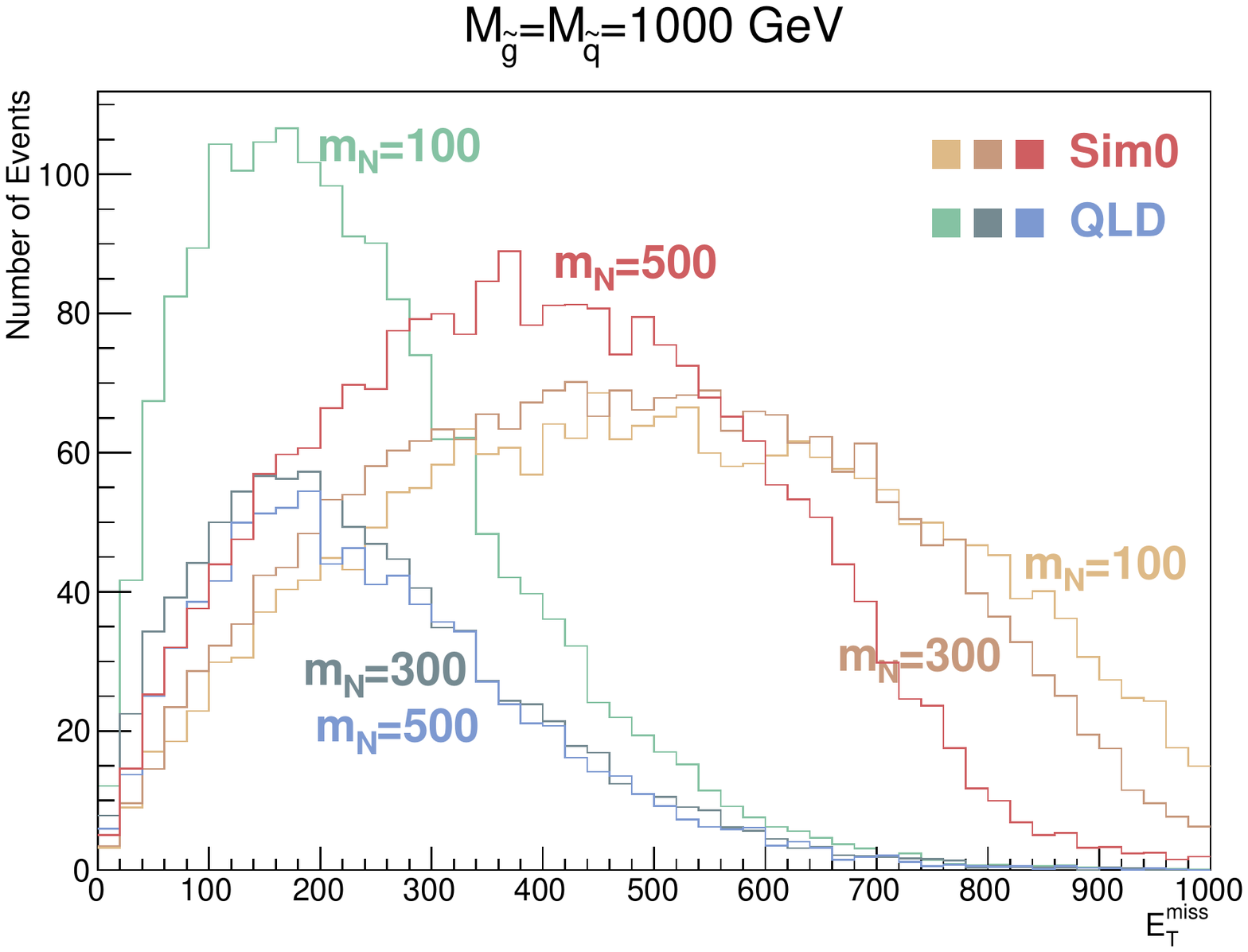}}
\quad
\subfigure[Hardest lepton $p_T$]{\includegraphics[width=7.5cm,height=5.5cm]{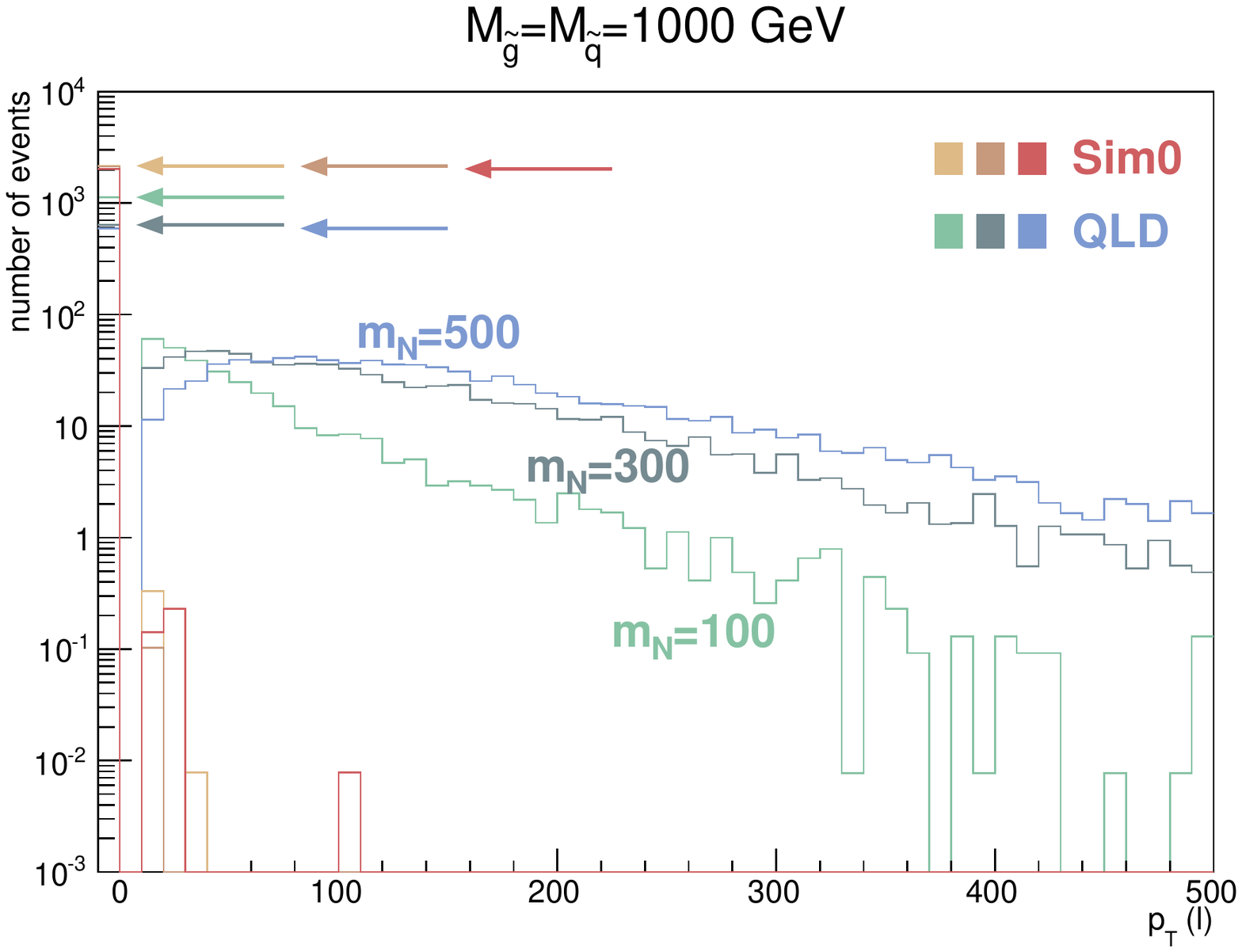}}
\caption{The MET ({\em left}) and transverse momentum $p_T$ ({\em right}) of the hardest lepton distributions in $q \ell d^c$ model (blue histograms) and Simplified Model {\bf Sim0} (red histograms) for $m_{\tilde{g}} = m_{\tilde{q}} = 1000$ GeV and $m_{\chi^0_1} =$ 100, 300 and 500 GeV.  In the right panel, the first bin shows the number of events that passes lepton veto cut of the 0 lepton analysis.  We indicate the first bin using arrows in the right panel. The color scheme for the neutralino mass is the same for both graphs.    }
\label{fig:QLD_NeutralinoLOSP_distribution}
\end{figure}

\vspace{0.1in}

\underline{\textit{Neutralino LOSP}}

\vspace{0.1in}

Next, we present the constraints for the neutralino LOSP case of the $q \ell d^c$ model via the diagrams of Fig.~\ref{fig:collider_for_0lep_xudd_neutralinolsp}. In this case, we do not have to assume a splitting between squarks since squarks decay promptly into the neutralino. The $(m_{\tilde{g}},m_{\tilde{q}})$ scan results of the ATLAS 0 lepton and 1-2 lepton analyses are shown in Fig.~\ref{fig:QLD_neutlosp_scan} for three different neutralino mass choices: $m_{\chi^0_1}=$ 100, 300 and 500 GeV. Again, we compare the result of the $q \ell d^c$ model with the 0 lepton analysis of the Simplified Model {\bf Sim0} with the same neutralino mass parameters. The contours of ${\rm Maximum}_i ( S_i / S^{95}_{{\rm exp},i} )= 1$ for $q\ell d^c$ and {\bf Sim0} are drawn as solid and dashed curves, respectively.

The constraints for the neutralino LOSP $q\ell d^c$ model are generically weaker than the Simplified Model {\bf Sim0} for small $m_{\chi^0_1}$ (100 GeV and 300 GeV), but reveal more complicated behavior in the $m_{\chi^0_1}=$ 500 GeV case. Several factors contribute to these results.  
One obvious factor that tends to give weaker constraints on the ADM model in the 0 lepton analysis is the branching fraction to charged leptons, which we have already seen in the squark LOSP case. More importantly, the missing energy of the neutralino is reduced as it decays to two additional jets plus a lepton. This feature is transparently comparable with the Simplified Model {\bf Sim0} since both models share the same event topology before the neutralino decay.  On the other hand, as the neutralino mass is set heavier, the energy of the jets from gluino/squark decay into the neutralino becomes smaller as the mass difference shrinks.  Therefore, the experimental sensitivity to the Simplified Model {\bf Sim0} (and ordinary R-partiy conserving MSSM scenarios generically) is reduced for a heavier neutralino mass, while the ADM models are subject to more severe constraints since a massive neutralino is able to ``store'' and transfer energy to the ADM particle. 
Therefore, for large neutralino mass, the ADM model can actually become substantially more constrained than the Simplified Model.

In Fig.~\ref{fig:QLD_NeutralinoLOSP_distribution}, we compare the MET distribution and the hardest lepton $p_T$ distribution of the neutralino LOSP $q\ell d^c$ model and the Simplified Model {\bf Sim0}  for $m_{\tilde{g}} = m_{\tilde{q}} = 1000$ GeV and $m_{\chi^0_1}$ = 100, 300 and 500 GeV.  Here, we use the same cuts as in Fig.~\ref{fig:QLD_SquarkLOSP_distribution}. Note that $E_T^{\rm miss}$ is distinctively smaller for the $q \ell d^c$ case. For the lepton $p_T$ distribution, the first bin implies events that pass the lepton veto cut. Note the significant difference among different $m_{\chi^0_1}$'s in the lepton veto and $p_T$ for the $q \ell d^c$ models. The acceptance of the lepton veto is 78.3\%, 47.8\% and 41.9\% for 100 GeV, 300 GeV and 500 GeV neutralino, respectively, for the $q \ell d^c$ model, while the acceptance of the lepton veto is nearly 100\% for the Simplified Model {\bf Sim0}. This implies leptons from light $\chi^0_1$ decay often fails the lepton veto cut ($p_T^{\ell} < 10$ GeV in this case).

\begin{figure}
\centering
\subfigure[~gluino-neutralino scan]{\includegraphics[width=7.5cm]{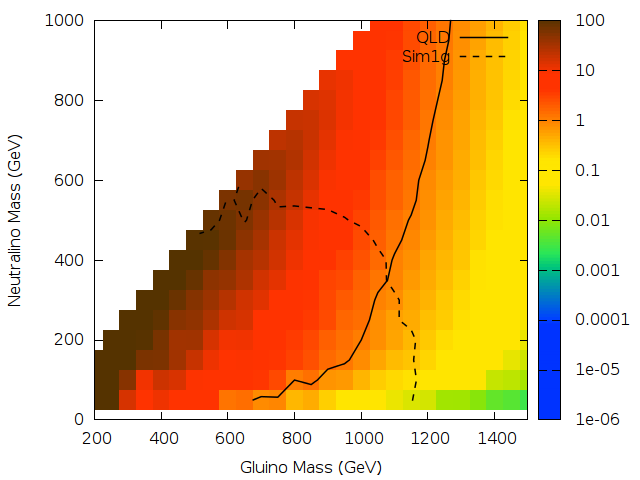} }
\quad
\subfigure[~squark-neutralino scan]{\includegraphics[width=7.5cm]{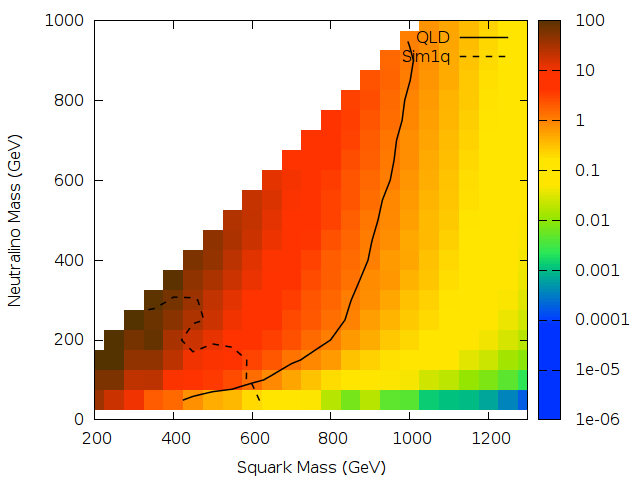}}
\caption{ATLAS 1-2 lepton+3-6 jet+MET SUSY search analysis  for the $q \ell d^c$ model.  Here the constraints on the MSSM Simplified Model of Fig.~\ref{fig:collider_for_1_2lep_simplified_SUSY} are compared against the $q \ell d^c$ model with a neutralino LOSP decaying through Fig.~\ref{fig:collider_for_0lep_xudd_neutralinolsp}.  The squarks have been decoupled in the left panel, while in the right panel the gluinos have been decoupled.}
\label{fig:multilep_scan}
\end{figure}

\begin{figure}
\centering
\subfigure[~$E_T^{\rm miss}$]{\includegraphics[width=8cm,height=6.5cm]{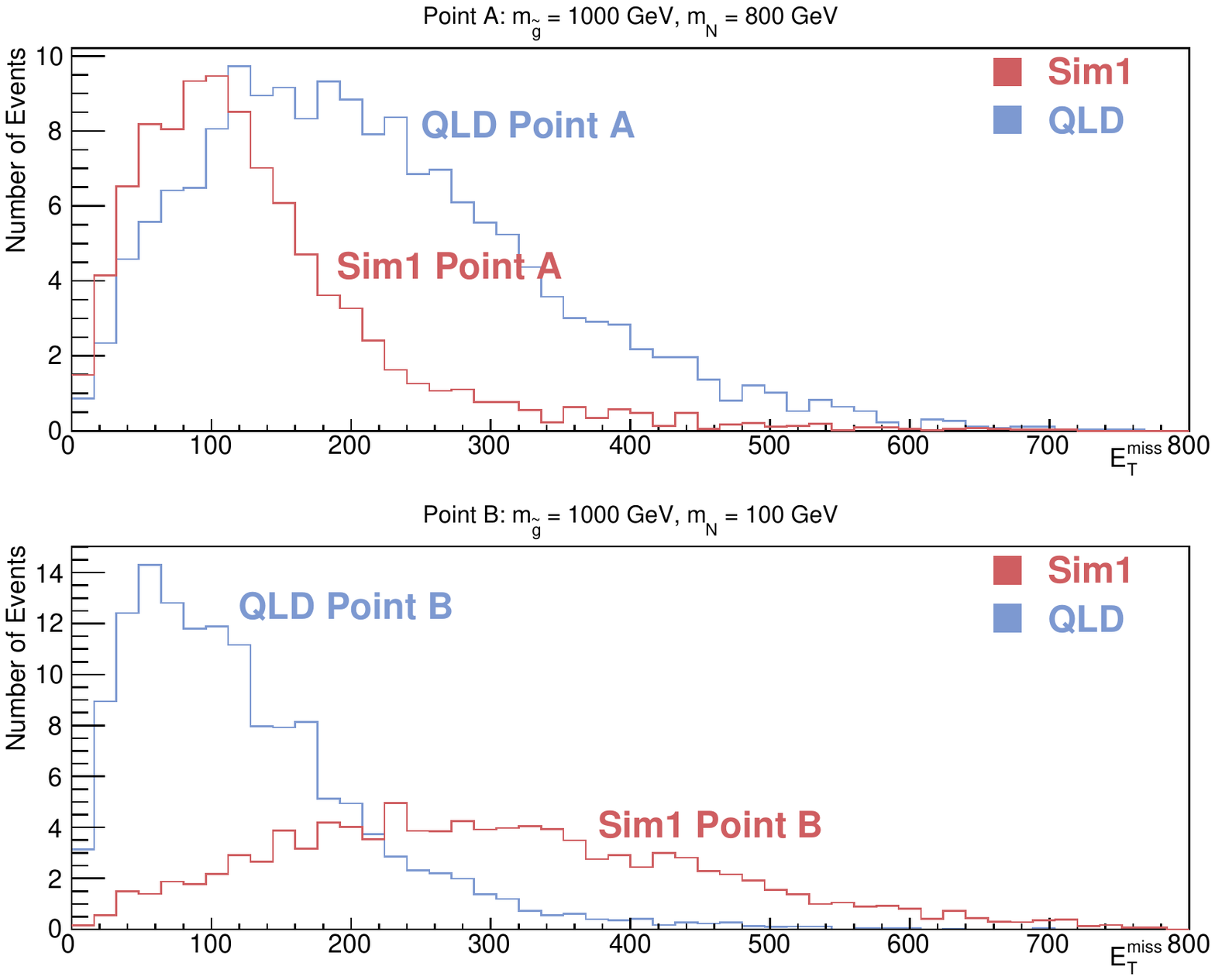}}
\subfigure[~$p_T (j1)$]{\includegraphics[width=8cm,height=6.5cm]{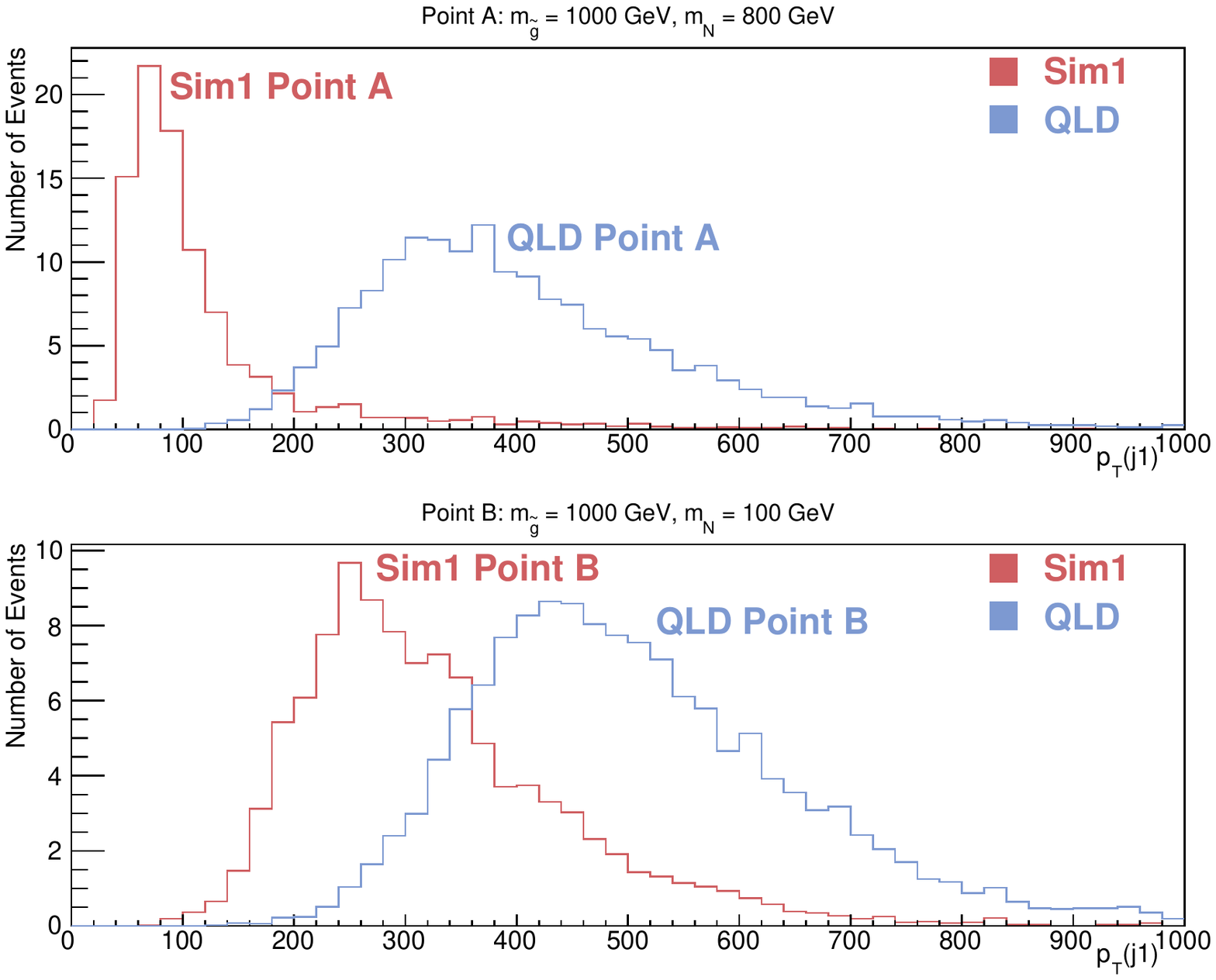}}

\caption{The missing transverse energy distribution and lepton $p_T$ distribution of the neutralino LOSP $q \ell d^c$ model with squark decoupled and the Simplified Model {\bf Sim1g} in the 1-2 lepton analysis.  Here, we chose two points of mass parameter: (A) $m_{\tilde{g}} =$ 1000 GeV, $m_{\chi^0_1}$ = 800 GeV and (B) $m_{\tilde{g}} =$ 1000 GeV, $m_{\chi^0_1}$ = 100 GeV.  }
\label{fig:1lep_MET_pTj1}
\end{figure}

Lastly, in Fig. \ref{fig:multilep_scan}, we compare the constraints from the ATLAS 1-2 lepton anlaysis for  
the neutralino LOSP $q \ell d^c$ model and the Simplified Model {\bf Sim1g} ({\bf Sim1q})  in the gluino(squark)-neutralino plane. The constraints have  completely different behaviors for each model from the same analysis, because the decay of a massive neutralino results in high $p_T$ and MET in $q \ell d^c$ while a smaller gap between the neutralino and the gluino (squark) tends to give softer jets and MET in the Simplified Models. This feature is illustrated clearly in the MET and the hardest jet $p_T$ distributions, shown in Fig. \ref{fig:1lep_MET_pTj1} for two benchmark points: (A) $m_{\tilde{g}} = 1000$ GeV and $m_{\chi^0_1} = 800$ GeV, (B) $m_{\tilde{g}} = 1000$ GeV and $m_{\chi^0_1} = 100$ GeV. To obtain Fig. \ref{fig:1lep_MET_pTj1}, we applied the $p_T$ cut of the ``Soft 1-$\ell$'' class (in Table \ref{table:ATLAS_1_2lep_3_6jet_MET_part1}) to the hardest three jets for soft lepton events, and applied the $p_T$ cut of the ``Hard 1-$\ell$ 3 jet'' class (in Table \ref{table:ATLAS_1_2lep_3_6jet_MET_part1}) to the hardest three jets for hard lepton events. One can easily see that the MET and $p_T(j_1)$ of $q \ell d^c$ model (blue color-coded) is higher for benchmark point A (above), but the MET and $p_T (j_1)$ of the Simplified Model (red color-coded) is higher for point B (below).

Before moving on to the $u^c d^c d^c$ model, we comment on observing the states which UV complete the ADM operators.  In principle these states, $Q,~L$ and $D$, can be directly produced at the collider.  When these states decay to the DM, $Q,~D \rightarrow X + q$ and $L \rightarrow X + \ell$, the signatures look similar to squark or stop signatures of jet or top quark plus missing energy, or slepton and sneutrino decays to lepton plus missing energy.  On the other hand, these states may have more exotic decays, for example, to a lepton and jet, or to flavor violating pairs of quarks such as a top and a light flavor jet.  For example, we may have
\begin{eqnarray}
D & \rightarrow & u \ell^- \\ \nonumber
Q_{u,d} & \rightarrow & (\ell^+,\nu) d, 
\end{eqnarray}
leading to the possibility of spectacular decay modes at the LHC, which are similar in spirit to leptoquark searches at the LHC.  The study of such signatures could give rise to interesting further constraints on ADM models.

\subsubsection{$W = X u^c d^c d^c$}

\vspace{0.1in}

\underline{\textit{Squark LOSP}}

\vspace{0.1in}

\begin{figure}
\centering
\includegraphics[width=7.5cm]{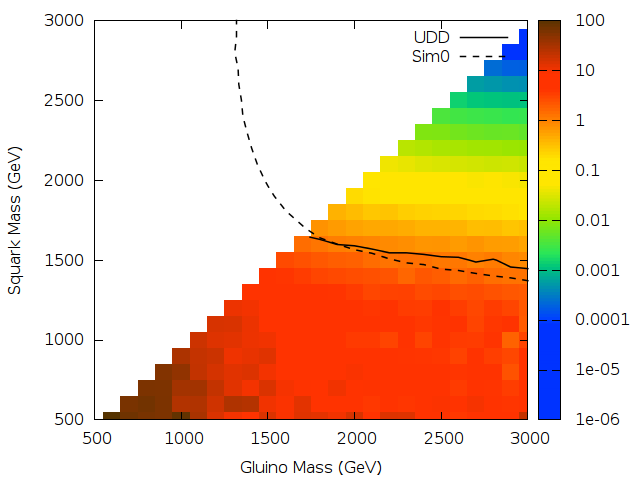}
\caption{ATLAS 0 lepton+2-6 jet+MET analysis  for the $u^c d^c d^c$ model with squark LOSP (solid curve) and Simplified Model {\bf Sim0} (dashed curve).   The solid curve ends for the $u^c d^c d^c$ model when the squark is no longer the LOSP. }
\label{fig:UDD_squarklosp_scan}
\end{figure}

Now we carry out the ATLAS 0 lepton analysis for the $u^c d^c d^c$ models and compare the result with the Simpified Model {\bf Sim0}.
First, we consider the squark LOSP case of the $u^c d^c d^c$ models. 
As in the $q \ell d^c$ model case, we assume that squarks have a large enough mass splitting for prompt decay to the lightest squark in the $u^c d^c d^c$ model, which is again implemented by a 5 GeV splitting between the lightest and other squarks. The relevant processes at the LHC are given in Fig. \ref{fig:collider_for_0lep_xudd_squarklsp} with a lepton/neutrino replaced by a jet in the lightest squark decay. The ADM mass is 10 GeV here, and for {\bf Sim0}, we set $m_{\chi^0_1} = 10$ GeV for a fair comparison.

Fig. \ref{fig:UDD_squarklosp_scan} shows the constraints from the 0 lepton analysis at $\sqrt{s} = 8$ TeV, 20.3 fb$^{-1}$ for the squark LOSP $u^c d^c d^c$ and the {\bf Sim0} model. The color level represents ${\rm max}_i (S_i / S^{95}_{{\rm exp,}i})$ for the $u^c d^c d^c$ model, similarly to the $q \ell d^c$ case, and thus the contour at 1 corresponds to 95\% C.L. roughly.  Interestingly, the constraints for both the $u^c d^c d^c$ model (solid) and the {\bf Sim0} model (dashed) are quite close, while the detailed distribution of relevant observables are much different for each model, as we see next. 

\begin{figure}
\centering
\subfigure[$E_T^{\rm miss}$]{\includegraphics[width=7.5cm]{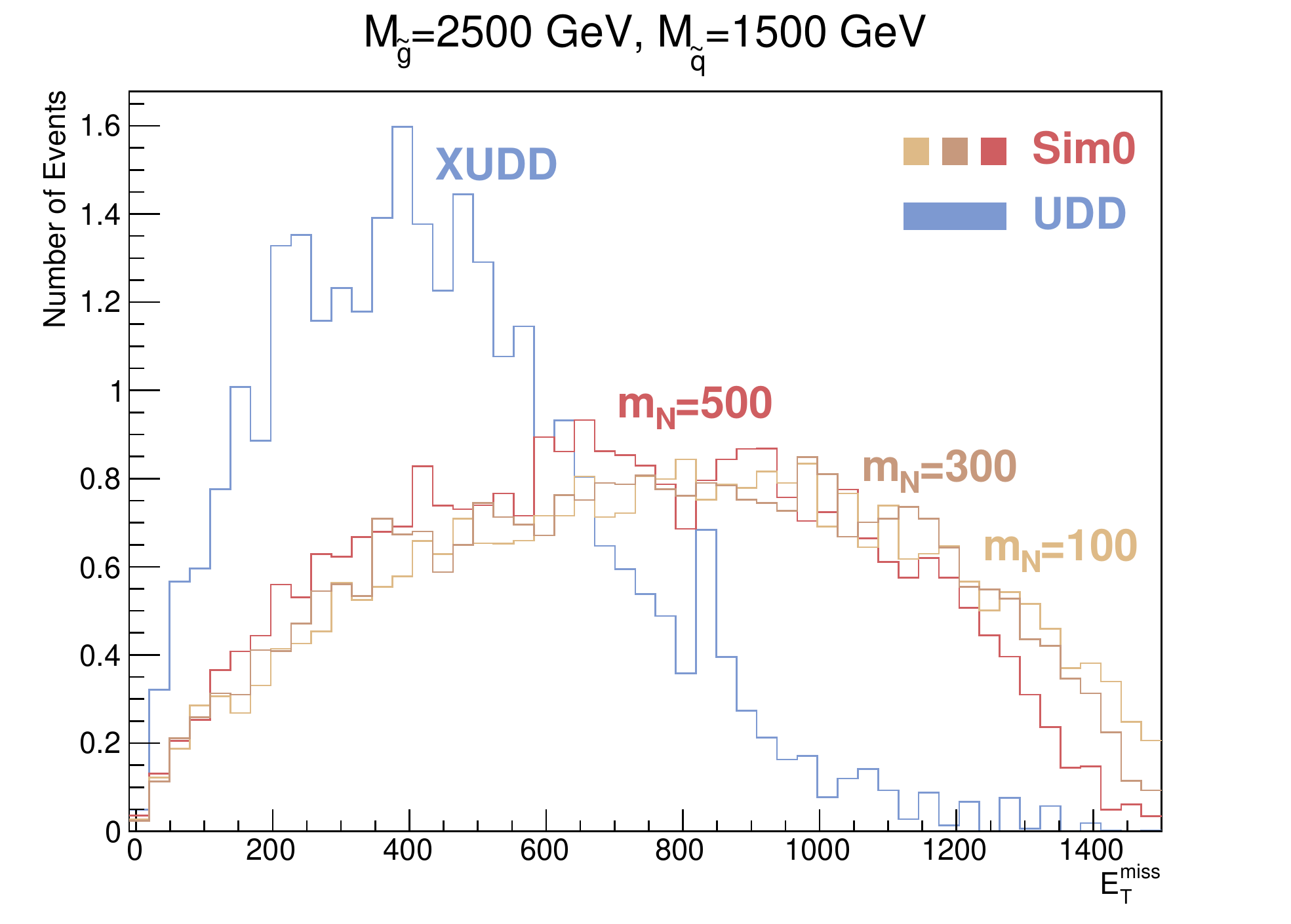} }
\quad
\subfigure[$N_{\rm jet}$]{\includegraphics[width=7.5cm]{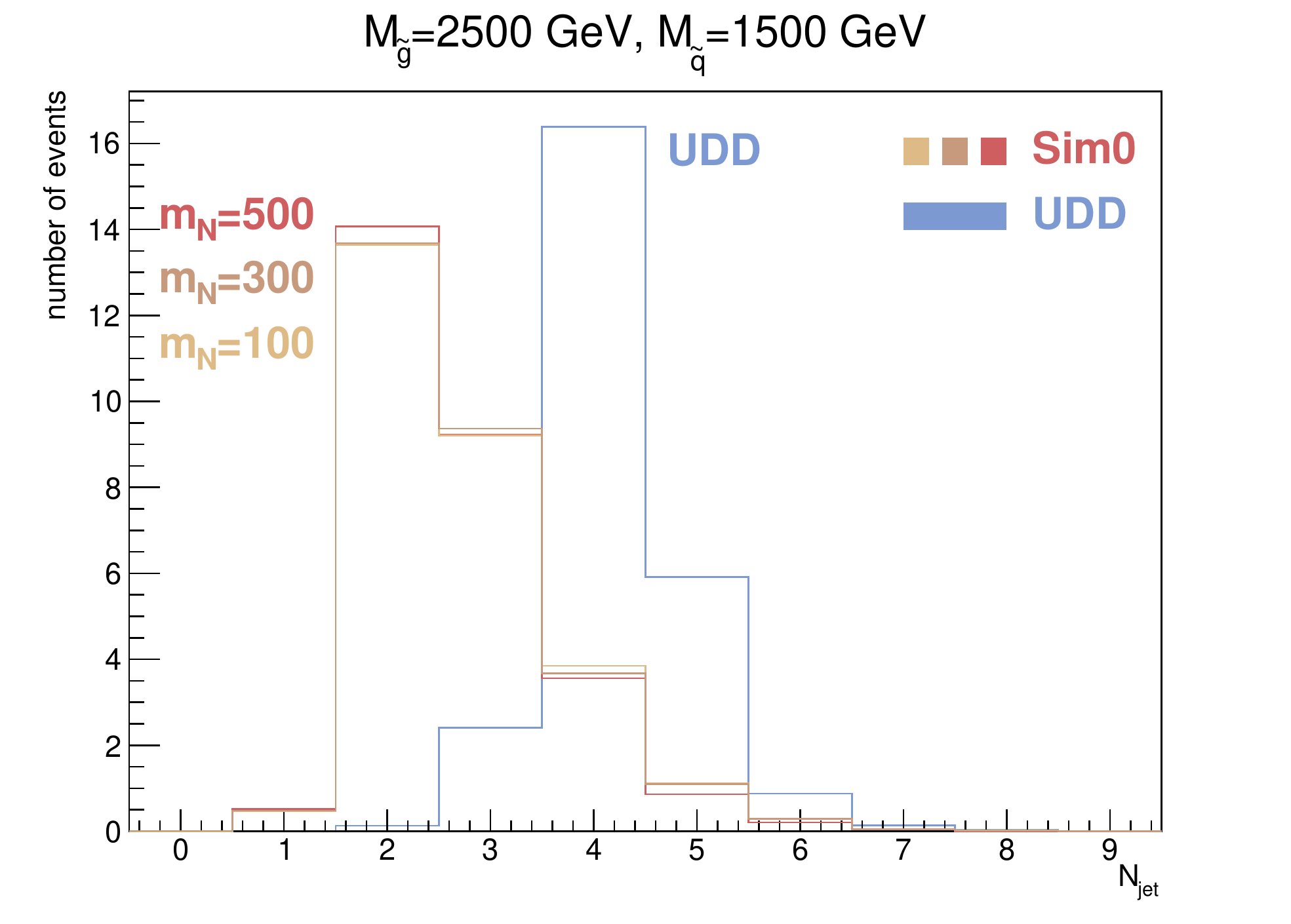} }

\subfigure[~$m_{\rm eff}$(incl.)]{\includegraphics[width=7.5cm]{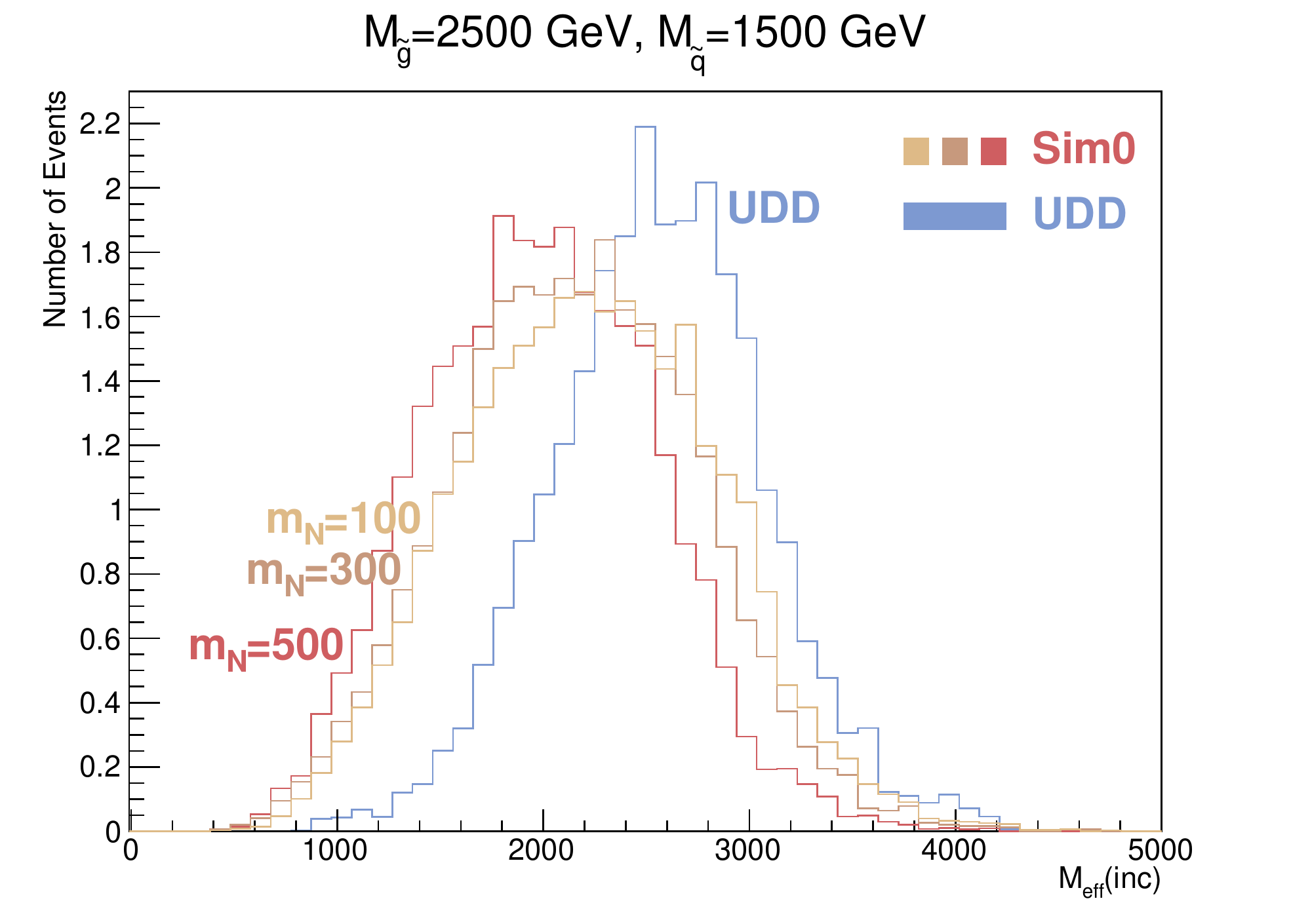} }
\quad
\subfigure[~$E_T^{\rm miss} / m_{\rm eff} (2j)$]{\includegraphics[width=7.5cm]{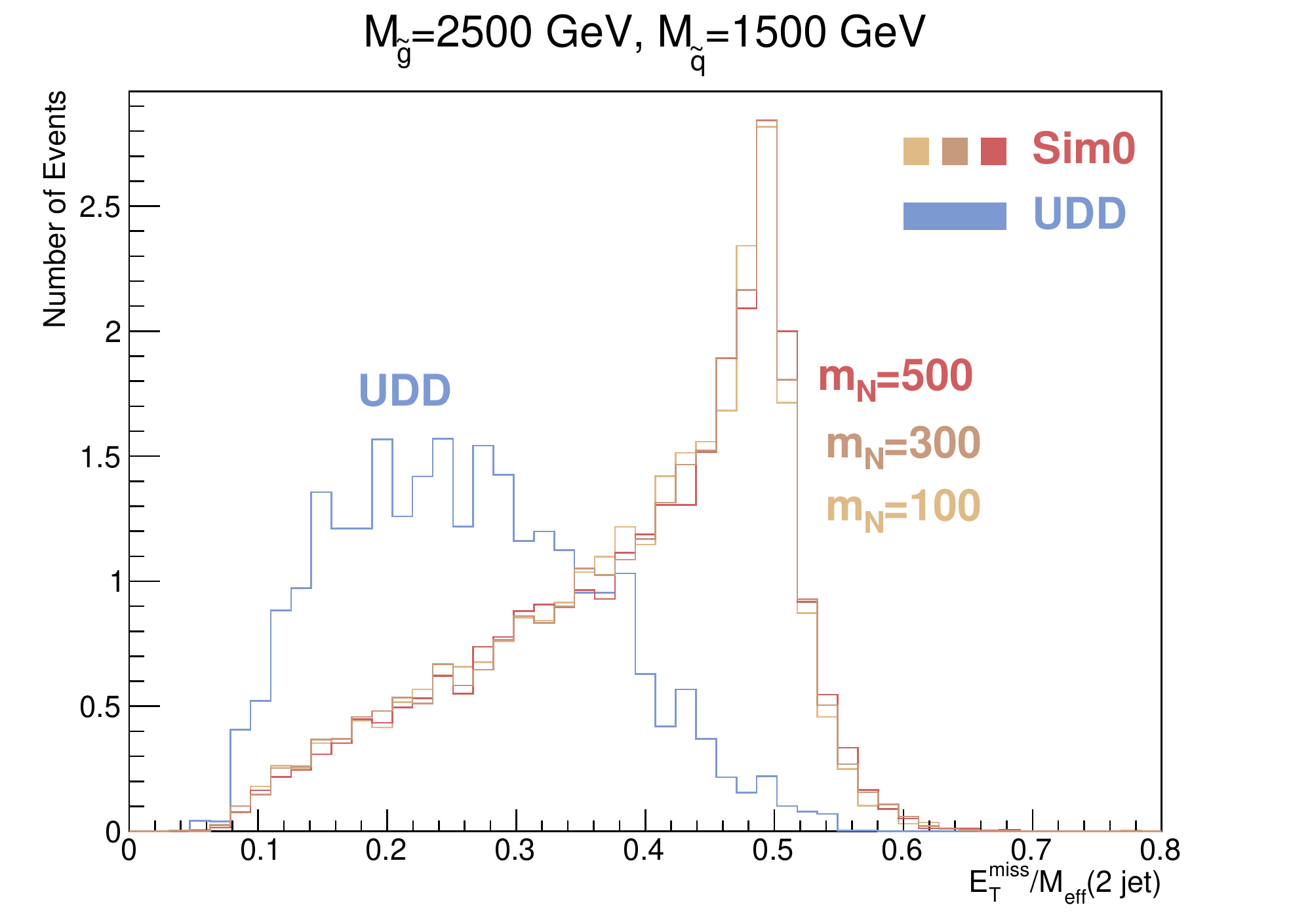} }
\caption{Various distributions for $u^cd^cd^c$ model with squark LOSP (blue histograms) compared with the Simplified Model {\bf Sim0} (red histograms): missing transverse energy $E_T^{\rm miss}$, number of jets $N_{\rm jets}$ with $p_T > 60$ GeV (for the hardest jet, $p_T > 130$ GeV), inclusive effective mass $m_{\rm eff}$(incl.) and $E_T^{\rm miss} / m_{\rm eff} (2j)$.  The mass parameters here are $m_{\tilde{g}}=2500$ GeV, $m_{\tilde{q}}=1500$ GeV. For the Simplified Model {\bf Sim0}, we show three different neutralino masses $m_{\chi^0_1} =$ 100, 300 and 500 GeV.  }
\label{fig:met_Njet_meff_xudd_vs_simplified_squarklosp}
\end{figure}

In Fig.~\ref{fig:met_Njet_meff_xudd_vs_simplified_squarklosp}, we show the $E_T^{\rm miss}$, $N_{\rm jet}$, $m_{\rm eff} ({\rm incl.})$ and $E_T^{\rm miss} / m_{\rm eff} (2j)$ distributions for both models at $m_{\tilde g} = 2500$ GeV and $m_{\tilde{q}} = 1500$ GeV. The chosen mass parameter set is near the limit of the experimental sensitivity. Here, $m_{\rm eff} ({\rm incl.})$ and $m_{\rm eff} (2j)$ are the effective mass defined inclusively, and exclusively with two hardest jets, respectively.
For {\bf Sim0}, we show three different neutralino masses: $m_{\chi^0_1} =$ 100, 300 and 500 GeV. For each histogram, we apply cuts in the 0 lepton analysis similarly to the case of Fig.~\ref{fig:QLD_SquarkLOSP_distribution}: after signal object identification/isolation, we apply the lepton veto, and the two hardest jet $p_T$ cuts: $p_T(j_1) > 130$ GeV, $p_T(j_2) > 60$ GeV for the MET distribution, and additionally the MET cut $E_T^{\rm miss} > 160$ GeV for the other distributions. 

One sees that the actual kinematic distributions are much different between the Simplified and $u^c d^c d^c$ models. Nonetheless, the reason why the $u^c d^c d^c$ model and {\bf Sim0} have similar constraints is due to saturation of cut acceptance. Near the 95\% C.L. experimental sensitivity, the cuts in Table \ref{table:ATLAS_0lep_2_6jet_MET} are not very effective in distinguishing one model from other since the $p_T$ of relevant objects and the MET are already very high.  The channels with harder cuts (for example, BT and CT) do not dominate the constraints, and hence do not distinguish between models. 
 For example, the acceptance of the AL channel cut is saturated above $m_{\tilde{q}} = 1000$ GeV for a fixed gluino mass $m_{\tilde{g}} = 2500$ GeV, to $\sim 0.5$ for $u^c d^c d^c$ and $\sim 0.75$ for {\bf Sim0}. Then, the constraints are simply determined by the production cross section, which is identical for both models. 

It is clear, however, that additional shape information from the kinematic distributions in Fig.~\ref{fig:met_Njet_meff_xudd_vs_simplified_squarklosp} is available for discrimination between the Simplified Model and ADM, so that the analysis could be better targeted to ADM models.  

\vspace{0.1in}

\underline{\textit{Neutralino LOSP}}

\vspace{0.1in}

\begin{figure}
\centering
\quad
\subfigure[~0 lepton analysis for $m_{\chi^0_1}$ = 100 GeV]{\includegraphics[width=7.5cm]{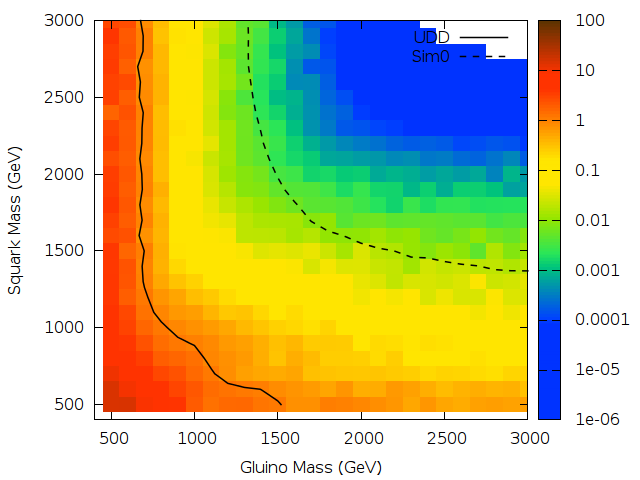}}
\quad
\subfigure[~0 lepton analysis for $m_{\chi^0_1}$ = 300 GeV]{\includegraphics[width=7.5cm]{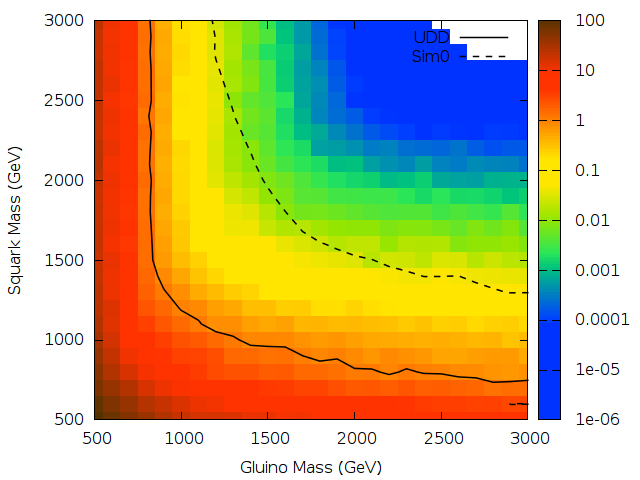} }
\quad
\subfigure[~0 lepton analysis for $m_{\chi^0_1}$ = 500 GeV]{\includegraphics[width=7.5cm]{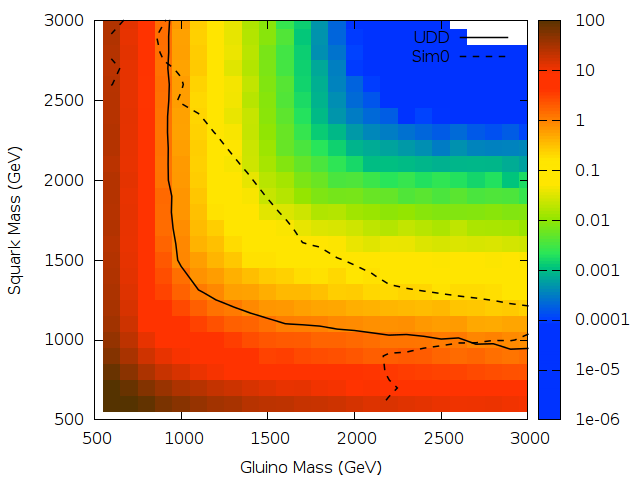}}
\caption{Constraint from ATLAS 0 lepton+2-6 jet+MET analysis on the $u^c d^c d^c$ model with neutralino LOSP (solid curve), compared with the Simplified Model {\bf Sim0} (dashed curve).    }
\label{UDDSearchB}
\end{figure}

Next we consider the $u^c d^c d^c$ model with a neutralino LOSP via the diagrams of Fig.~\ref{fig:collider_for_0lep_xudd_neutralinolsp} with the lepton/neutrino replaced with a jet.  The results are shown in Fig.~\ref{UDDSearchB}.  In this case, as for the $q \ell d^c$ model with neutralino LOSP, we do not have to assume a splitting between squarks since squarks decay promptly into the neutralino. The $(m_{\tilde{g}},m_{\tilde{q}})$ scan results of the ATLAS 0 lepton analysis is shown in Fig.~\ref{UDDSearchB} for three different neutralino mass choices: $m_{\chi^0_1}=$ 100, 300 and 500 GeV. Again, we compare the result of the $u^c d^c d^c$ model with the 0 lepton analysis of the Simplified Model {\bf Sim0} with the same neutralino mass parameters. The contours of ${\rm Maximum}_i ( S_i / S^{95}_{{\rm exp},i} )= 1$ for $u^c d^c d^c$ and {\bf Sim0} are drawn as solid and dashed curves, respectively. 

\begin{figure}
\centering
\subfigure[$E_T^{\rm miss}$]{\includegraphics[width=7.5cm]{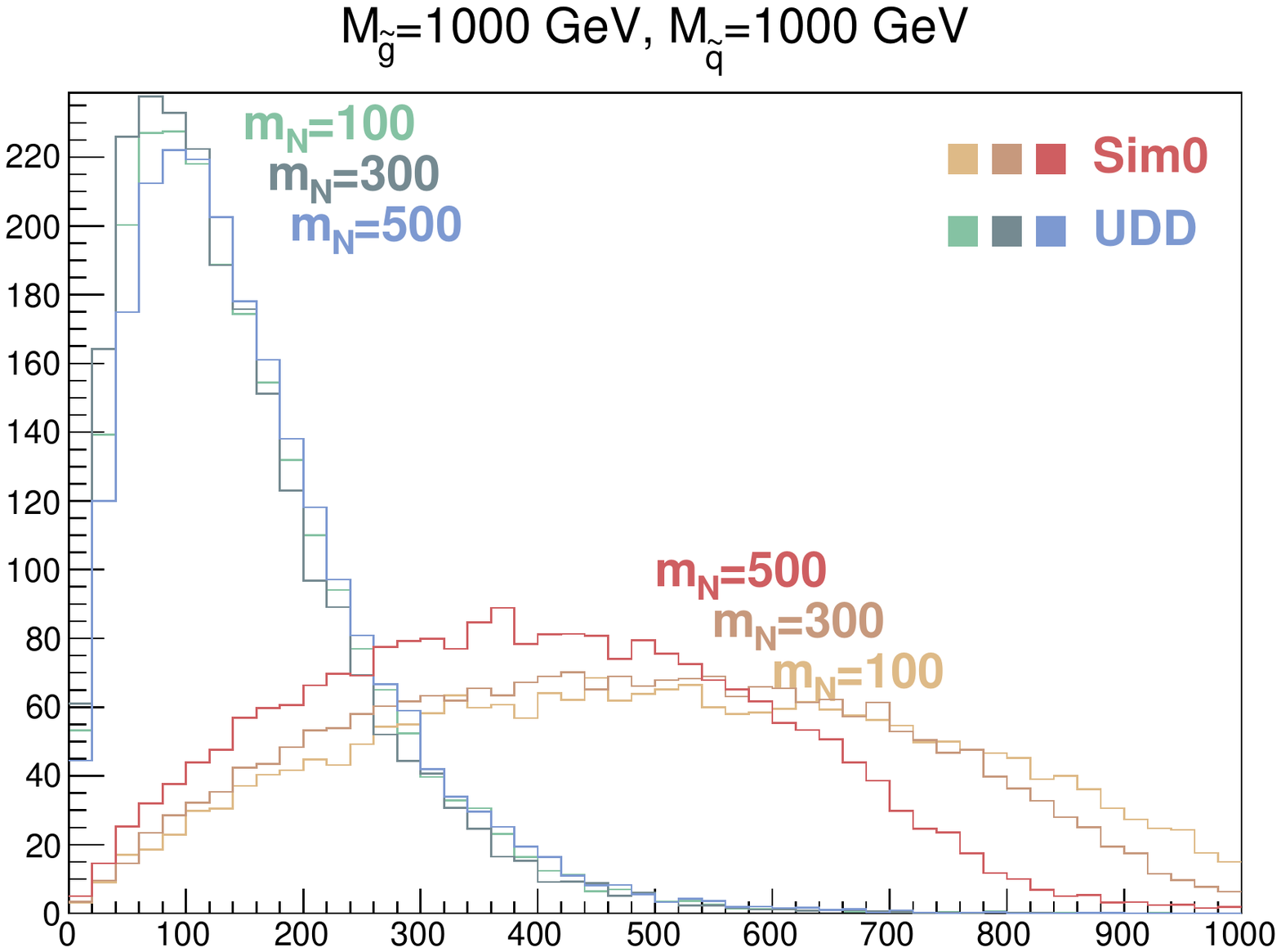} }
\quad
\subfigure[$N_{\rm jet}$]{\includegraphics[width=7.5cm]{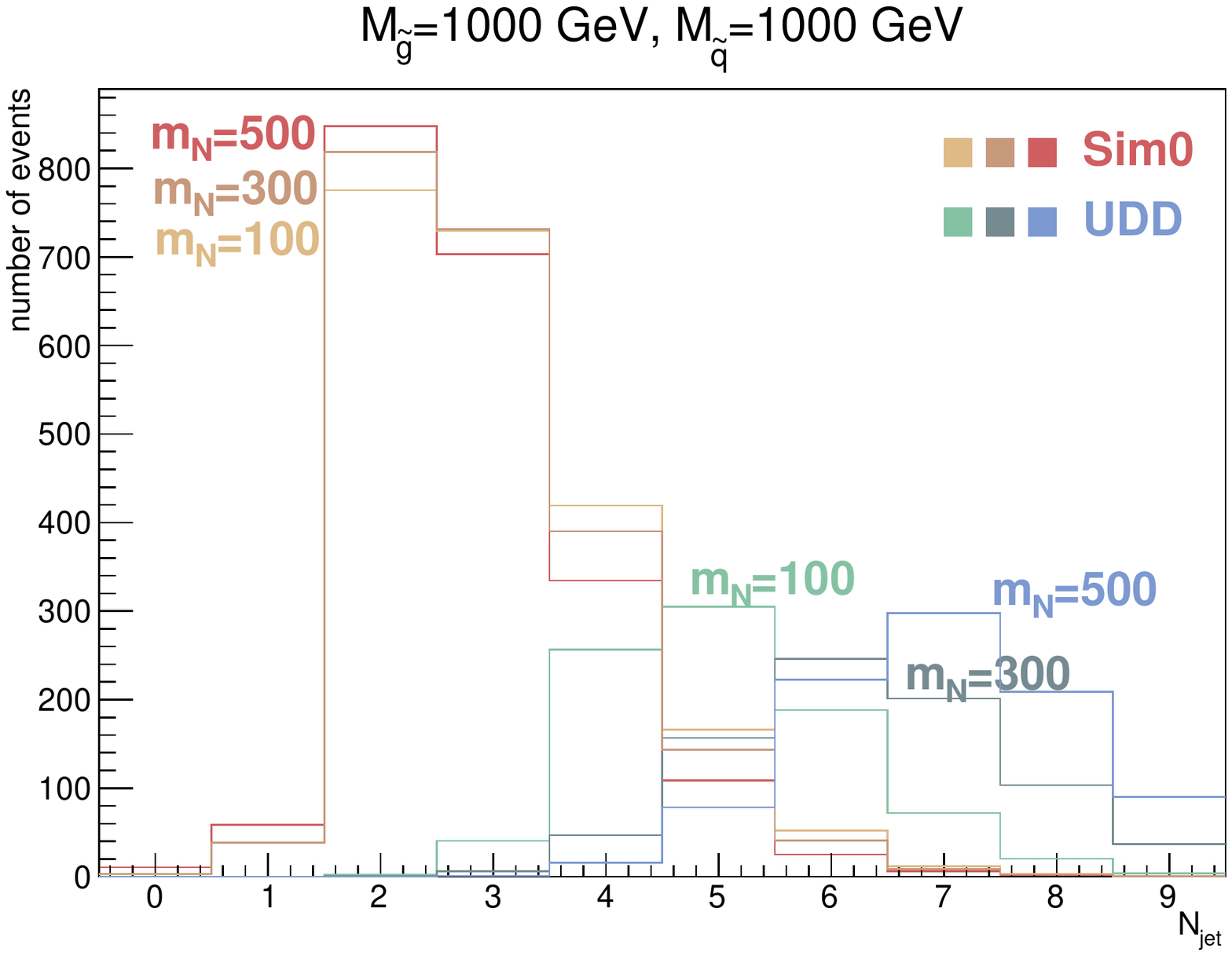} }

\subfigure[~$m_{\rm eff}$(incl.)]{\includegraphics[width=7.5cm]{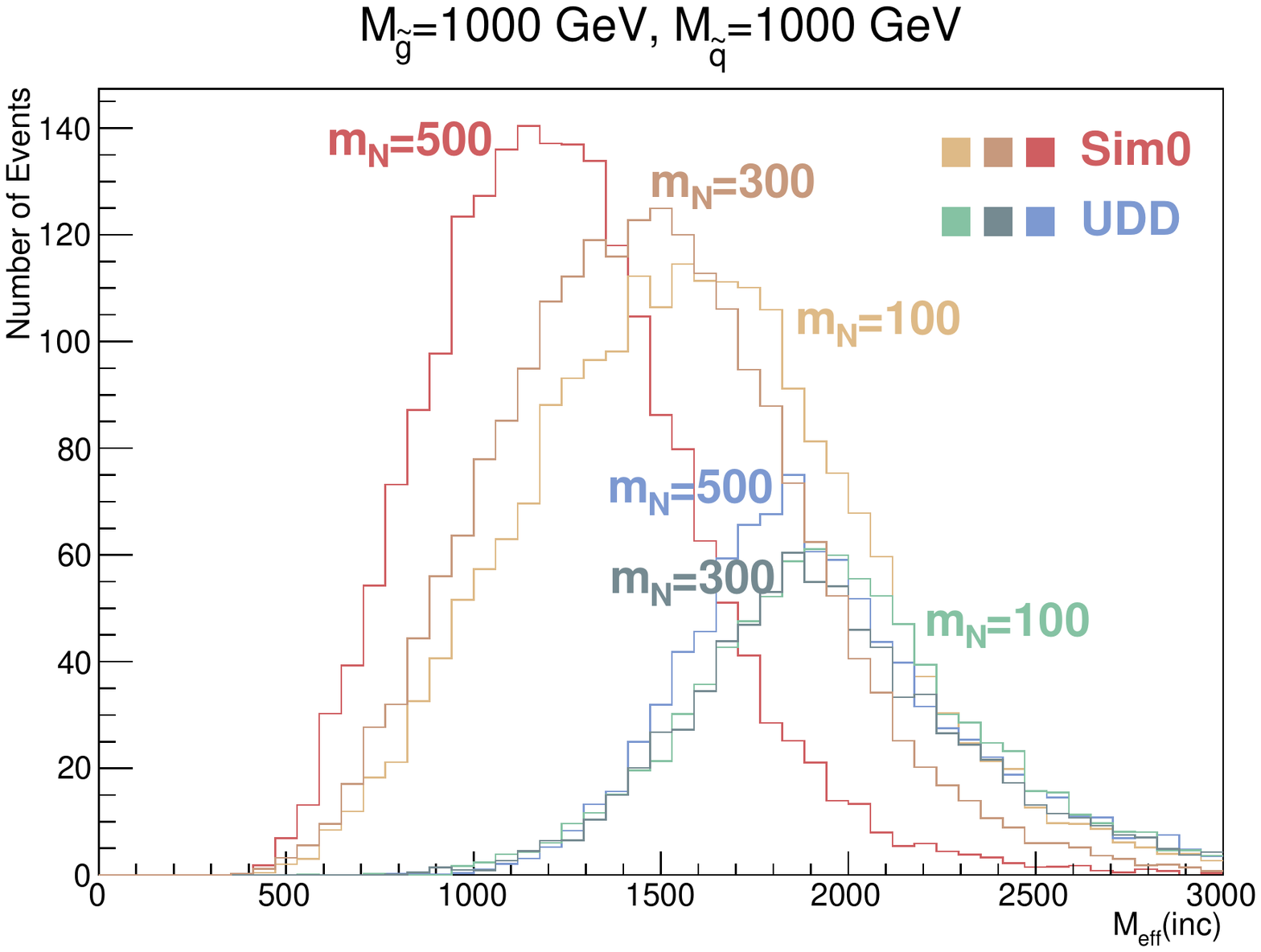} }
\quad
\subfigure[~$E_T^{\rm miss} / m_{\rm eff} (2j)$]{\includegraphics[width=7.5cm]{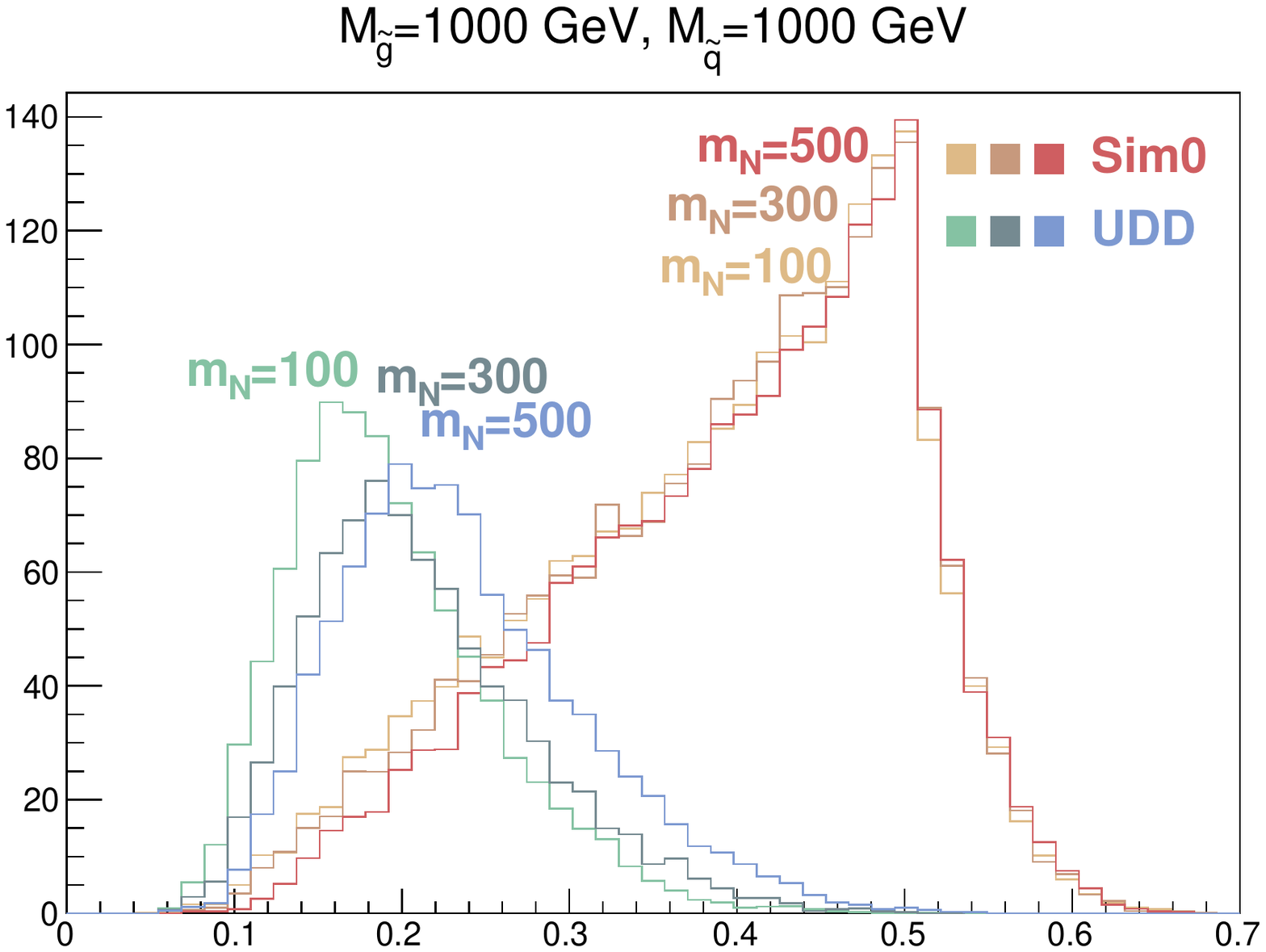} }
\caption{Various distributions for $u^cd^cd^c$ model with neutralino LOSP (blue histograms) compared with the Simplified Model {\bf Sim0} (red histograms): missing transverse energy $E_T^{\rm miss}$, number of jets $N_{\rm jets}$ with $p_T > 60$ GeV (for the hardest jet, $p_T > 130$ GeV), inclusive effective mass $m_{\rm eff}$(incl.) and $E_T^{\rm miss} / m_{\rm eff} (2j)$.  The mass parameters here are $m_{\tilde{g}}=m_{\tilde{q}}=1000$ GeV. For each model, we show three different neutralino masses $m_{\chi^0_1} =$ 100, 300 and 500 GeV.
}
\label{fig:met_Njet_xudd_vs_simplified_neutralinolosp}
\end{figure}

The constraints for the neutralino LOSP $u^c d^c d^c$ model are generically weaker than the Simplified Model {\bf Sim0} for small $m_{\chi^0_1}$ (100 GeV and 300 GeV), but reveal more complicated behavior in the $m_{\chi^0_1}=$ 500 GeV case. Several factors contribute to these results.  
One obvious factor that tends to give weaker constraints on the ADM model in the 0 lepton analysis is that the missing energy of the neutralino is reduced as it decays to three additional jets, as shown in Fig.~\ref{fig:met_Njet_xudd_vs_simplified_neutralinolosp}. This feature is transparently comparable with the Simplified Model {\bf Sim0} since both models share the same event topology before the neutralino decay.  On the other hand, as the neutralino mass is set heavier, the energy of the jets from gluino/squark decay into the neutralino becomes smaller as the mass difference shrinks.  Therefore, the experimental sensitivity to the Simplified Model {\bf Sim0} (and ordinary R-parity conserving MSSM scenarios generically) is reduced for a heavier neutralino mass, while the ADM models are subject to more severe constraints since a massive neutralino is able to ``store'' and transfer energy to the ADM particle. 
Therefore, for large neutralino mass, the ADM model can actually become substantially more constrained than the Simplified Model.

In Fig.~\ref{fig:met_Njet_xudd_vs_simplified_neutralinolosp}, we compare the the $E_T^{\rm miss}$, $N_{\rm jet}$, $m_{\rm eff} ({\rm incl.})$ and $E_T^{\rm miss} / m_{\rm eff} (2j)$ distributions of the neutralino LOSP $u^c d^c d^c$ model and the Simplified Model {\bf Sim0}  for $m_{\tilde{g}} = m_{\tilde{q}} = 1000$ GeV and $m_{\chi^0_1}$ = 100, 300 and 500 GeV.  Here, we use the same cuts as in Fig.~\ref{fig:met_Njet_meff_xudd_vs_simplified_squarklosp}. Note that $E_T^{\rm miss}$ is distinctively smaller and $m_{\rm eff}$ is significantly higher for the $u^c d^c d^c$ ADM model than for the simplified model, indicating that the simplified model more easily passes the $m_{\rm eff}$ requirement.  The net effect is that the constraints on the ADM model are weaker than for the simplified model, though the ADM model becomes more constrained relative to the simplified model as $m_{\chi^0_1}$ increases.  There are a couple of reasons that the ADM model constraints become stronger at larger neutralino mass.  First, the number of hard jets in the ADM model increases, improving the sensitivity to the model for the channels which require a high multiplicity of jets.  Second, the acceptance on $E_T^{\rm miss} / m_{\rm eff} (2j)$ cut improves markedly as the neutralino mass increases: $m_{\rm eff} (2j)$ decreases as the energy stored in the neutralino increases. 

Lastly, as we did for the $q \ell d^c$ operator, we comment on detecting the states of the UV completion of the ADM operator -- flavor violating signatures can also result from prompt decays of the new states $U$ and $D$.  When these states decay to the DM, $U,~D \rightarrow X + q$, the signatures look similar to squark or stop signatures of jet or top quark plus missing energy.  On the other hand, these states may have flavor violating decays to pairs of quarks, which may include only the light quarks, but also may result in flavor violating decays $U \rightarrow t j$ or $ D \rightarrow b j$.  A study of these signatures could give rise to additional constraints on ADM sectors.

\section{Conclusion and Outlook} 
\label{sec:conclusion}

We have carried out the first detailed study of flavor constraints and collider signatures of Asymmetric Dark Matter.  We found that while flavor constraints from meson oscillations and lepton flavor conservation place significant requirements on the scale $M$ of the ADM operators, this scale $M$ is not so high that a variety of collider prompt decays of the lightest ordinary supersymmetric particle (LOSP) into the $X$-sector, including exotic flavor combinations, could not arise.  We applied two standard 8 TeV LHC searches for SUSY to LOSP decays to ADM plus additional jets and leptons. These analyses involved 2-6 jets plus missing energy, or 1-2 leptons plus 3-6 jets and missing energy.  We found that the constraints from these analyses, whether the LOSP is a squark, slepton, or neutralino, are somewhat weakened, depending on the spectrum, in comparison to the standard searches.  However, the detailed kinematic distributions show significant difference between the conventional SUSY models and the ADM models.  This suggests that other SUSY searches at the LHC might be sensitive to the ADM-extended MSSM, in particular searches which involve an extremely high multiplicity of jets \cite{Aad:2013wta,CMS-PAS-SUS-13-012}.  It also suggests that dedicated searches tuned to ADM could significantly extend the reach at the LHC.

One of the interesting conclusions of this work is that the source of large flavor violation may not be much beyond our current reach.  The suppression scale of the ADM operator could be as low as 10 TeV, and the leptoquark-type states being integrated out could be as low as 1 TeV.  These states, when they decay to the ADM sector or to the visible sector, could give rise to exotic flavor-violating signatures.  
Performing ADM model analyses for other SUSY searches, {\it e.g.} high jet multiplicity searches, third-generation focused searches, and exotica searches ({\it e.g.} leptoquark searches) will provide a better understanding of the current status of ADM models.  We aim to carry out this study in the future.  It will also be interesting to design searches for ADM to learn how much the LHC reach can be extended. The well-motivated, simple extension of an ADM sector shows interesting interplay between flavor physics and collider physics, and opens new unexplored directions for LHC phenomenology.

\begin{acknowledgements}
We thank Yuval Grossman,  Michele Papucci and Lian-Tao Wang for discussions and Anson Hook for pointing out a problem in one of the Monte Carlo tools that we employ.  We also thank the theory group at Lawrence Berkeley National Laboratory for hospitality while part of this work was being completed. This work is supported by by NSF CAREER award PHY 1049896 and by the DoE under contract de-sc0007859.
\end{acknowledgements}

\appendix 

\section{One-Loop Box Diagram Correction to Flavor Violation } 

In this section, we present the full one-loop corrections through box diagrams for flavor violating processes in {\it (i)} meson oscillations, {\it (ii)} $\mu-e$ conversion, {\it (iii)} $B_s \rightarrow \ell^+ \ell^-,~b\rightarrow s \ell^+ \ell^-$ and {\it (iv)} $\mu \rightarrow 3 e$.  

\begin{figure}
\subfigure[~$B_O$ ]{\includegraphics[width=5cm]{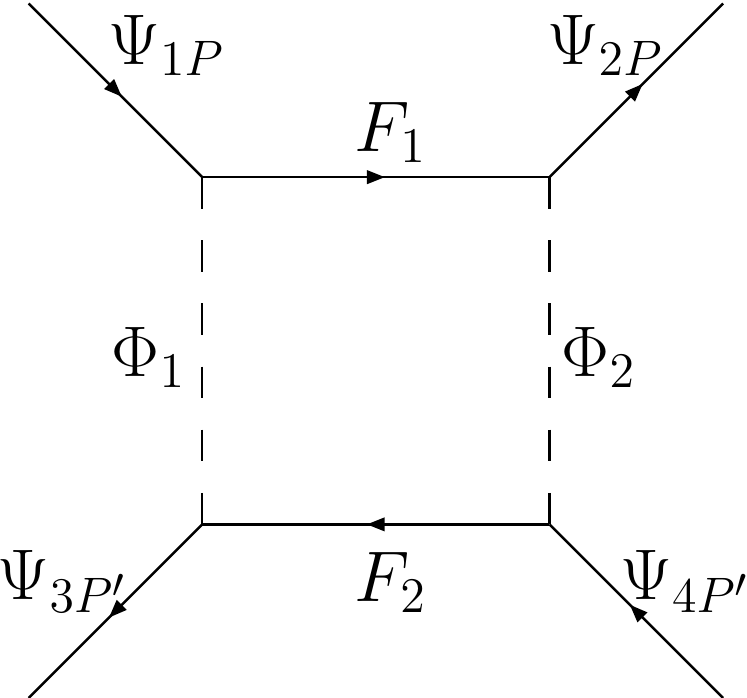}} 
\qquad\qquad
\subfigure[~$B_M$ ]{\includegraphics[width=5cm]{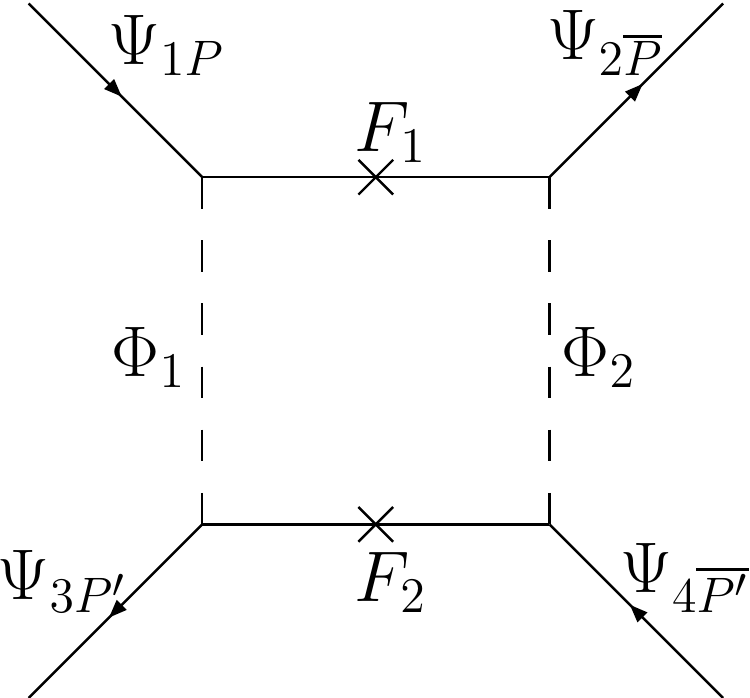}}
\caption{Box diagrams with fermion line (a) without and (b) with a mass insertion. $P$ and $P'$ denote chirality $L$ or $R$, and $\overline{P}$ and $\overline{P'}$ are opposite chirality to $P$ and $P'$, respectively. Here, we use 4-component notation to easily match with Feynman amplitude expressions. }
\label{fig:generalbox}
\end{figure}

First we begin with the general loop functions that will be useful for calculating the diagrams for all relevant processes.  For the box diagrams, it is convenient to split an internal fermion propagator into chirality preserving part ($\propto \gamma^\mu$) and chirality flipping part in the fermion line. From the diagrams (a) and (b) shown in Fig. \ref{fig:generalbox},  we have effective operators 
\barray
B_o &=& ({\rm Sym}) \left(\overline{\Psi}_{2 P} \gamma^\mu \Psi_{1P}  \right) 
\left( \overline{\Psi}_{3P'}\gamma_\mu \Psi_{4P'} \right) 
\times \frac{\lambda_1 \lambda_2 \lambda_3 \lambda_4}{64\pi^2} H (m_{F_1}, m_{F_2}, m_{\phi_1}, m_{\phi_2})
\label{eq:genboxresult}
\\
B_m &=& ({\rm Sym}) \left(\overline{\Psi}_{2 \overline{P}} \Psi_{1P} \right)
\left(\overline{\Psi}_{3P'} \Psi_{4\overline{P'}} \right)
\times \frac{\lambda_1 \lambda_2 \lambda_3 \lambda_4}{16\pi^2} m_{F_1} m_{F_2} 
K ( m_{F_1}, m_{F_2}, m_{\phi_1}, m_{\phi_2} ), 
\nonumber
\earray
where $m_A$ is the mass of the particle A in the loop, $\lambda_i$'s are four couplings involved in the diagram and ${(\rm Sym})$ is an appropriate symmetry factor if there are identical particles in final states. We denote the chirality of each particle by $P,P' = L,R$ and $\overline{P},\overline{P'}$ for the opposite chirality, as shown in Fig.~\ref{fig:generalbox}. Note that the contribution $B_m$ can be reinterpreted as vector-vector current interaction due to Fierz identities: 
\barray
\left(\overline{\Psi}_{1R} \Psi_{2L}\right) \left( \overline{\Psi}_{3L} \Psi_{4R} \right)
= -\frac{1}{2} \left(\overline\Psi_{1R} \gamma^\mu \Psi_{4R} \right) \left( \Psi_{3L} \gamma_\mu \Psi_{2L} \right).
\earray
The loop functions in Eq.~(\ref{eq:genboxresult}) are defined by  
\barray
&&H(m_{F_1},m_{F_2},m_{\phi_1},m_{\phi_2}) \equiv \label{eq:Hfunc}\\
&& \qquad \frac{m_{F_1}^4 \log (m_{F_1}^2) }{(m_{F_1}^2 - m_{F_2}^2)(m_{F_1}^2 -  m_{\phi_1}^2) (m_{F_1}^2 - m_{\phi_2}^2)} 
 -\frac{m_{F_2}^4 \log (m_{F_2}^2)}{(m_{F_1}^2 - m_{F_2}^2)(m_{F_2}^2 - m_{\phi_1}^2) (m_{F_2}^2 - m_{\phi_2}^2)} 
\nonumber \\
&& \qquad~ + \frac{m_{\phi_1}^4 \log m_{\phi_1}^2}{(m_{F_1}^2 - m_{\phi_2}^2)(m_{F_2}^2-m_{\phi_1}^2)(m_{\phi_1}^2-m_{\phi_2}^2)} 
 - \frac{ m_{\phi_2}^4 \log (m_{\phi_2}^2)}{(m_{F_1}^2-m_{\phi_2}^2)(m_{F_2}^2-m_{\phi_2}^2)(m_{\phi_1}^2-m_{\phi_2}^2)}\,,
\nonumber\\
&&K(m_{F_1},m_{F_2},m_{\phi_1},m_{\phi_2}) \equiv \label{eq:Kfunc}\\
&&\qquad -\frac{m_{F_1}^2\log (m_{F_1}^2)}{(m_{F_1}^2-m_{F_2}^2)(m_{F_1}^2-m_{\phi_1}^2)(m_{F_1}^2-m_{\phi_2}^2)}
+\frac{m_{F_2}^2\log (m_{F_2}^2)}{(m_{F_1}^2-m_{F_2}^2)(m_{F_2}^2-m_{\phi_1}^2)(m_{F_2}^2-m_{\phi_2}^2)}
\nonumber \\
&& \qquad~ -\frac{m_{\phi_1}^2\log (m_{\phi_1}^2)}{(m_{F_1}^2-m_{\phi_1}^2)(m_{F_2}^2-m_{\phi_1}^2)(m_{\phi_1}^2-m_{\phi_2}^2)}
+\frac{m_{\phi_2}^2\log (m_{\phi_2}^2)}{(m_{F_1}^2-m_{\phi_2}^2)(m_{F_2}^2-m_{\phi_2}^2)(m_{\phi_1}^2-m_{\phi_2}^2)}\,. \nonumber
\earray 
For the loop contributions under consideration, we have $m_{F_1} = m_{F_2}$ or $m_{\phi_1} = m_{\phi_2}$ in most cases. If $m_{F_1}$ and $m_{F_2}$ are the same, $H$ and $K$ are given by  
\barray 
&& H (m_F ; m_{\phi_1}, m_{\phi_2} ) =\nonumber \\  
&& \qquad  \frac{ m_F^2}{(m_F^2 - m_{\phi_1}^2)(m_F^2 - m_{\phi_2}^2)}
- \frac{m_F^2 \log (m_F^2)}{(m_F^2 - m_{\phi_1}^2)(m_F^2-m_{\phi_2}^2)} 
\left[ \frac{ m_{\phi_1}^2}{m_F^2 - m_{\phi_1}^2} + \frac{ m_{\phi_2}^2}{m_F^2 - m_{\phi_2}^2}  \right] 
\nonumber \\
&&\qquad\qquad
+ \frac{m_{\phi_1}^4 \log (m_{\phi_1}^2)}{(m_{\phi_1}^2 - m_{\phi_2}^2)(m_F^2 - m_{\phi_1}^2)^2}
- \frac{m_{\phi_2}^4 \log (m_{\phi_2}^2)}{(m_{\phi_1}^2 - m_{\phi_2}^2)(m_F^2 - m_{\phi_2}^2)^2} \,, 
\\
&&K (m_F ; m_{\phi_1}, m_{\phi_2}) = \nonumber \\
&& \qquad -\frac{1}{(m_F^2 - m_{\phi_1}^2)(m_F^2 - m_{\phi_2}^2)}
+ \frac{(m_F^4- m_{\phi_1}^2 m_{\phi_2}^2) \log (m_F^2)}{(m_F^2 - m_{\phi_1}^2)^2 (m_F^2 - m_{\phi_2}^2)^2 } 
\nonumber \\
&& \qquad\qquad
- \frac{ m_{\phi_1}^2 \log (m_{\phi_1}^2) }{(m_{\phi_1}^2 - m_{\phi_2}^2)(m_F^2 - m_{\phi_1}^2)^2}
+ \frac{ m_{\phi_2}^2 \log (m_{\phi_2}^2) }{(m_{\phi_1}^2 - m_{\phi_2}^2)(m_F^2 - m_{\phi_2}^2)^2}
\earray
and similarly for $m_{\phi_1}=m_{\phi_2}$. For $m_{F_1}=m_{F_2}$ and $m_{\phi_1}=m_{\phi_2}$, the loop functions are reduced to  
\barray
H(m_{F}; m_{\phi}) = \frac{m_F^2+m_\phi^2}{(m_F^2-m_\phi^2)^2} - \frac{2 m_F^2 m_\phi^2}{(m_F^2-m_\phi^2)^3} \log \left(\frac{m_F^2}{m_\phi^2}\right), \\
K(m_{F}; m_{\phi}) = - \frac{2}{(m_F^2-m_\phi^2)^2} + \frac{m_F^2+m_\phi^2}{(m_F^2-m_\phi^2)^3} \log \left( \frac{m_F^2}{m_\phi^2}\right).
\earray

In the following subsections, we present the corresponding expressions in the UV completemodels for various flavor constraints. 

\subsection{Meson mixing constraints}

\begin{table}
\begin{tabular}{|c||ccc||cc|}
\hline
 Operator &  \multicolumn{3}{c||}{$\lambda^4$ ($\lambda^2$ for tree level) in $X q \ell d^c$}  & \multicolumn{2}{|c|}{$X u^c d^c d^c$} \\ \hline
 Limit (TeV) & $D$ & $L$ & $Q$ & $U$ & $D$ \\ \hline \hline
$(\bar{s}_R \gamma^\mu d_R)^2$  & $(\lambda_{XD}^1 \lambda_{XD}^2)^2$ & $( \lambda_L^{i1}\lambda_L^{i2} )^2$ & $(\lambda_Q^{i 1}\lambda_Q^{i 2})^2$  & $(\lambda_U^{13} \lambda_U^{2 3})^2$ & (*)\\
 980 &  78 & 110 & 110 & 78  & 78  \\
$(\bar{s}_L \gamma^\mu d_L)^2$ &  $(\lambda_D^{i 1}\lambda_D^{i 2})^2$ & $(\lambda_L^{1 i}\lambda_L^{2 i})^2$ & $(\lambda_{XQ}^1 \lambda_{XQ}^2)^2$ &  & \\
 980 & 78 & 78 & 78 &  &  \\
 $(\bar{s}_L d_R) (\bar{s}_R d_L)$ & ~~$\lambda_{XD}^1 \lambda_{XD}^2 \lambda_D^{1 i}\lambda_D^{2 i} R_D$~~ &   \fbox{$\lambda_L^{12}\lambda_L^{21}$}, $\lambda^{1i}_L \lambda^{2i}_L \lambda_L^{j1} \lambda_L^{j2} $  & ~~$\lambda_{XQ}^1 \lambda_{XQ}^2 \lambda_Q^{i1}\lambda_Q^{i2} R_Q$~~ &  & \\
 18000 & 1400 & \fbox{18000}\,,~~\qquad 990 & 1400 &  & \\ \hline \hline
$(\bar{c}_R \gamma^\mu u_R)^2$  &  & & & $(\lambda_{XU}^1 \lambda_{XU}^2)^2$ & $(\lambda_D^{1 i} \lambda_D^{2 i})^2$ \\
 1200 & & & & 95 & 95 \\
 $(\bar{c}_L \gamma^\mu u_L)^2$ & $(\lambda_D^{1 i} \lambda_D^{2 i})^2$ & $(\lambda_L^{1 i} \lambda_L^{2 i})^2$ & $(\lambda_{XQ}^1 \lambda_{XQ}^2)^2$ &   &  \\
 1200 & 95 & 95 & 95 & &  \\ \hline \hline
$(\bar{b}_R \gamma^\mu d_R)^2$ & $(\lambda_{XD}^1 \lambda_{XD}^3)^2$ & $(\lambda_L^{i1}\lambda_L^{i3})^2$ & $(\lambda_Q^{i 1}\lambda_Q^{i 3})^2$ &  $(\lambda_U^{12} \lambda_U^{2 3})^2$ & (**) \\
 510 & 41 & 57 & 57 & 41 & 41 \\
 $(\bar{b}_L \gamma^\mu d_L)^2$  & $(\lambda_D^{i 1}\lambda_D^{i 3})^2$ & $(\lambda_L^{1 i}\lambda_L^{3 i})^2$ & $(\lambda_{XQ}^1 \lambda_{XQ}^3)^2$ &  & \\
 510 & 41 & 41 & 41 & & \\
 $(\bar{b}_L d_R) (\bar{b}_R d_L)$ & $\lambda_{XD}^1 \lambda_{XD}^3 \lambda_D^{1 i}\lambda_D^{3 i}  R_D$ & \fbox{$\lambda_L^{13}\lambda_L^{31}$}\,, $\lambda_L^{1i} \lambda_L^{3i} \lambda_L^{j1} \lambda_L^{j3}$ & $\lambda_{XQ}^1 \lambda_{XQ}^3 \lambda_Q^{i 1}\lambda_Q^{i 3} R_Q$ &  & \\
 1900 & 151 & \fbox{1900}\,,~~\qquad 110 & 151 & & \\  \hline\hline
$(\bar{b}_R \gamma^\mu s_R)^2$ & $(\lambda_{XD}^2 \lambda_{XD}^3)^2$ & $(\lambda_L^{i2}\lambda_L^{i3})^2$ & $(\lambda_Q^{i 2}\lambda_Q^{i 3})^2$ &  $(\lambda_U^{12} \lambda_U^{1 3})^2$ & (***) \\
 110 & 8.7 & 12 & 12 & 8.7 & 8.7 \\
 $(\bar{b}_L \gamma^\mu s_L)^2$  & $(\lambda_D^{i 2}\lambda_D^{i 3})^2$ & $(\lambda_L^{2 i}\lambda_L^{3 i})^2$ & $(\lambda_{XQ}^2 \lambda_{XQ}^3)^2$ &  & \\
 110 & 8.7 & 8.7 & 8.7 & & \\
 $(\bar{b}_L s_R) (\bar{b}_R s_L)$ & $\lambda_{XD}^2 \lambda_{XD}^3 \lambda_D^{2 i}\lambda_D^{3 i}  R_D$ & \fbox{$\lambda_L^{23}\lambda_L^{32}$}\,, $\lambda_L^{2i} \lambda_L^{3i} \lambda_L^{j2} \lambda_L^{j3}$ & $\lambda_{XQ}^2 \lambda_{XQ}^3 \lambda_Q^{i 2}\lambda_Q^{i 3} R_Q$ &  & \\
 370 & 29 & \fbox{370}\,,~~\qquad 21 & 29 & & \\  
\hline
\end{tabular}
\caption{Flavor constraints from meson oscillations. The numbers are in TeV. The operator which is constrained is shown, along with the constraint on $\Lambda$ \cite{Isidori:2010kg}. For the model $\Phi$ where $\Phi$ denotes a pair $(\Phi, \Phi^c)$ in the UV completion, $m_\Phi / \sqrt{\lambda^4}$ is constrained as shown in the table. Here, $R_\Phi = \log (m^2_\Phi/ m^2_{\rm soft}) -1 $. For the model $L$ in $Xq\ell d^c$, we show the tree level contribution (boxed) for $LR$ mixing operator for $K-\bar{K}$, $B_{d,s}-\bar{B}_{d,s}$. The constraints for tree level operator is implied for $m_L / \sqrt{\lambda^2}$. For the model $D$ in $Xu^cd^c d^c$, the coupling combination (*),(**), and (***) are presented in Eq.~(\ref{eq:coupling_in_D}).
} 
\label{table: flavor constraints}
\end{table}

 Experimental constraints from $K$-, $D$-, $B$-meson mixing put stringent constraints on the UV models for the $Xq\ell d^c$ and $Xu^c d^c d^c$ operators. The effective operators generated from the models are summarized by the following effective Lagrangian: 
\barray
{\mathcal L}_{\rm eff} &=& \frac{1}{4} K_{RR} (\bar{d}_R \gamma^\mu s_R)(\bar{d}_R \gamma_\mu s_R)
+ \frac{1}{4} K_{LL} (\bar{d}_L \gamma^\mu s_L)(\bar{d}_L \gamma_\mu s_L)
+ K_{LR} (\bar{s}_L d_R)(\bar{s}_R d_L) \nonumber \\ 
&&+ \frac{1}{4} D_{RR} (\bar{c}_R \gamma^\mu u_R)(\bar{c}_R \gamma_\mu u_R)
+ \frac{1}{4} D_{LL} (\bar{c}_L \gamma^\mu u_L)(\bar{c}_L \gamma_\mu u_L)  \\
&&+ \frac{1}{4} B_{dRR} (\bar{d}_R \gamma^\mu b_R)(\bar{d}_R \gamma_\mu b_R)
+ \frac{1}{4} B_{dLL} (\bar{d}_L \gamma^\mu b_L)(\bar{d}_L \gamma_\mu b_L)
+ B_{dLR} (\bar{b}_L d_R)(\bar{b}_R d_L) \nonumber \\
&&+ \frac{1}{4} B_{sRR} (\bar{s}_R \gamma^\mu b_R)(\bar{s}_R \gamma_\mu b_R)
+ \frac{1}{4} B_{sLL} (\bar{s}_L \gamma^\mu b_L)(\bar{s}_L \gamma_\mu b_L)
+ B_{sLR} (\bar{b}_L s_R)(\bar{b}_R s_L)\,, \nonumber 
\earray
where $K_{PP'}$, $D_{PP'}$, $B_{dPP'}$ and $B_{sPP'}$ are the coefficients of the corresponding operators. 
For $B_{dPP'}$ and $B_{sPP'}$, the results can be easily read from $K_{PP'}$ by changing the generation index to $b$-quark, so we will omit them in the following.

Under the assumption that $m_X \sim m_{\tilde{x}} \ll m_{\rm soft} \ll m_D, m_L, m_Q, m_U$, 
we summarize the tree-level and one-loop-level constraints on the mass and the coupling from meson mixing in Table \ref{table: flavor constraints}. In the table, $R_\Phi$ denotes $\log (m_{\Phi}^2 / m_{\rm soft}^2) -1$. The coupling combinations for $RR$ operators for $K$- and $B$-meson mixing in the $D$ UV completion for $X u^c d^c d^c$ are given by 
\barray
(*) &=& (\lambda_{XD}^1 \lambda_{XD}^2 )^2 + (\lambda_D^{i1} \lambda_D^{i2})^2 
- 2 \lambda^1_{XD}\lambda^2_{XD} \lambda_D^{i1} \lambda_D^{i2} R_D\,, \label{eq:coupling_in_D} \\
(**) &=& (\lambda_{XD}^1 \lambda_{XD}^2 )^3 + (\lambda_D^{i1} \lambda_D^{i3})^2 
- 2 \lambda^1_{XD}\lambda^3_{XD} \lambda_D^{i1} \lambda_D^{i3} R_D\,, \nonumber \\
(***) &=& (\lambda_{XD}^2 \lambda_{XD}^3 )^2 + (\lambda_D^{i2} \lambda_D^{i3})^2 
- 2 \lambda^2_{XD}\lambda^3_{XD} \lambda_D^{i2} \lambda_D^{i3} R_D\,. \nonumber 
\earray

\subsubsection{$X q \ell d^c$}
Since we have three classes of UV completions for the operator $Xq\ell d^c$, we specify the contribution from the model $M$ by putting a superscript $(M)$ in the following. First, we present the contributions from the UV completion of the $Xq\ell d^c$ operator.   
  
For the model $(D)$ defined by Eq.~(\ref{eq:modelDinQLD}), we obtain the operators for Kaon physics, 
\barray
K_{RR}^{(D)} &=& \frac{(\lambda_{XD}^1 \lambda_{XD}^2)^2}{64\pi^2} \big[  2 H (m_X, m_X, m_{\tilde{D}}, m_{\tilde{D}}) + 2 H (m_D, m_D, m_{\tilde{x}}, m_{\tilde{x}}) \big]\,, \\ 
K_{LL}^{(D)} &=& \frac{\lambda_{D}^{i1}\lambda_{D}^{i2}\lambda_{D}^{j1}\lambda_{D}^{j2}}{64\pi^2} 
\big[ 2H(m_D,m_D,m_{\tilde{\nu}^i},m_{\tilde{\nu}^j}) 
+ 2H(m_{\nu^i},m_{\nu^j},m_{\tilde{D}},m_{\tilde{D}})\big]\,, \nonumber \\
K_{LR}^{(D)} &=& \frac{\lambda_{D}^{1i}\lambda_{D}^{2i}\lambda_{XD}^1\lambda_{XD}^2}{16\pi^2} m_D^2 K (m_D, m_D, m_{\tilde{\nu}^i}, m_{\tilde{x}}) \,, \nonumber
\earray
and for $D$-meson physics,
\barray
D_{RR}^{(D)} &=& 0\,, \\
D_{LL}^{(D)} &=& \frac{\lambda_D^{1i} \lambda_D^{2i} \lambda_D^{1j} \lambda_D^{2j}}{64\pi^2}
\big[
2 H (m_D, m_D, m_{\tilde{e}^i}, m_{\tilde{e}^j} )
+ 2 H (m_{e^i}, m_{e^j}, m_{\tilde{D}}, m_{\tilde{D}} )
\big]\,. \nonumber 
\earray

For the model $(L)$ from Eq.~(\ref{eq:modelLinQLD}), we have
\barray
K_{RR}^{(L)} &=& \frac{\lambda_{L}^{i1}\lambda_{L}^{i2}\lambda_{L}^{j1}\lambda_L^{j2}}{64\pi^2} 
\Bigg[
2 H (m_{d^i}, m_{d^j}, m_{\tilde{L}}, m_{\tilde{L}} )
+ 2 H (m_L, m_L, m_{\tilde{d}^i}, m_{\tilde{d}^j} ) \nonumber \\ 
&& \qquad\qquad\qquad
+ 2 H (m_{u^i}, m_{u^j}, m_{\tilde{L}}, m_{\tilde{L}})
+ 2 H (m_L, m_L, m_{\tilde{u}^i}, m_{\tilde{u}^j})
\Bigg] \,, \\
K_{LL}^{(L)} &=& \frac{\lambda_{L}^{1i}\lambda_{L}^{2i}\lambda_{L}^{1j}\lambda_{L}^{2j}}{64\pi^2}
\big[
2H(m_L,m_L,m_{\tilde{d}^{c\,i}},m_{\tilde{d}^{c\,j}})
+2H(m_{d^i},m_{d^j},m_{\tilde{L}},m_{\tilde{L}})
\big] \,, \nonumber \\
K_{LR}^{(L)}&=&\frac{\lambda_L^{1i}\lambda_L^{2i}\lambda_L^{j1}\lambda_L^{j2}}{64\pi^2} 
\big[
H(m_{d^i},m_{d^j},m_{\tilde{L}},m_{\tilde{L}})
+H(m_L,m_L,m_{\tilde{d}^{c\,i}},m_{\tilde{d}^{c\,j}})
\big] \,,\nonumber \\
D_{RR}^{(L)} &=& 0\,, \nonumber \\
D_{LL}^{(L)} &=& \frac{\lambda_L^{1i}\lambda_L^{2i}\lambda_L^{1j}\lambda_L^{2j}}{64\pi^2}
\big[
2 H (m_L,m_L,m_{\tilde{d}^{c\,i}},m_{\tilde{d}^{c\,j}})
+ 2 H (m_{d^i}, m_{d^j}, m_{\tilde{L}}, m_{\tilde{L}} )
\big] \,. \nonumber 
\earray

Now, we show the result for the model $(Q)$: 
\barray
K_{RR}^{(Q)} &=& \frac{\lambda_Q^{i1} \lambda_Q^{i2} \lambda_Q^{j1} \lambda_Q^{j2}}{64\pi^2}
\Bigg[
2 H (m_{e^i}, m_{e^j}, m_{\tilde{Q}}, m_{\tilde{Q}} )
+ 2 H (m_Q, m_Q, m_{\tilde{e}^i} , m_{\tilde{e}^j} ) 
\\
&& \qquad \qquad\qquad
+2 H (m_{\nu^i}, m_{\nu^j}, m_{\tilde{Q}}, m_{\tilde{Q}})
+ 2 H (m_Q, m_Q, m_{\tilde{\nu}^i}, m_{\tilde{\nu}^j} )
\Bigg]\,, \nonumber \\
K_{LL}^{(Q)} &=& \frac{(\lambda_{XQ}^1\lambda_{XQ}^2)^2}{64\pi^2}
\big[
2H(m_X,m_X,m_{\tilde{Q}},m_{\tilde{Q}})
+2H(m_Q,m_Q,m_{\tilde{x}},m_{\tilde{x}})
\big] \,,  \nonumber \\
K_{LR}^{(Q)} &=& \frac{\lambda_{XQ}^1\lambda_{XQ}^2\lambda_Q^{i1}\lambda_Q^{i2}}{16\pi^2}
m_Q^2 K (m_Q, m_Q, m_{\tilde{x}}, m_{\tilde{\nu}^i} )\,, \nonumber \\
D_{RR}^{(Q)} &=& 0 \,, \nonumber \\
D_{LL}^{(Q)} &=& \frac{(\lambda_{XQ}^1\lambda_{XQ}^2)^2}{64\pi^2}
\big[
2 H (m_X,m_X,m_{\tilde{Q}},m_{\tilde{Q}})
+ 2 H (m_Q, m_Q, m_{\tilde{x}}, m_{\tilde{x}})
\big].
\earray 

\subsubsection{$X u^c d^c d^c$}
As in the $X q \ell d^c$ case, we specify each model by the superscripts $(U)$ and $(D)$. For the model $(U)$, 
\barray
K_{RR}^{(U)} &=& \frac{(\lambda_U^{13}\lambda_U^{23})^2}{64\pi^2}
\big[
2 H (m_U,m_U,m_{\tilde{d}^{c\,3}},m_{\tilde{d}^{c\,3}})
+ 2 H (m_b, m_b, m_{\tilde{U}}, m_{\tilde{U}} )
\big] \,, \\
D_{RR}^{(U)} &=& \frac{(\lambda_{XU}^1\lambda_{XU}^2)^2}{64\pi^2}
\big[
2 H (m_X, m_X, m_{\tilde{U}}, m_{\tilde{U}})
+ 2 H (m_U, m_U, m_{\tilde{x}}, m_{\tilde{x}})
\big] \,, \nonumber \\
K_{LL}^{(U)} &=& K_{LR}^{(U)} = D_{LL}^{(U)} = 0, \nonumber \\
\earray 
and for the model $(D)$,
\barray
K_{RR}^{(D)} &=& \frac{(\lambda_{XD}^1\lambda_{XD}^2)^2}{64\pi^2}
\big[
2H(m_X,m_X,m_{\tilde{D}},m_{\tilde{D}})+2H(m_D,m_D,m_{\tilde{x}},m_{\tilde{x}})
\big] \\
&& + \frac{\lambda_D^{i1}\lambda_D^{i2}\lambda_D^{j1}\lambda_D^{j2}}{64\pi^2}
\big[
2H(m_{u^i},m_{u^j},m_{\tilde{D}},m_{\tilde{D}})
+2H(m_D,m_D,m_{\tilde{u}^{c\,i}},m_{\tilde{u}^{c\,j}})
\big] \nonumber \\
&& - \frac{\lambda_{XD}^1\lambda_{XD}^2\lambda_D^{i1}\lambda_D^{i2}}{8\pi^2}
m_D^2 K (m_D, m_D, m_{\tilde{x}}, m_{\tilde{u}^{c\,i}})\,,\nonumber \\
D_{RR}^{(D)}&=& \frac{\lambda_D^{1i} \lambda_D^{2i} \lambda_D^{1j} \lambda_D^{2j}}{64\pi^2}
\big[
2 H(m_D,m_D,m_{\tilde{d}^{c\,i}},m_{\tilde{d}^{c\,j}})
+ 2 H(m_{d^i},m_{d^j}, m_{\tilde{D}}, m_{\tilde{D}} )
\big]\,, \nonumber\\
K_{LL}^{(D)} &=& K_{LR}^{(D)} = D_{LL}^{(D)} = 0. \nonumber 
\earray

\subsection{$\mu-e$ conversion}
\label{mu-e conversion}

\begin{figure}
\centering
\subfigure[]{\includegraphics[width=2.5cm]{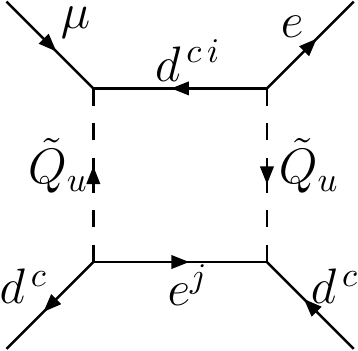}}
\quad
\subfigure[]{\includegraphics[width=2.5cm]{boxMuEConv2.pdf}}
\quad
\subfigure[]{\includegraphics[width=2.5cm]{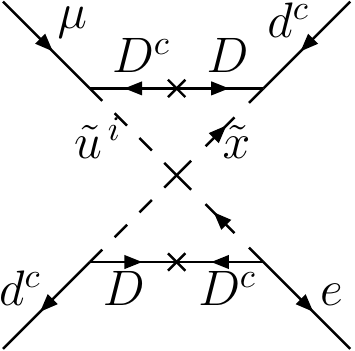}}
\quad
\subfigure[]{\includegraphics[width=2.5cm]{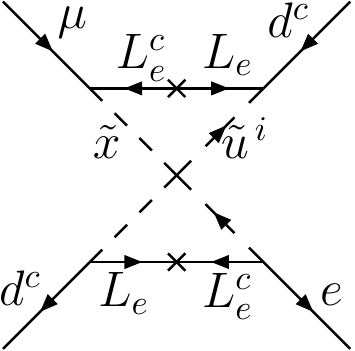}}
\quad
\subfigure[]{\includegraphics[width=2.5cm]{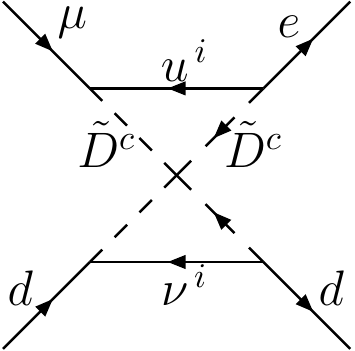}}

\subfigure[]{\includegraphics[width=2.5cm]{boxMuEConv6.pdf}}
\quad
\subfigure[]{\includegraphics[width=2.5cm]{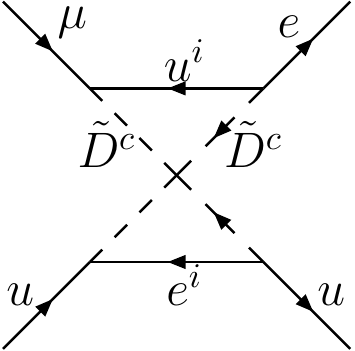}}
\quad
\subfigure[]{\includegraphics[width=2.5cm]{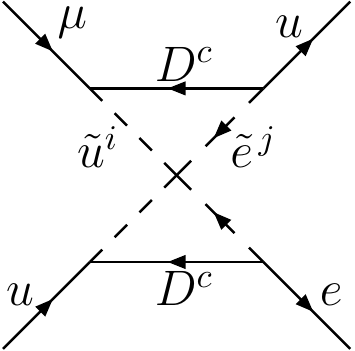}}
\quad
\subfigure[]{\includegraphics[width=2.5cm]{boxMuEConv9.pdf}}
\quad
\subfigure[]{\includegraphics[width=2.5cm]{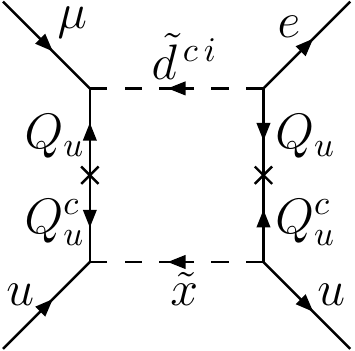}}

\caption{ \label{fig:mu e conversion full} One-loop box diagrams contributing to $\mu-e$ conversion in $X q \ell d^c$ models. Here, we use 2-component spinor notation. }
\end{figure}

\begin{table}
\begin{tabular}{|c|c|c|}
\hline
 & Tree $\lambda^2$ & One loop $\lambda^4$ \\ \hline\hline 
~~$D$~~ & $\lambda_D^{11} \lambda_D^{12}$ & $\frac{3}{2} \lambda_D^{i1} \lambda_D^{i2} \lambda_D^{1j} \lambda_D^{1j} -\frac{1}{2} \lambda_D^{i1} \lambda_D^{i2} \lambda_{XD}^1 \lambda_{XD}^2 R_D $  \\
    & 290 TeV &  23 TeV \\ \hline  
~~$L$~~ & & $  \lambda_{XL}^1 \lambda_{XL}^2 (\frac{1}{2} \lambda_L^{i1} \lambda_L^{i1} + \lambda_L^{1i} \lambda_L^{i1} ) R_L $ \\
   &  & 23 TeV \\ \hline 
~~$Q$~~ & $\lambda_Q^{11} \lambda_Q^{21}$ & $\frac{1}{2} \lambda_Q^{1i} \lambda_Q^{2i} \lambda_Q^{j1} \lambda_Q^{j1} -  \lambda_Q^{1i}\lambda_Q^{2i} \lambda_{XQ}^1 \lambda_{XQ}^2 R_Q$ \\ 
  & 210 TeV &  23 TeV \\
\hline 
\end{tabular}
\caption{Flavor constraints from $\mu-e$ conversion for the $Xq\ell d^c$ models. Each row represents the UV completion. The numbers below the couplings are the constraints on $\Lambda$ in TeV which $ m_\Phi / \sqrt{\lambda^2}$ for tree level contribution and $m_\Phi / \sqrt{\lambda^4}$ for the one loop contribution for the model $\Phi$.   } 
\label{table: mu e conversion}
\end{table}

Among the models under consideration, only the $Xq\ell d^c$-type model is subject to the constraint from $\mu$-$e$ conversion \cite{Dohmen:1993mp, Bertl:2006up}. Box diagrams can contribute only to the following vector-vector current interactions: 
\barray
{\mathcal L}_{\rm eff} = C^d_{LR} (\bar e_L \gamma^\rho \mu_L)(\bar d_R \gamma_\rho d_R)
+ C^d_{LL} (\bar e_L \gamma^\rho mu_L)(\bar d_L \gamma_\rho d_L)
+ C^u_{LL} (\bar e_L \gamma^\rho mu_L)(\bar u_L \gamma_\rho u_L),
\earray 
where $C^q_{PP'}$ is the coefficient of the corresponding diagram. From the effective operators, we obtain the $\mu-e$ conversion branching ratio for $_{13}$Al \cite{Kitano:2002mt}:
\barray
B_{\mu N \rightarrow e N} (Z=13) \approx 2.0 \times \frac{1}{2G_F^2} \left| 2 C^u_{LL} + C^d_{LL} + C^d_{LR}  \right|^2,
\earray
where $G_F$ is the Fermi constant.

For the model $(D)$,   
\barray
C^{d\,(D)}_{LR} &=& -\frac{\lambda_D^{i1}\lambda_D^{i2}\lambda_{XD}^1\lambda_{XD}^2}{32\pi^2} m_D^2 K (m_D, m_D, m_{\tilde{u}^i} m_{\tilde{x}} )\,, \nonumber \\ 
C^{d\,(D)}_{LL} &=& \frac{\lambda_D^{i1}\lambda_D^{i2}\lambda_D^{1j}\lambda_D^{1j}}{64\pi^2} \big[ H(m_{u^i}, m_{\nu^i}, m_{\tilde{D}}, m_{\tilde{D}}) 
+ H(m_D, m_D, m_{\tilde{u}^i}, m_{\tilde{\nu}^i}) \big]\,, \label{eq:mueconv_result_modelD} \\
C^{u\,(D)}_{LL} &=& \frac{\lambda_D^{i1} \lambda_D^{i2} \lambda_D^{1j} \lambda_D^{1j} }{ 64\pi^2 } 
\big[ H(m_{u^i}, m_{e^i} , m_{\tilde{D}}, m_{\tilde{D}} ) 
+ H( m_D, m_D, m_{\tilde{u}^i}, m_{\tilde{e}^i} ) \big]\,, \nonumber 
\earray
where $i,j$ are flavor indices and the tilde over a particle name implies its supersymmetric scalar partner with odd $R$-parity. Note that we can safely ignore the masses of quarks and leptons except the top quark mass although we show generic results for the purpose of completeness.  

Similarly, for models $(L)$ and $(Q)$ from Eq.~(\ref{eq:modelLinQLD}) and (\ref{eq:modelQinQLD}), 
\barray
C^{d\,(L)}_{LR} &=& - \frac{\lambda_{XL}^1 \lambda_{XL}^2 \lambda_L^{i1} \lambda_L^{i1}}{32\pi^2} 
m_L^2 K (m_L, m_L, m_{\tilde{x}}, m_{\tilde{u}^i} ) \nonumber \\
C^{d\,(L)}_{LL} &=& 0, \\
C^{u\,(L)}_{LL} &=& - \frac{\lambda_{XL}^1 \lambda_{XL}^2 \lambda_L^{1i} \lambda_L^{1i}}{32\pi^2}
m_L^2 K(m_L, m_L, m_{\tilde{x}}, m_{\tilde{d}^{c\,i}} )\,, \nonumber
\earray
and
\barray
C^{d\,(Q)}_{LR} &=& \frac{\lambda_Q^{1i} \lambda_Q^{2i} \lambda_Q^{j1} \lambda_Q^{j1}}{64\pi^2} 
\big[
H(m_{d^i},m_{e^i}, m_{\tilde{Q}}, m_{\tilde{Q}})
+ H(m_Q, m_Q, m_{\tilde{d}^{c\,i}}, m_{\tilde{e}^i})
\big]\,, \nonumber \\
C^{d\,(Q)}_{LL} &=& 0, \\
C^{u\,(Q)}_{LL} &=& - \frac{\lambda_Q^{1i} \lambda_Q^{2i} \lambda_{XQ}^1 \lambda_{XQ}^1 }{32\pi^2} m_Q^2 K (m_Q,m_Q,m_{\tilde{d}^{c\,i}}, m_{\tilde{x}} )\,. \nonumber
\earray

\subsection{\label{sec:appendix_b_s_transition} $B_s \rightarrow \ell^+ \ell^-$, $b \rightarrow s \ell^+ \ell^-$ transition }

\begin{figure}
\centering
\subfigure[]{\includegraphics[width=2.5cm]{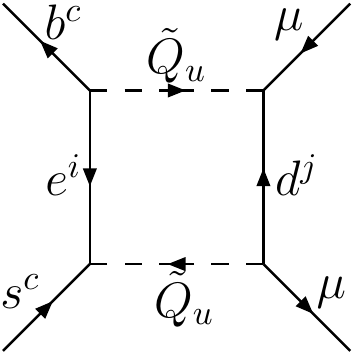}}
\quad
\subfigure[]{\includegraphics[width=2.5cm]{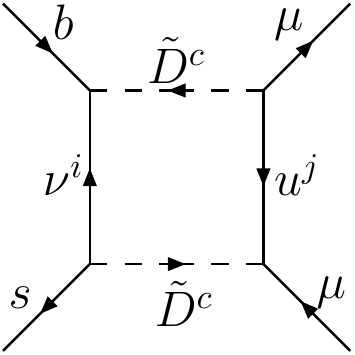}}
\quad
\subfigure[]{\includegraphics[width=2.5cm]{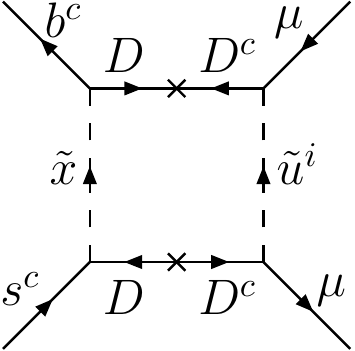}}
\caption{\label{fig:additional_box_diagram_for_b_to_s_transition} Remaining box diagrams that contribute to $B_s \rightarrow \mu^+ \mu^-$, in addition to Fig. \ref{fig:Bstomumu}. Generic $b-s$ transitions can be easily deduced by replacing the external states in the diagrams. }
\end{figure}

\begin{table}
\begin{tabular}{|c||c|c|c|}
\hline 
Operator & \multicolumn{3}{|c|}{$\lambda^4$ ($\lambda^2$ for tree level) in $X q \ell d^c$}  \\
\hline
 Limit(TeV) & $D$ & $L$ & $Q$ \\
\hline\hline
 $(\bar{s}_L \gamma^\mu b_L)(\bar{\ell}_L \gamma_\mu \ell_L)$ & \fbox{$\lambda_D^{3\ell} \lambda_D^{2\ell}$} $\lambda_D^{3i} \lambda_D^{2i} \lambda_D^{j \ell} \lambda_D^{j \ell}$ & & $\lambda_Q^{i3} \lambda_Q^{i2} \lambda_Q^{\ell j} \lambda_Q^{\ell j}$ \\
  45 (16) & \fbox{32 (11)} ~~2.5 (0.9) & & 1.8 (0.64) \\ \hline
 $(\bar{s}_R \gamma^\mu b_R)(\bar{\ell}_L \gamma_\mu \ell_L)$ & $\lambda^3_{XD} \lambda^2_{XD} \lambda_D^{i\ell} \lambda_D^{i\ell} R_D$ & 
$\lambda_L^{i3} \lambda_L^{i2} \lambda_{XL}^\ell \lambda_{XL}^\ell R_L$
& \fbox{$\lambda_Q^{\ell 3} \lambda_Q^{2\ell}$}  $\lambda_Q^{i3} \lambda_Q^{i2} \lambda_Q^{\ell j} \lambda_Q^{\ell j}$  \\
  63 (16) & 3.5 (0.9) & 3.5 (0.9) & \fbox{45 (11)}  ~~2.5 (0.64) \\ \hline
\hline
\end{tabular}
\caption{\label{table:b_s_transition_constraint} Flavor constraints from $b - s$ transition for the $X q \ell d^c$ models.
The operator which is constrained is shown, along with the constraint on $\Lambda$ in TeV according to \cite{Altmannshofer:2012az}. The numbers outside the parentheses are the strongest constraints and the numbers in the parentheses are the weakest constraints. For the UV completion model $\Phi$,  $m_\Phi / \sqrt{\lambda^2}$ is constrained for tree level (boxed), and $m_\Phi / \sqrt{\lambda^4}$ is constrained for one-loop level (unboxed). Here, $R_\Phi = \log (m^2_\Phi/ m^2_{\rm soft}) -1 $.  $i,j$ are flavor indices that runs over 1,2 and 3 and the summation is implied, but $\ell$ denotes an electron or muon external state, and thus is not summed over. }

\end{table}

As with $\mu-e$ conversion described in the previous section, the $X q \ell d^c$ model is also subject to 
constraints from $b \rightarrow s$ transition measurements. At one-loop level, the contributing 
Feynman diagrams are listed in Fig. \ref{fig:Bstomumu} and Fig. \ref{fig:additional_box_diagram_for_b_to_s_transition}. The one-loop contributions lead to 
\barray
{\mathcal L}_{\rm eff} &=& Z_{LL} (\bar{s}_L \gamma^\rho b_L) (\bar{\ell}_L \gamma_\rho \ell_L)
+ Z_{LR} (\bar{s}_L \gamma^\rho s_L) (\bar{\ell}_R \gamma_\rho \ell_R) \nonumber \\
&& + Z_{RL} (\bar{s}_R \gamma^\rho s_R) (\bar{\ell}_L \gamma_\rho \ell_L)
+ Z_{RR} (\bar{s}_R \gamma^\rho s_R) (\bar{\ell}_R \gamma_\rho \ell_R),
\earray
where $\ell$ denotes electron and muon. 
Note that there are no contributions to (scalar current)-(scalar current) interactions, such as $\bar{s}_R b_L \ell_R \ell_L$, since the UV completions of the $X q \ell d^c$ model involve only left-handed leptons.   
These induced effective couplings are constrained from various rare $B$-meson decays, for which the constraints on the scales of the effective operators are computed in \cite{Altmannshofer:2012az}. 

We present the full one-loop $Z_{PP'}$ from each UV completion $(D)$, $(L)$ and $(Q)$. For the model $(D)$, 
\barray
Z^{(D)}_{LL} &=& \frac{\lambda_D^{3i} \lambda_D^{2i} \lambda_D^{j\ell} \lambda_D^{j\ell}}{64\pi^2} 
\left[H(m_{\nu^i}, m_{u^j} , m_{\tilde{D}} , m_{\tilde{D}}) + H(m_D, m_D, m_{\tilde{\nu}^i}, m_{\tilde{u}^j})\right]\,, \nonumber \\
Z^{(D)}_{RL} &=& -\frac{\lambda_{XD}^2\lambda_{XD}^3\lambda_D^{i\ell}\lambda_D^{i\ell}}{32\pi^2}
m_D^2 K (m_D, m_D, m_{\tilde{x}}, m_{\tilde{u}^i} )\,, \nonumber \\
Z^{(D)}_{LR} &=& Z^{(D)}_{RR} = 0,
\earray
where $\ell$ denotes the generation index of the external leptons, such that $\ell = 1$ for electron and 2 for muon. For the $(L)$ UV completion, 
\barray
Z^{(L)}_{RL} &=& -\frac{\lambda_L^{i3} \lambda_L^{i2} \lambda^{\ell}_{XL} \lambda^{\ell}_{XL}}{32\pi^2} 
m_L^2 K (m_L, m_L , m_{\tilde{u}^i}, m_{\tilde{x}} )\,, \nonumber \\
Z^{(L)}_{LL} &=& Z^{(L)}_{LR} = Z^{(L)}_{RR}
\earray
For $(Q)$, 
\barray
Z^{(Q)}_{RL} &=& \frac{\lambda_Q^{i3} \lambda_Q^{i2} \lambda_Q^{\ell j} \lambda_Q^{\ell j}}{64\pi^2} \left[
H(m_{e^i}, m_{d^j}, m_{\tilde{Q}}, m_{\tilde{Q}}) + H(m_Q, m_Q, m_{\tilde{e}^i}, m_{\tilde{d}^{c\,j}}) 
\right]\,,
\nonumber \\
Z^{(Q)}_{LL} &=& Z^{(Q)}_{LR} = Z^{(Q)}_{RR}.
\earray

We summarize the $b-s$ transition constraints in Table \ref{table:b_s_transition_constraint} assuming $m_X \sim m_{\tilde{x}} \ll m_{\rm soft} \ll m_D, m_L, m_Q, m_E$, using the result from \cite{Altmannshofer:2012az}. Since the constraints depend on whether the couplings are real or complex, we show both the strongest and weakest lower bounds in the table.

\subsection{$\mu^\pm \rightarrow e^\pm e^\mp e^\pm$ }

\begin{table}
\begin{tabular}{|cc|c|c|}
\hline
\multicolumn{2}{|c|}{Model} & Tree $\lambda^2$ & One loop $\lambda^4$ \\ \hline\hline
 $X q\ell d^c$ & ~~$D$~~ & & $(\lambda_D^{i1} \lambda_D^{i2})^2$ \\ 
  & & & 9.8 TeV \\
 & ~~$L$~~ & & $\lambda_{XL}^2 (\lambda_{XL}^1)^3$ \\
 & & & 9.8 TeV \\
 & ~~$Q$~~ & & $(\lambda_Q^{i1} \lambda_Q^{2i})^2$ \\
 & & & 9.8 TeV \\ \hline\hline
 $X \ell \ell e^c$ & ~~$L$~~ & $\lambda_L^{11} \sqrt{ (\lambda_L^{12})^2 + (\lambda_L^{21})^2 }$   & Eq.~(\ref{eq:xlle_coupling1}) \\
 & & 87 TeV & 9.8 TeV \\ 
 & ~~$E$~~ & & Eq.~(\ref{eq:xlle_coupling2}) \\
 & & & 9.8 TeV  \\ \hline 
\end{tabular}
\caption{Flavor contraints from $\mu^- \rightarrow e^- e^+ e^-$. Each row represents the UV completion $\Phi$. The numbers below the couplings are the constraints $\Lambda$ in TeV on $ m_\Phi / \sqrt{\lambda^2}$ for a tree level contribution and $m_\Phi / \sqrt{\lambda^4}$ for a loop contribution. } 
\label{table:mu_to_eee}
\end{table}

For the $X q \ell d^c$ and $X \ell \ell e^c$ models, we have constraints from rare muon decays, with box diagrams contributing to $\mu \rightarrow 3 e$ decay. The relevant effective operators are 
\barray
{\mathcal L}_{\rm eff} &=& 
\frac{1}{2} A_{RR} (\bar{e}_R \gamma^\rho e_R)(\bar{e}_R \gamma_\rho \mu_R) 
+ \frac{1}{2} A_{LL} (\bar{e}_L \gamma^\rho e_L)(\bar{e}_L \gamma_\rho \mu_L) \nonumber \\
&& + A_{RL} (\bar{e}_R \gamma^\rho e_R)(\bar{e}_L \gamma_\rho \mu_L) 
 + A_{LR} (\bar{e}_L \gamma^\rho e_L)(\bar{e}_R \gamma_\rho \mu_R)\,,
\earray
where $A_{PP'}$'s are the coefficient generated from the one-loop contribution. Note that we have symmetry factors for the RR and LL couplings. 

The partial decay width of a muon to three electrons is 
\barray
\Gamma_{\mu\rightarrow e e e} 
=\frac{m_\mu^5}{3 \cdot 2^{11} \pi^3} \left( \left| A_{RR} \right|^2 + \left| A_{LL} \right|^2 + \left| A_{RL} \right|^2 + \left| A_{LR} \right|^2 \right)\,,
\earray
where $m_\mu$ is the muon mass.  The branching fraction is currently constrained to be $\lesssim 10^{-12}$, with the total muon width being $\Gamma_\mu = G_F^2 m_\mu^5/(192 \pi^3)$.

For the $X q \ell d^c$ operator, we have only a contribution to $A_{LL}$ since only left-handed leptons take part in the new physics couplings. We obtain the following effective operators for the UV completion models $(D)$, $(L)$ and $(Q)$, respectively: 
\barray
A_{LL}^{(D)} &=& \frac{\lambda_D^{i1} \lambda_D^{i2} \lambda_D^{j1} \lambda_D^{j1}}{64\pi^2}
\big[ 2 H (m_{u^i}, m_{u_j} , m_{\tilde{D}}, m_{\tilde{D}} ) 
+ 2 H (m_D, m_D, m_{\tilde{u}^i}, m_{\tilde{u}^j} ) 
\big]\,, \\
A_{LL}^{(L)} &=& \frac{\lambda_{XL}^2(\lambda_{XL}^1)^3}{64\pi^2}
\big[ 2 H (m_X, m_X, m_{\tilde{L}}, m_{\tilde{L}} ) 
+ 2 H (m_L, m_L, m_{\tilde{x}}, m_{\tilde{x}} )
\big]\,,\\
A_{LL}^{(Q)} &=& \frac{\lambda_{Q}^{1i} \lambda_{Q}^{2i} \lambda_{Q}^{1j} \lambda_{Q}^{1j} }{64\pi^2}
\big[ 2 H (m_Q, m_Q, m_{\tilde{d}^{c\,i}}, m_{\tilde{d}^{c\,j}} )
+ 2 H ( m_{d^i}, m_{d^j}, m_{\tilde{Q}}, m_{\tilde{Q}} ).
\big]
\earray
Note that we generally a factor of two larger contribution since two electrons are identical. 

We now summarize the results for $X \ell \ell e^c$ here. We have two UV completion models $(L)$ and $(E)$, as defined in Eq.~(\ref{eq:modelLinXLLE}) and (\ref{eq:modelEinXLLE}), respectively. For the model $(L)$, we have
\barray
A_{LL}^{(L)} &=& \frac{1}{64\pi^2} \Bigg[ 
\lambda_{XL}^2 (\lambda_{XL}^1 )^3 \big[ 2 H ( m_X, m_X , m_{\tilde{L}} , m_{\tilde{L}} )
+ 2 H ( m_L , m_L , m_{\tilde{x}} , m_{\tilde{x}} \big] \\
&& \qquad\qquad
+ \lambda_L^{1i} \lambda_L^{2i} \lambda_L^{1j} \lambda_L^{1j}
\big[
2H(m_L,m_L,m_{\tilde{e}^{c\,i}},m_{\tilde{e}^{c\,j}})
+2H(m_{e^i},m_{e^j},m_{\tilde{L}},m_{\tilde{L}})
\big]
\Bigg]\,,
\nonumber \\
A_{RR}^{(L)} &=& \frac{\lambda_L^{i1} \lambda_L^{i2} \lambda_L^{j1} \lambda_L^{j1}}{64\pi^2}
\Bigg[
2 H (m_L,m_L,m_{\tilde{e}^i},m_{\tilde{e}^j})
+ 2 H (m_{e^i}, m_{e^j}, m_{\tilde{L}}, m_{\tilde{L}}) \nonumber \\
&& \qquad \qquad \qquad + 2 H (m_{L}, m_{L}, m_{\tilde{\nu}^i}, m_{\tilde{\nu}^j} ) 
+ 2 H (m_{\nu^i}, m_{\nu^j}, m_{\tilde{L}}, m_{\tilde{L}} )
\Bigg]\,, 
\nonumber \\
A_{RL}^{(L)} &=& \frac{\lambda_L^{1i} \lambda_L^{2i} \lambda_L^{j1} \lambda^{j1}}{64\pi^2}
\big[ H(m_{e^i}, m_{e^j}, m_{\tilde{L}}, m_{\tilde{L}})
+ H(m_L,m_L, m_{\tilde{e}^{c\,i}}, m_{\tilde{e}^{c\,j}}) \big] \nonumber \\
&& \quad -\frac{\lambda_{XL}^1 \lambda_{XL}^2 \lambda_L^{i1} \lambda_L^{i1} }{32\pi^2} m_L^2 K (m_L, m_L, m_{\tilde{x}}, m_{\tilde{\nu}^i} )\,, \nonumber \\
A_{LR}^{(L)} &=& \frac{\lambda_L^{i1}\lambda_L^{i2}\lambda_L^{1j}\lambda_L^{1j}}{64\pi^2}
\big[ H(m_{e^i},m_{e^j},m_{\tilde{L}},m_{\tilde{L}}) 
+H(m_L,m_L,m_{\tilde{e}^i},m_{\tilde{e}^j})\big] \nonumber \\
&& \quad -\frac{\lambda_L^{i1}\lambda_L^{i2}\lambda_{XL}^1\lambda_{XL}^1}{32\pi^2} m_L^2 K (m_L, m_L, m_{\tilde{\nu}^i},m_{\tilde{x}})\,. \nonumber
\earray   
For the model $(E)$, 
\barray
A_{LL}^{(E)} &=& \frac{\lambda_{E}^{13} \lambda_{E}^{23} \lambda_E^{1j} \lambda_E^{1j} }{64\pi^2}
\big[ 2 H (m_{\nu^3}, m_{\nu^j}, m_{\tilde{E}}, m_{\tilde{E}} )
+ 2 H ( m_E, m_E, m_{\tilde{\nu}^3}, m_{\tilde{\nu}^i} ) \big]\,, \\
A_{RR}^{(E)} &=& \frac{\lambda_{XE}^2 (\lambda_{XE}^1)^3 }{64\pi^2}
( 2 H ( m_X, m_X, m_{\tilde{E}}, m_{\tilde{E}}) + 2 H ( m_E, m_E, m_{\tilde{x}}, m_{\tilde{x}} ) )\,, \nonumber \\
A_{RL}^{(E)} &=& -\frac{\lambda_E^{13} \lambda_E^{23} \lambda_{XE}^1 \lambda_{XE}^1}{32\pi^2} m_E^2 K ( m_E, m_E, m_{\tilde{x}}, m_{\tilde{\nu}^3} )\,, \nonumber \\
A_{LR}^{(E)} &=& -\frac{\lambda_{XE}^1 \lambda_{XE}^2 \lambda_E^{1j} \lambda_E^{1j}}{32\pi^2} m_E^2 K ( m_E, m_E, m_{\tilde{x}}, m_{\tilde{\nu}^j} )\,, \nonumber
\earray
where the index $j$ runs over only the second and the third generation since the indices in $\lambda_E$ couplings are antisymmetric.

We summarize the $\mu^- \rightarrow e^- e^+ e^-$ constraints in Table \ref{table:mu_to_eee} under the assumption $m_X \sim m_{\tilde{x}} \ll m_{\rm soft} \ll m_D, m_L, m_Q, m_E$. For the $X \ell \ell e^c$ model, the couplings in the table are
\barray
&& \Bigg[ \left(\lambda_{XL}^2 (\lambda_{XL}^1)^3 + (\lambda_L^{1i}\lambda_L^{2i})^2 \right)^2 
 + 4 \left(\lambda_L^{i1}\lambda_L^{i2} \lambda_L^{j1} \lambda_L^{j1} \right)^2 \nonumber \\ 
&&  + \frac{1}{4} \left(\lambda_L^{1i}\lambda_L^{2i} \lambda_L^{j1} \lambda_L^{j1} 
- \lambda_{XL}^1 \lambda_{XL}^2 \lambda_L^{i1} \lambda_L^{i1} R_L \right)^2 
+ \frac{1}{4} \left( \lambda_L^{i1} \lambda_L^{i2} \lambda_L^{1j} \lambda_L^{1j}
  - \lambda_{L}^{i1} \lambda_{L}^{i2} \lambda_{XL}^1 \lambda_{XL}^1 R_L \right)^2 \Bigg]^{1/2}
\label{eq:xlle_coupling1} 
\earray 
for the model $(L)$ and 
\barray 
&& \Bigg[ 
\left(\lambda_E^{13} \lambda_E^{23} \lambda_E^{1j} \lambda_E^{1j} \right)^2  
+ \left(\lambda_{XE}^2 (\lambda_{XE}^1)^3 ) \right)^2
 \nonumber \\
&& + \frac{1}{4} \left( \left( \lambda_E^{13} \lambda_E^{23} \lambda_{XE}^1 \lambda_{XE}^1 \right)^2  
+ \left( \lambda_{XE}^1 \lambda_{XE}^2 \lambda_E^{1i} \lambda_E^{1i} \right)^2 \right) R^2_E
\Bigg]^{1/2}
\label{eq:xlle_coupling2}
\earray   
for the model $(E)$.

\section{Decay Through Dimension Five Effective Operators \label{sec:decay_formula}} 

\begin{figure}
\centering
\subfigure[~Three-body decay]{\includegraphics[width=5cm]
{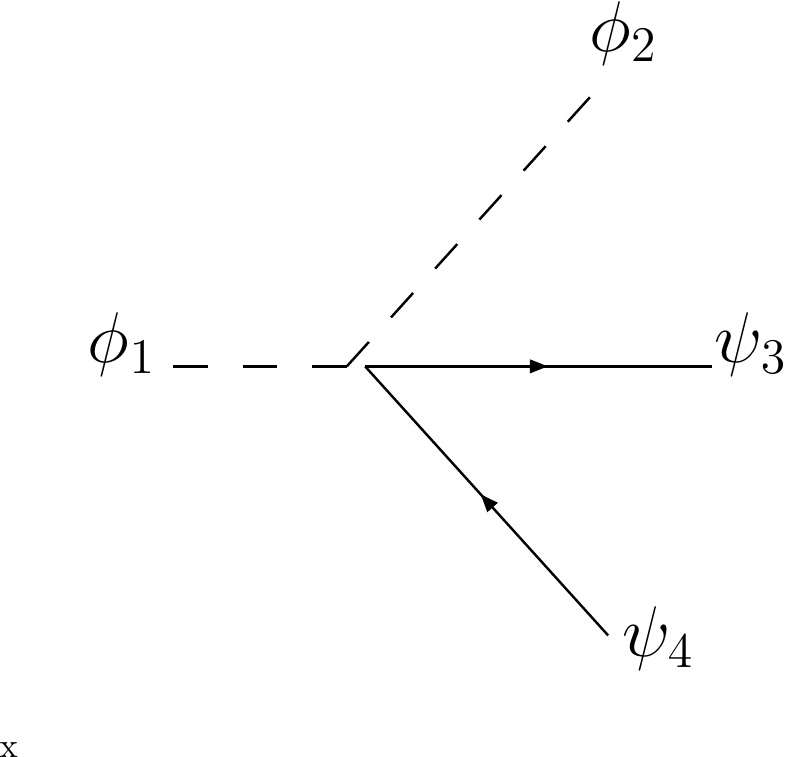}}
\qquad\qquad
\subfigure[~Four-body decay]{\includegraphics[width=5cm]{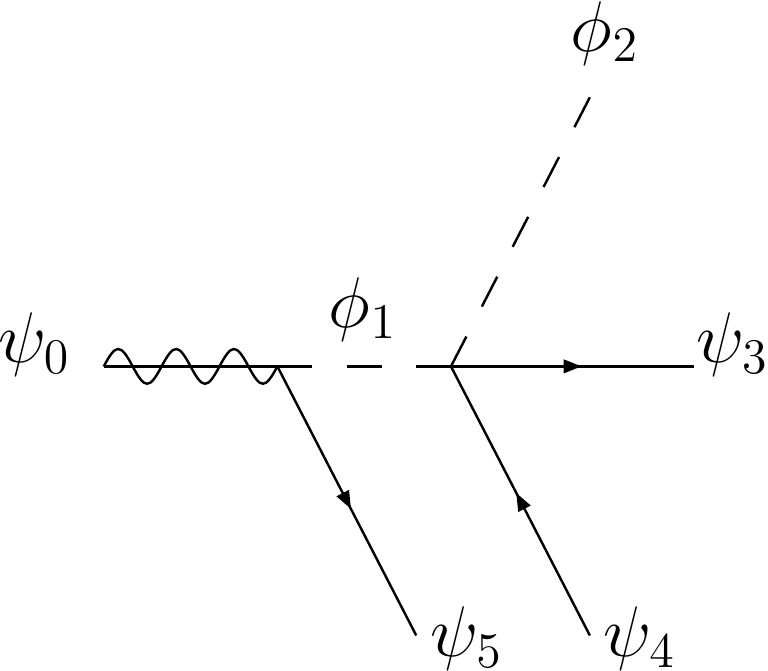}}
\caption{\label{fig:feynman_general_decay} General 3-body and 4-body decays of the LOSP in ADM models.  Special cases are depicted in Fig. \ref{fig:feynman_xqld_decay}. Here, we denote scalar fields by $\phi$ and fermion fields by $\psi$. } 
\end{figure}

This appendix summarizes the calculation of the LOSP decay width through dimension-five operators in the effective Lagrangian in ADM models. We consider three-body decays through a contact interaction and four-body decays through an off-shell intermediate particle as shown in Fig.~\ref{fig:feynman_general_decay}. 

The effective dimension-four superpotential operators of ADM models, which yields a dimension-five Lagrangian, are generically of the form 
\begin{eqnarray}
W_{\rm eff} = \frac{d_{IJKL}}{\Lambda} \Phi^I_1 \Phi^J_2 \Phi^L_3 \Phi^L_4
= \frac{\lambda_{ijkl} c_{abcd} }{\Lambda} \Phi^{ia}_1 \Phi^{jb}_2 \Phi^{kc}_3 \Phi^{ld}_4, 
\label{effectiveoperator}
\end{eqnarray}
where $I=(i,a)$, $J=(j,b)$, $K=(k,c)$ and $L=(l,d)$ represent the flavor indices $i,j,k,l$ and 
the gauge indices $a,b,c,d$ respectively for the corresponding chiral superfields\footnote{In this generic calculation, we treat $\Phi_1$, $\Phi_2$, $\Phi_3$ and $\Phi_4$ as distinct fields. If some fields are the same, we need to compensate the resultant formulae by an appropriate symmetry factor in the definition of $W_{\rm eff}$.}. $d_{IJKL}$'s are the coefficient of the superpotential term, which is factorized by a flavor-dependent coefficient $\lambda_{ijkl}$ and a purely gauge-group dependent Clebsh-Gordon coefficient $c_{abcd}$.

The three- and four-body decays of an $R$-parity odd scalar $\phi_1$ or an $R$-parity odd fermion $\psi_0$ are shown in Fig. \ref{fig:feynman_general_decay} (identical to the diagram Fig.~\ref{fig:feynman_xqld_decay}, though with the particles labeled now with numerical subscripts for notational clarity in what follows). We parameterize the ordinary MSSM interaction
among $\psi_0$, $\phi_1$ and $\psi_5$ by
\begin{eqnarray}
\Delta \mathcal L &=& -(g_1)_{IJK} \, \phi^I_1 \overline{ \Psi_0}^J P_L \Psi^K_5 - (g_2)_{IJK} \, \phi^I_2 \overline {\Psi_0}^J P_R \Psi^K_5 + \{\rm h.c.\} \nonumber \\
 &=&  -(y_1)_{ijk} \tilde{c}_{abc} \, \phi^{ia}_1 \overline{ \Psi_0}^{jb} P_L \Psi^{kc}_5 - (y_2)_{ijk} \tilde{c}_{abc} \, \phi^{ia}_2 \overline {\Psi_0}^{jb} P_R \Psi^{kc}_5 + \{\rm h.c.\}  
\label{eq:Psi0interaction}
\end{eqnarray}
where $\psi_3$, $\psi_4$ and $\psi_5$ are SM fermions, and $g_1$ and $g_2$ are the coefficients.  Here, we again use the collective notation for gauge and flavor indices using $I = (i,a)$, $J=(j,b)$ and $K=(k,c)$. $g_1$ and $g_2$ are factorized into a flavor-dependent coupling $y_1$ and $y_2$, and a Clebsch-Gordon coefficient $\tilde{c}$ for the gauge group.  
We will take $\psi_3$, $\psi_4$ and $\psi_5$  to be massless since they are SM fermions, ignoring top quarks in the final state. $\phi_2$ will be the scalar particle of an ADM chiral multiplet $X$. In natural ADM scenarios, the mass of $\phi_2$ will be around 10 GeV. We can additionally simplify resultant expressions if we treat $\phi_2$ as being massless.

\subsection{\label{sec:directdecay}Three-Body Decay Through a Contact Interaction} 

The spin-averaged amplitude square for the process of Fig. \ref{fig:feynman_general_decay}a is given by 
\begin{eqnarray}
\overline{ |{\mathcal M}|^2 } = \frac{1}{N_1} \sum_{I,J,K,L} 
\frac{|d_{IJKL}|^2}{\Lambda^2} ( 4 p_3 \cdot p_4 ), 
\end{eqnarray}
where $d_{IJKL}$ is the coefficient of the effective superpotential operator in Eq.~(\ref{effectiveoperator}),  and $I, J, K$ and $L$ denote the gauge and flavor indices collectively for $\phi_1$, $\phi_2$, $\psi_3$ and $\psi_4$, respectively. Here, the notation $p_X$ implies the momentum of particle $X$.  We average the amplitude squared over the initial states of the decayed particle, so we have  the number of internal degrees of freedom $N_1$ of $\phi_1$ in the denominator. For example, if $\phi_1$ is a color-triplet $SU(2)$-doublet scalar particle, $N_1$ is $3\times 2 = 6$. Then, the differential decay width for the three-body decay process can be expressed in terms of invariant masses:
\begin{eqnarray}
d\Gamma = \frac{1}{(2\pi)^3} \frac{1}{32 m_1^3} \, \overline {|{\mathcal M}|^2}
\, dm_{23}^2 \, dm_{34}^2, 
\end{eqnarray} 
where $m_{23} = (p_2 + p_3)^2$ and $m_{34} = (p_3 + p_4)^2$. The limits of integration for obtaining the total decay width is determined by the kinematic constraints on the system. In general cases with arbitrary masses, the integration domain is represented by a Dalitz plot.  Considering only $m_3 = m_4 = 0$, the domain for $m_{12}$ and $m_{34}$ are given by 
\begin{eqnarray}
&& \left(m_{23}^2\right)_{\rm min}=m_2^2, \qquad
\left(m_{23}^2\right)_{\rm max}=m_1^2.  \\
&& \left(m_{34}^2\right)_{\rm min} = 0, \qquad
\left(m_{34}^2\right)_{\rm max} = \frac{ (m_{23}^2 - m_2^2) (m_1^2 - m_{23}^2 ) }{m_{23}^2}, 
\end{eqnarray}
By integrating over the domain, we obtain the total decay width 
\begin{eqnarray}
&& \Gamma = \frac{C_{SU(2)} C_{SU(3)} }{(128 \pi^3 N_1) \, (m_1^3 \Lambda^2)} 
\left( \sum_{i,j,k,l} \left|\lambda_{ijkl}\right|^2  \right)
\times   \label{eq:Gamma_3_body} \\
&& \times 
\left[\frac16 (m_1^2 - m_2^2 ) (m_1^4 + 10 m_1^2 m_2^2 + m_2^4 ) 
- m_1^2 m_2^2 (m_1^2 + m_2^2 ) \log (m_1^2/m_2^2) \right]. \nonumber
\end{eqnarray}
Note that we factorize the coupling factor $\sum_{i,j,k,l} \left|d_{ijkl}\right|^2$ into the flavor-dependent coupling squared $\left( \sum_{i,j,k,l} \left|\lambda_{ijkl}\right|^2  \right)$ and the group theoretical factor $C_{SU(2)} C_{SU(3)}$ assuming only $SU(2)$ and $SU(3)$ groups are relevant.  In Table \ref{groupfactor1}, we summarize $C_{SU(2)}$ and $C_{SU(3)}$ for various possible combinations of the representations of participating particles. 

\begin{table} 
\begin{tabular}{|c|c||c|c|}
\hline 
  $(R_1,R_2,R_3,R_4)^{SU(2)}$ & $C_{SU(2)}$ &
  $(R_1,R_2,R_3,R_4)^{SU(3)}$ & $C_{SU(3)}$ \\
 \hline
 $(1,1,1,1)$ & 1 & $(1,1,1,1)$ & 1 \\ 

 $(1,1,2,2)$ \& perm & $\epsilon_{ab} \epsilon^{ab} = 2$ &  $(3,\overline{3},1,1)$ \& perm & $\delta_a^{\;b} \delta_b^{\;a} = 3$ 
\\
 $(2,2,2,2)$ \& perm & $\epsilon_{ab} \epsilon_{cd} \epsilon^{ab} \epsilon^{cd} = 4$ &  
  $(3,\overline{3},3,\overline{3})$ \& perm & $\delta_a^{\;b} \delta_c^{\;d} \delta_b^{\;a} \delta_d^{\;c} = 9$ \\
 & & $(3,3,3,1)$ \& perm & $\epsilon_{abc} \epsilon^{abc} =6$ \\
\hline 
\end{tabular}
\caption{\label{groupfactor1} Group theoretical factors for 3-body decay through a contact interaction. Here, $\delta^a_b$ is the Kronecker delta, $\epsilon_{ab}$ and  $\epsilon_{abc}$ are Levi-Civita symbols. Note that the overall switch between complex representations $3$ and $\overline{3}$ does not change the factor.}
\end{table}

\subsection{Four-Body Decay Through An Intermediate Off-Shell Particle}

\begin{table} 
\begin{tabular}{|c|c||c|c|}
\hline
  \multicolumn{2}{|c||}{$SU(2)$}  &  \multicolumn{2}{|c|}{$SU(3)$} \\
\hline 
  $(R_0,R_5,R_1,R_2,R_3,R_4)$ & $D_{SU(2)}$ &
  $(R_0,R_5,R_1,R_2,R_3,R_4)$ & $D_{SU(3)}$ \\
 \hline
 $(1,1,1,1,1,1)$ & 1 & $(1,1,1,1,1,1)$ & 1 \\
 $(1,1,1,\underline{2,2,1})$ \& perm & $\epsilon_{bc} \epsilon^{bc} = 2$ & 
$(1,1,1,\underline{3,\overline{3},1})$ \& perm & $\delta_b^{\;c} \delta_c^{\;b} = 3$ \\
 $(2,2,1,1,1,1)$ & $\epsilon_{ef} \epsilon^{ef} = 2$ & 
$(1,1,1,3,3,3)$ & $\epsilon_{bcd} \epsilon_{bcd} = 6$ \\
 $(2,2,1,\underline{2,2,1})$ \& perm & $\epsilon_{ef} \epsilon^{ef} \epsilon_{bc} \epsilon^{bc} = 4$ &
 $(3,\overline{3},1,1,1,1)$ & $\delta_e^{\;f} \delta_f^{\;e} = 3$ \\
 $(\underline{2,1},2,\underline{2,1,1})$ \& perm & $\epsilon_{fa} \epsilon^{fa'} \epsilon^{ab} \epsilon_{a'b} = 2$ & 
$(3,\overline{3},1,\underline{3,\overline{3},1})$  \& perm &
$\delta_e^{\;f} \delta_f^{\;e} \delta_b^{\;c} \delta_c^{\;b} = 9 $  \\
 $(\underline{2,1},\underline{2,2,2,2})$ \& perm & $\epsilon_{fa} \epsilon^{fa'} \epsilon^{ab} \epsilon^{cd} \epsilon_{a'b} \epsilon_{cd} = 4$ & 
$(3,\overline{3},1,3,3,3)$ & $\delta_e^{\;f} \delta_f^{\;e} \epsilon_{bcd} \epsilon^{bcd} = 18$ \\
$(3,\overline{2},2,\underline{2,1,1})$ \& perm & $(T^e)_a^{\;f} (T^e)_f^{\;a'}  \epsilon^{ab} \epsilon_{a'b} = \frac{3}{2} $ & 
$(\underline{3,1},\overline{3},\underline{3,1,1})$ \& perm & 
$\delta_{f}^{\;a} \delta_{a'}^{;f} \delta_a^{\;b} \delta_b^{\;a'} = 3$ \\
$(3,\overline{2},\underline{2,2,2,2})$ \& perm & 
$(T^e)_a^{\;f} (T^e)_f^{\;a'} \times ~~~~~~ $  & 
 $(\underline{3,1},\overline{3},\underline{\overline{3},\overline{3},1})$ \& perm &
$\delta_{f}^{\;a} \delta_{a'}^{\;f} \epsilon_{abc} \epsilon^{a'bc} = 6$ \\
& $\times \epsilon^{ab} \epsilon^{cd} \epsilon_{a'b} \epsilon_{cd} = 3$ & 
$(\underline{3,1},\underline{\overline{3},3,\overline{3},3})$ \& perm & 
$\delta_f^{\;a} \delta_{a'}^{\;f} \delta_a^{\;b} \delta_c^{\;d} \delta_b^{\;a'} \delta_d^{\;c} = 9$ \\
 & & $(\overline{3},\overline{3},\overline{3},\underline{3,1,1})$ \& perm & 
$\epsilon^{fea} \epsilon_{fea'} \delta_a^{\;b} \delta_b^{\;a'} = 6$ \\
 & & $(\overline{3},\overline{3},\overline{3},\underline{\overline{3},\overline{3},1})$ \& perm & 
$\epsilon^{fea} \epsilon_{fea'} \epsilon_{abc} \epsilon^{a'bc} = 12$ \\
 & & $(\overline{3},\overline{3},\underline{\overline{3},3,\overline{3},3})$ \& perm &
$\epsilon^{fea} \epsilon_{fea'} \delta_a^{\;b} \delta_c^{\;d} \delta_b^{\;a'} \delta_d^{\;c} = 18$ \\
 & & $(8,3,\overline{3},\underline{3,1,1})$ \& perm & 
 $(T^e)_a^{\;f} (T^e)_f^{\;a'} \delta_b^{\;a} \delta_{a'}^{\;b} = 4$ \\
 & &  $(8,3,\overline{3},\underline{\overline{3},\overline{3},1})$ \& perm  & 
$(T^e)_a^{\;f} (T^e)_f^{\;a'} \epsilon^{abc} \epsilon_{a'bc} = 8$ \\
 & & $(8,3, \underline{\overline{3},3,\overline{3},3})$ \& perm  &
$(T^e)_a^{\;f} (T^e)_f^{\;a'} \delta_b^{\;a} \delta_c^{\;d} \delta_{a'}^{\;b} \delta_d^{\;c} = 12$ \\
\hline
\end{tabular}
\caption{\label{groupfactor2} Group theoretical factors for 4-body decay through an intermediate (off-shell) particle.  $(T^A)_{ab}$ is $\sigma^A_{ab} /2$ for $SU(2)$ and  $\lambda^A_{ab} / 2$  for $SU(3)$, where $\sigma^a$'s and $\lambda^a$'s are Pauli and Gell-Mann matrices, respectively.
 Permutations are defined only within underlined items (if underline is disconnected, they are two separate permutation sets). 
Switching $3$ with $\overline{3}$ altogether leads to the same factor.   }
\end{table}

Next we consider the case of Fig. \ref{fig:feynman_general_decay}b.  
The spin-averaged amplitude square is 
\begin{eqnarray}
\overline{|{\mathcal M}|^2}  = 
\frac{1}{2N_0} \sum_{I,I',J,K,L,M,N} \left[ 
(g_1^* g'_1 + g_2^* g'_2) \frac{d d^{\prime *}}{\Lambda^2} \right]
\left( \frac{1}{q^2 - m_1^2} \right)^2 (4 p_0 \cdot p_5 ) (4 p_3 \cdot p_4), 
\end{eqnarray}
where the summation over $I = (i,a)$ and $I'=(i',a')$ is from the intermediate $\phi_1$ exchange, and  $J=(j,b),K=(k,c),L=(l,d),M=(m,e),N=(n,f)$ denote the 
 collective flavor indices $i,j,\dots$ and gauge indices $a,b,\dots$ of the external particles $\phi_2$, $\psi_3$, $\psi_4$, $\psi_0$ and $\psi_5$, respectively. $q = p_2+p_3+p_4$ is the momentum of the intermediate $\phi_1$. Here, we use abbreviations $g_1 = (g_1)_{IMN}$, $g'_1 = (g_1)_{I'MN}$ (and $g_2$ and $g'_2$ in the same way), $d =d_{IJKL}$ and $d' = d_{I'JKL}$, where $d$'s and $g$'s are defined in Eq.~(\ref{effectiveoperator}) and Eq.~(\ref{eq:Psi0interaction}). $N_0$ represents the number of internal degrees of freedom of $\psi_0$, as in the 3-body decay case.

With similar tricks to the 3-body decay case by using the invariant masses of subsystems as integration variables and decomposing the 4-body phase space (PS) integration into the 3-body and the 2-body PS integrations, one can get the analytic formula for the full decay width in the following integral form: 
\begin{eqnarray}
& & \Gamma = \frac{1}{3 \cdot 2^{10} \pi^5 N_0}\frac{1}{ \Lambda^2 m_0^3} \left( \sum_{I,I',J,K,L,M,N} (g_1^* g'_1 + g_2^* g_2' ) (d d^{\prime *})\right)
 \times  \\
& & \times 
 \int_{m_2^2}^{m_0^2} dq^2 
\; \frac{(m_0^2 - q^2)^2}{(q^2 - m_1^2)^2} \frac{1}{q^2} 
 \left[ 
 \frac16 (q^2- m_2^2) (q^4 + 10 q^2 m_2^2 + m_2^4 ) 
 - q^2 m_2^2 ( q^2 + m_2^2 ) \log (q^2 / m_2^2) 
 \right]. \nonumber 
\label{phasespaceint} 
\end{eqnarray} 

For nonzero $m_2$, the analytic result from the integration in Eq.~(\ref{phasespaceint}) is rather complicated.  For the most non-degenerate cases where SUSY particle mass difference is larger than the typical ADM mass ($\sim 10$ GeV), we can safely assume that $m_2 = 0$. In such cases,  the total decay width has a simplified form: 
\begin{eqnarray}
&& \Gamma  =  \frac{D_{SU(2)} D_{SU(3)}}{3 \cdot 2^{10} \pi^5  N_0} \frac{1}{\Lambda^2 m_0^3}
\left[
\sum_{i,i',j,k,l,m,n} 
\left( (y_1^*)_{i m n} (y_1)_{i' m n} + (y_2^*)_{i m n} (y_2)_{i' m n} \right) 
\left( \lambda_{i j k l} \lambda^*_{i' j k l} \right)
\right]
 \times \nonumber \\
&& 
\times \left[ \frac13 m_0^2 (m_0^4 - 12 m_0^2  m_1^2 + 12 m_1^4 ) 
+ 2 m_1^2 (m_1^2 - m_0^2 ) (2 m_1^2 - m_0^2) 
\log \left( \frac{m_1^2 - m_0^2}{m_1^2}  \right) \right], 
\label{eq:Gamma_4_body}
\end{eqnarray}
where we factorize the coupling factor $\sum (g_1^* g'_1 + g_2^* g'_2) (d d^{\prime *})$ into the flavor-dependent coupling squared in terms of $y_1$ and $y_2$ defined in Eq.~(\ref{eq:Psi0interaction}) with explicit flavor indices and the group theoretical factor $D_{SU(2)} D_{SU(3)}$, assuming only $SU(2)$ and $SU(3)$ groups are relevant again. We summarize $D_{SU(2)}$ and $D_{SU(3)}$ for various combinations in Table \ref{groupfactor2}.  One can check that $\Gamma \sim \frac{m_0^7}{\Lambda^2 m_1^4} $ in the limit $m_1 \gg m_0$, as expected from a simple dimensional argument.

\section{The ATLAS Analysis Observables}
\label{sec:analysis_observable}

We summarize the experimental observables used for the ATLAS 0 lepton+2-6 jet+MET and 1-2 lepton + 3-6 jet + MET analyses in the following. 

\begin{itemize}
\item $\vec{p}_T^{\,\rm miss}$ : Missing Transverse Momentum, the negative vector sum of the transverse momentum $p_T$'s of identifiable objects.
\item $E_T^{\rm miss} = \left| \vec{p}_T^{\,\rm miss} \right|$ : Missing Transverse Energy. 
\item $\vec{p}_T (j_i)$ : the transverse momentum $\vec{p}_T$ of the $i$-th hardest jet in $p_T$ size ordering. Without $\rightarrow$, it implies the magnitude.  
\item $\Delta \phi ({\rm obj}, E_T^{\rm miss})$ : Azimuthal angle between $\vec{p}_T$ of a given object (jet or lepton) and $\vec{p}_T^{\rm miss}$. 
\item $m_{\rm eff} (n j)$ : Effective Mass with the hardest $n$ jets in $p_T$ size ordering. $m_{\rm eff} (n j) = \sum_{\ell}  p_T (\ell) + \sum_{i = 1, \dots, n} p_T (j_i) + E_T^{\rm miss}$ including all signal leptons. In the 0 lepton analysis, $m_{\rm eff} (N j)$ means $N =2,3,4,5,6$ for channel A,B,C,D and E, respectively. In the 1-2 lepton analysis, $m_{\rm eff}^{\rm excl.}$ is defined similarly.

\item $m_{\rm eff} ({\rm incl.})$ : Inclusive Effective Mass. Effective mass defined with all jets with $p_T > 40 $ GeV for the 0 lepton analysis, or with all signal jets for the 1 lepton analysis. 
\item $N_{\rm b-tag}$ : Number of $b$-tagged jets. 
\item $m_T$ : Transverse Mass of the lepton $\ell$ (for a single lepton event) and $\vec{p}_T^{\rm miss}$, 
$m_T = \sqrt{ 2\,  p_T^{\ell} \, E_T^{\rm miss} \, \left[ 1 - \cos \Delta \phi (\ell , E_T^{\rm miss})  \right]}$ 
\item $\Delta R_{\rm min} ({\rm jet},\ell)$ : The minimum of $\Delta R = \sqrt { \Delta \eta^2 + \Delta \phi^2 }$ between the lepton $\ell$ (for a single lepton event) and each signal jet. 
\item $\Delta \phi_{\rm min} = {\rm min} ( \Delta \phi (j_1, E_T^{\rm miss}) , \Delta \phi (j_2, E_T^{\rm miss} ))$, where $j_1$ and $j_2$ are the first and second hardest jets, respectively. 
\item $m_{CT}$ : Contransverse Mass of the two $b$-jets (for 2 $b$-jet events) defined by $m_{CT}^2 (b_1, b_2) = [ E_T^{b_1} + E_T^{b_2} ]^2 - [ \vec{p}_T^{\,b_1} - \vec{p}_T^{\,b_2} ]^2 $.

\end{itemize}


\section{\texttt{evchain}: Subprocess Chaining for Event Generation}
\label{sec:analysispipeline}

The event generation that is required for the analyses we have carried out in this paper has a few technical challenges. As we see in Figs. \ref{fig:collider_for_0lep_xudd_squarklsp} and \ref{fig:collider_for_0lep_xudd_neutralinolsp}, the cascade decay from the gluino and/or the squarks gives rise to a large number of outgoing particles at parton level, so that the phase space becomes very high-dimensional.
 In addition, the LOSP decays through a non-renormalizable interaction, and especially the neutralino LOSP leads to a 4-body decay through an intermediate off-shell squark or slepton. We also have exotic color vertices that involve color index contraction with the invariant tensor $\epsilon^{ijk}$ ($i,j,k$'s are color indices) in the case of the $u^cd^cd^c$ models. 

Such technical challenges strongly restrict the choice of available tools. As of now, non-renormalizable interactions and exotic color vertices can be treated successfully by using \texttt{MadGraph5} \cite{Alwall:2011uj}. However, \texttt{MadGraph5} generates events by a Monte Carlo integration of the matrix element over the full phase space, and with a higher multiplicity of final state particles, the integration often leads to unbearably slow performance and a big accumulation of error. This problem becomes worse if we have several on-shell particles in the process with narrow decay widths since more careful sampling near on-shell poles is needed for a given required accuracy, which will take more sampling iterations and will thus be more prone to numerical errors.

Therefore, it is much more desirable to generate events by splitting a process into a few subprocesses: production channels and decay modes, and to connect the subprocesses into a single big event by doing appropriate transformation. Many Monte Carlo event generators indeed do the job in this way. For \texttt{MadGraph5}, an external tool called \texttt{BRIDGE} is designed to address this issue \cite{Meade:2007js}. However, as far as we know, the \texttt{BRIDGE} tool is restricted to 2-body or 3-body decays for each decay subprocess, and it is not clear whether the tool has been actively maintained with the recent rapid changes of \texttt{MadGraph5}.

\begin{figure}
\centering 
\includegraphics[width=16cm]{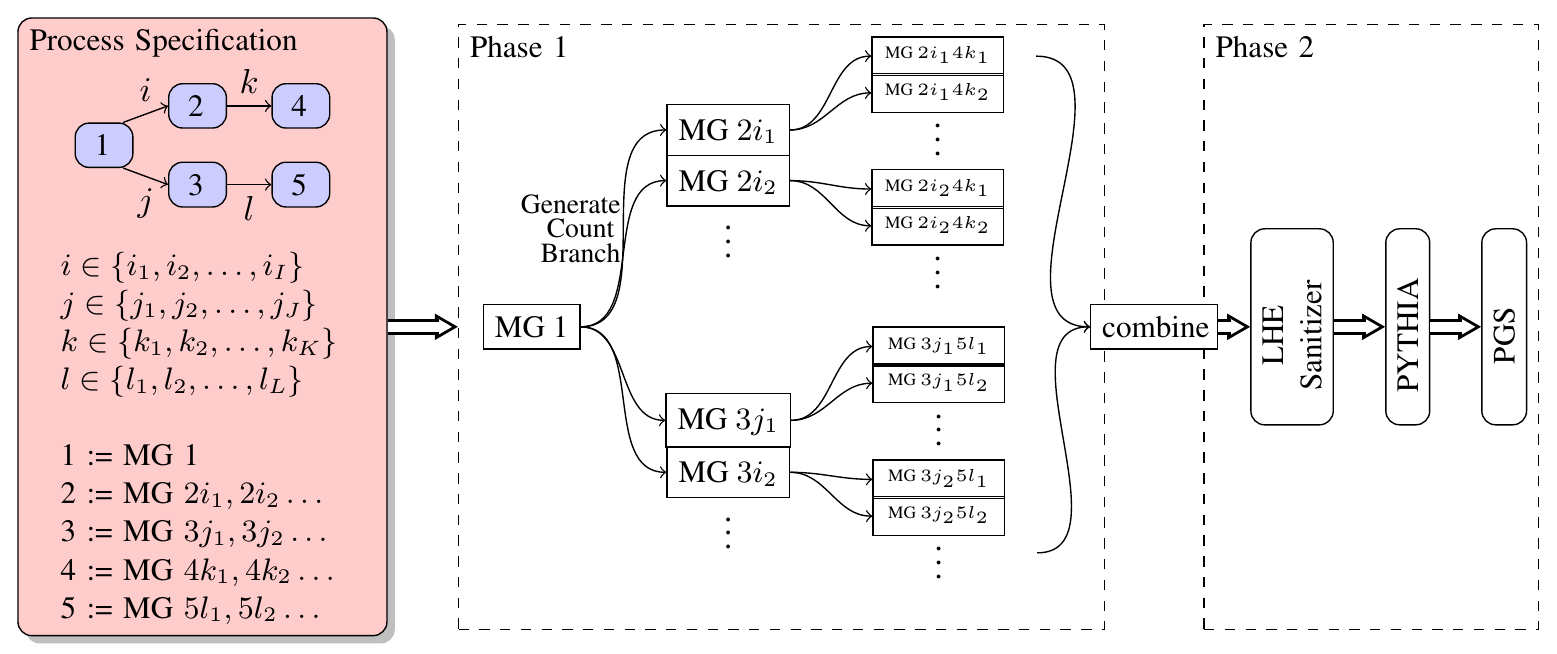}
\caption{\texttt{evchain} pipeline} 
\label{fig:evchain_pipeline}
\end{figure}

We address this difficulty by creating our own in-house tool called \texttt{evchain} \cite{evchain}: an event chaining tool that automatically orders \texttt{MadGraph5} event generation for each subprocess and combines resultant Les Houches Event (LHE) format files \cite{Alwall:2006yp} into a single LHE format file by doing appropriate Lorentz transformations and color flow number adjustments. Although the current version is tightly incorporated with \texttt{MadGraph5}, the general idea of \texttt{evchain} is not restricted to \texttt{MadGraph5} since we treat each subprocess as a module with an interface of incoming and outgoing particles. Insofar as incoming and outgoing particle types are matched, any event generator with any specific process can be used for generating each subprocess. We also note that we do not aim to provide an automatic decay width calculation, differently from \texttt{BRIDGE}. The total decay width must be provided by \texttt{MadGraph5} or equivalent, while a relative branching ratio in one specific subprocess is automatically given by actual event generation. By this design choice, we simplified program requirements and we were able to generalize easily to any N-body decay processes. In the following, we describe the tool in more detail. 

\begin{figure}
\centering
\begin{code}
gluino = [1000021]
neutralino = [1000022]
jets = [1,2,3,4,-1,-2,-3,-4,21]
lepton_and_neutrino = [11,12,13,14,-11,-12,-13,-14]
adms = [9000201,-9000201,9000202,-9000202]

decay_gluino :: DDecay 
decay_gluino = d (gluino, [decay_neutralino, t jets, t jets])

decay_neutralino :: DDecay 
decay_neutralino = d (neutralino, [t lepton_and_neutrino, t jets, t jets, t adms])

total_process :: DCross 
total_process = x (t proton, t proton, [decay_gluino, decay_gluino])

madgraph_process_map :: ProcSpecMap
madgraph_process_map = 
    fromList [ (Nothing             , MGProc [] [ "p p > go go QED=0" ])
             , (Just (3,1000021,[]) , MGProc [] [ "go > n1 j j " ] ) 
             , (Just (4,1000021,[]) , MGProc [] [ "go > n1 j j " ] )
             , (Just (1,1000022,[3]), MGProc [ "define lep = e+ e- mu+ mu- ve ve~ vm vm~ "
                                             , "define sxx = sxxp sxxp~ " ] 
                                             [ "n1 > sxx lep j j " ] )
             , (Just (1,1000022,[4]), MGProc [ "define lep = e+ e- mu+ mu- ve ve~ vm vm~ "
                                             , "define sxx = sxxp sxxp~ " ]
                                             [ "n1 > sxx lep j j " ] ) ] 
\end{code}
\caption{A haskell code example of \texttt{evchain} process specification for the $\tilde{g}-\tilde{g}$ production of the $q \ell d^c$ model with neutralino LOSP as shown in Fig. \ref{fig:collider_for_0lep_xudd_neutralinolsp}a. Note that we define multiparticles \texttt{jets}, \texttt{lepton\_and\_neutrino} and \texttt{adms} for the sake of convenience. 
}
\label{fig:evchain_specification}
\end{figure}

\texttt{evchain} works as a ``meta-event-generator'' that supervises \texttt{MadGraph5} event generation for subprocesses. In Fig. \ref{fig:evchain_pipeline}, we show the overall pipeline of \texttt{evchain} event generation.  The tool is written in haskell~\cite{haskell}, which is buildable by using \texttt{ghc 7.4} or higher~\cite{ghc}. \texttt{evchain} currently exists in a library form, so a user makes a program executable linked with \texttt{evchain}.   

In the source code of the user's program, the total event process is specified as a haskell tree data structure. The specification language as an Embedded Domain Specific Language (EDSL) for \texttt{evchain} inside the haskell program is self-explanatory. We provide one example of such a specification description in Fig. \ref{fig:evchain_specification}, which is gluino pair-production of the $q \ell d^c$ model with neutralino LOSP as shown in Fig. \ref{fig:evchain_specification}. 
A total process is  a production process module with two incoming particles and arbitrary number of outgoing particles. Each outgoing particle can be either a terminal particle or a decay process, which is a module with one incoming particle and an arbitrary number of outgoing particles, where again an outgoing particle of a decay process is either a terminal particle or a decay process, recursively. 
An incoming/terminal particle is specified by a list of PDG codes, so that we can define a collection of particles as incoming or outgoing particles for convenience. Each subprocess is mapped into \texttt{MadGraph5} processes. In the example, the total process is defined in \texttt{total\_process}, which has \texttt{decay\_gluino}, and \texttt{decay\_gluino} is again defined by \texttt{decay\_neutralino}. 
\texttt{madgraph\_process\_map} defines actual \texttt{MadGraph5} commands for each subprocess.

 When running, the program will first prepare \texttt{MadGraph5} directories for each subprocess. As shown in Fig. \ref{fig:evchain_pipeline}, the on-shell particles (denoted as $i,j,k,l$ in the figure) that connect mother and daughter subprocesses can be multiple particles. \texttt{evchain} automatically prepares for all of the cases as different working directories and avoids a name clash by making different hash numbers for distinct subprocesses and particles. Since the same hash number is produced for the same process specification, the  preparation step can be efficiently done only once for repeating event generations with different parameter sets. \texttt{evchain} provides a configuration method for customizing the directory paths of relevant tools and working directories, which is adjustable for various cluster computing setups. 

After the preparation step, the event generation is done in two stages: (i) generating LHE event files for each subprocess in the order of subprocess dependency, (ii) combining LHE event files into a single LHE file to pass to the rest of the event generator (event file sanitization, parton shower and hadronization using \texttt{PYTHIA}, and detector simulation using \texttt{PGS}). \texttt{evchain} facilitates an event counter and classifier.  In every step after finishing each subprocess event generation, \texttt{evchain} counts the number of outgoing particles, and orders the next dependent subprocess event generation for only the required number as determined by the previous step. Once all of the subprocess event generation is done, the combining routine runs through all events of the root subprocess, and recursively finds events in daughter subprocesses and chains them by adjusting particle numbers and color flow numbers and transforming particle momenta in the daughter process from the rest frame of decayed particle to the mother frame. After this, the total number of events of the resultant LHE file are automatically matched with the number the user specifies, and the LHE file is fed into the rest of the pipeline. 

The \texttt{evchain} tool is designed to fit as a subsystem of \texttt{pipeline}, a cluster job coordinator for common high-energy physics tasks, described in Appendix B. 3 in \cite{Gresham:2011fx}. 
We plan to improve the system with better support for general cluster facilities, at the same time as implementing missing functionality such as spin correlations.


\end{document}